\documentclass[12pt]{article}

\pdfoutput=1
\usepackage{color}
\usepackage{epsfig, palatino}
\usepackage{pstricks,pst-node,pst-tree}
\usepackage{epic}
\usepackage{mathrsfs}
\usepackage{ae} 
\usepackage[T1]{fontenc}
\usepackage[ansinew]{inputenc}
\usepackage{amsmath}
\usepackage{amssymb}
\usepackage{dsfont}
\usepackage{graphicx}

\usepackage{color}
\definecolor{darkblue}{cmyk}{0.9,0.9,0,0}
\definecolor{OliveGreen}{rgb}{0,0.6,0}
\usepackage[colorlinks=true,linkcolor=darkblue,citecolor=darkblue,urlcolor=darkblue]{hyperref}
\usepackage{cite}
\usepackage{hyperref}
\usepackage{wasysym}
\usepackage{makeidx}
\usepackage[english]{babel}

\usepackage[font={small}]{caption}

\newcommand{\comment}[1]{}

\newcommand{\beq}{\begin{equation}}
\newcommand{\eeq}{\end{equation}}
\newcommand{\beqq}{\begin{equation*}}
\newcommand{\eeqq}{\end{equation*}}
\newcommand\beqa{\begin{eqnarray}}
\newcommand\eeqa{\end{eqnarray}}
\newcommand\beqaa{\begin{eqnarray*}}
\newcommand\eeqaa{\end{eqnarray*}}
\newcommand\bea{\begin{array}}
\newcommand\eea{\end{array}}

\newcommand{\neqa}{\nonumber\end{eqnarray}}

\renewcommand{\d}{\partial}

\newcommand{\<}{{\langle}}
\renewcommand{\>}{{\rangle}}

\newcommand{\re}{\relax{\rm I\kern-.18em R}}

\renewcommand{\sp}{p\hspace{-.40em}/}

\definecolor{darkgreen}{rgb}{0.0, 0.45, 0.0}

\def\XXint#1#2#3{{\setbox0=\hbox{$#1{#2#3}{\int}$}
\vcenter{\hbox{$#2#3$}}\kern-.5\wd0}}

\def\su2{{SU(2)}}

\def\[{\left[}
\def\]{\right]}

\def\({\left(}
\def\){\right)}
\def\[{\left[}
\def\]{\right]}

\def\<{\langle}
\def\>{\rangle}

\def\i2{\frac{i}{2}}

\def\spi{\relax{\rm \pi\kern-0.5em /}}
\def\sA{\relax{\rm A\kern-0.5em /}}
\def\sp{\relax{\rm p\kern-0.5em /}}
\def\sd{\relax{\rm \d\kern-0.5em /}}
\def\sk{\relax{\rm k\kern-0.5em /}}
\def\sn{\relax{\rm n\kern-0.5em /}}
\def\sl{\relax{\rm l\kern-0.5em /}}
\def\sP{\relax{\rm P\kern-0.7em /}}
\def\sBethe{\relax{\rm \Bethe\kern-0.5em /}}

\def\2F1{\,_2{\rm F}_1}

\makeindex

\begin{document}

\thispagestyle{empty}

\renewcommand{\thefootnote}{\fnsymbol{footnote}}
\setcounter{page}{1}
\setcounter{footnote}{0}
\setcounter{figure}{0}

\begin{center}
$$$$
{\Large\textbf{\mathversion{bold}
Lightcone Bootstrap at higher points
}\par}
\vspace{1.0cm}

\textrm{Ant\'{o}nio Antunes, Miguel S. Costa, Vasco Gon\c{c}alves,  Jo\~{a}o Vilas Boas}
\\
\vspace{8mm}
\footnotesize{\textit{Centro de F\'{i}sica do Porto e Departamento de F\'{i}sica e Astronomia, Faculdade de Ci\^{e}ncias da Universidade do Porto, Rua do Campo Alegre 687, 4169-007, Porto, Portugal  
}  
\vspace{4mm}
}
\end{center}

\noindent

\setcounter{page}{1}
\renewcommand{\thefootnote}{\arabic{footnote}}
\setcounter{footnote}{0}

\setcounter{tocdepth}{2}

 \def\nref#1{{(\ref{#1})}}

\begin{abstract}
Higher-point functions of scalar operators are a rich observable in CFTs, as they contain OPE data involving multiple spinning operators. 
We derive the lightcone blocks for  five- and six-point functions in the snowflake channel and use them to 
bootstrap these correlators in the lightcone limit. As a result we determine the large spin expansion of OPE
coefficients involving two or three spinning operators. We verify our results by comparing to the block decomposition 
of higher-point functions in generalized free theory and in theories with a cubic coupling.
\end{abstract}

\pagebreak

\tableofcontents

\parskip 5pt plus 1pt   \jot = 1.5ex

\section{Introduction}

Analytic bootstrap methods have given a structural understanding of CFTs by leveraging  the analytic structure of four-point functions \cite{Komargodski:2012ek,Fitzpatrick:2012yx,Simmons-Duffin:2016wlq,Alday:2015ewa,Alday:2015eya,Kaviraj:2015cxa,Kaviraj:2015xsa,Alday:2016njk,Costa:2017twz,Kulaxizi:2017ixa,Li:2017lmh,Meltzer:2017rtf}.
Typically such studies consider the four-point function of scalar operators. This fact  limits the data that can be accessed to scalar/scalar/symmetric traceless (of spin $J$)
OPE coefficients. However, it is important to consider OPE coefficients between multiple spinning operators, of which an important example is the OPE coefficient of three
stress tensors \cite{Hofman:2008ar,Camanho:2014apa}. A possibility would be to extend the analytic  bootstrap to the four-point function of operators with spin, but this approach  
is technically challenging and works mostly in a case by case basis. An alternative is to consider higher-point functions of scalar operators, which through the OPE contains information about
operators of arbitrary spin~\cite{Vieira:2020xfx, Bercini:2021jti}. In this case the technical challenge lies upon our knowledge of higher-point  conformal blocks, which is still incomplete \cite{Rosenhaus:2018zqn,Goncalves:2019znr,PolandValentina:2021,Bercini:2021jti}.

For the scalar four-point function, the lightcone bootstrap predicts the universal behaviour of scalar/scalar/spin $J$ 
OPE coefficients at large spin, which are of mean field type \cite{Komargodski:2012ek,Fitzpatrick:2012yx}.  Subsequent corrections, that include scaling dimensions  and OPE coefficients, 
are determined by the leading twist operators in the theory \cite{Komargodski:2012ek,Fitzpatrick:2012yx}. This large spin expansion is actually convergent up to a low spin value determined by the Regge 
behaviour of the four-point function \cite{Caron-Huot:2017vep,Simmons-Duffin:2017nub}.  A remarkable check of the accuracy of this method was done in the 3D Ising model where the numerical bootstrap 
provided the data for comparison \cite{Simmons-Duffin:2016wlq,Caron-Huot:2020ouj} (see also \cite{Liu:2020tpf} for the $O(2)$ model). Motivated by this success, our goal is to extend the lightcone bootstrap to the 
case of higher-point functions and therefore access OPE data involving  spinning operators.

More concretely, we bootstrap five- and six-point functions. In the five-point case there is an unique OPE topology which 
involves the exchange of two operators of spin $J_1$ and $J_2$ and therefore includes the scalar/spin $J_1$/spin $J_2$
OPE  coefficient, see (\ref{eq:OPEid(34)5-pt}) and (\ref{eq:OPE5ptlt}). In the six-point case we consider the snowflake OPE channel which 
involves the exchange of three operators of spin $J_1$, $J_2$ and $J_3$ and therefore includes the spin $J_1$/spin $J_2$/spin $J_3$
OPE  coefficient, see (\ref{eq:P6pt3indentities}), (\ref{eq:OPEsixpt1identity}) and (\ref{eq:OPE6p3leadtwist}). This bootstrap analysis is done in section \ref{sec:Bootstrap}, which follows section \ref{sec:KinematicsBlocks} where we review the kinematics and derive the lightcone 
conformal blocks for  five- and six-point functions. Our results are tested in section \ref{sec:examples} for the case of generalized free theory and of theories with a cubic coupling, whose
block decomposition we determine explicitly. We conclude with a discussion of  open problems in section \ref{sec:Discussion}. 

Additional technical details are given in the appendices:
appendix \ref{app:Blocks} gives more details on higher-point blocks, including some comments about the Euclidean expansion and the Mellin representation; appendix \ref{app:Dfuncs} 
discusses higher-point $D$-functions 
based on AdS techniques; appendix \ref{app:Harmonic} presents new results on conformal harmonic analysis relevant for higher-point functions and can be read mostly independently from the main text.

\section{Kinematics and conformal blocks}
\label{sec:KinematicsBlocks}

It is a well known property that $n$-point correlation functions in a conformal field theory depend nontrivially on $n(n-3)/2$ conformal invariant variables for high enough spacetime dimension\footnote{There are relations between conformal invariant cross-ratios for low dimensions ($d\leq n-2$) such that the number of independent variables is instead $nd -(d+1)(d+2)/2$.}. 
The choice of  conformal invariant cross-ratios usually depends on the problem one is analysing. In  a four-point function, that depends on two cross-ratios (say $u$ and $v$), there are several choices of cross-ratios used throughout the literature, for example
\begin{equation}
u=z\bar{z}=\frac{x_{12}^2x_{34}^2}{x_{13}^2x_{24}^2}\,, \quad  v=(1-z)(1-\bar{z})=\frac{x_{14}^2x_{23}^2}{x_{13}^2x_{24}^2}\,, \\
\end{equation}
or
\begin{equation}
 s=|z|\,, \quad \xi=\cos \theta=\frac{z+\bar{z}}{2|z|}\,.
 \label{eq:sxi4pt}
\end{equation}

This paper is focused on the analytic bootstrap of five- and six-point correlation functions, and therefore we will need to use appropriate sets of cross-ratios.
For the five-point function 
 it will be convenient to work with the five variables $u_1, \dots ,u_5$ given by
 \begin{equation}
u_1=\frac{x_{12}^2x_{35}^2}{x_{13}^2x_{25}^2}\,, \qquad  u_{i+1}=u_i\big|_{x_{j}\rightarrow x_{j+1} } \,, 
\end{equation}
where in this definition  the subscript in $x_j$ is taken modulo 5 (for example $x_6\equiv x_1$).
For the  six-point function we introduce the nine cross-rations $u_1,\dots u_6$ and $U_1,\dots, U_3$ 
defined by
\begin{align}
&u_1=\frac{x_{12}^2x_{35}^2}{x_{13}^2x_{25}^2}\,, \quad  u_{i+1}=u_i\big|_{x_{j}\rightarrow x_{j+1}}\,,\qquad 
U_1=\frac{x_{13}^2x_{46}^2}{x_{14}^2x_{36}^2}\,, \quad  U_{i+1}=U_{i}\big|_{x_j\rightarrow x_{j+1}}  \,,
\end{align}
where the subscript in $x_j$ is now taken modulo 6.

We will be interested in the Lorentzian lightcone expansion of correlation functions. The difference between the Lorentzian and Euclidean expansions  can be easily understood from the OPE of two operators. In the Euclidean case  the operators are taken to be   coincident ($x_{ij}\to0$) while in the Lorentzian case the operators  approach the lightcone of each other ($x_{ij}^2\to0$). As is well known, the Euclidean limit is dominated by the operators with lowest scaling dimension, in contrast with the 
Lorentzian case that is dominated by the operator with lowest twist $\tau=\Delta-J$. 
This is evident from the leading term of the formula for the  OPE
\begin{align}
&\phi(x_1)\,\phi(x_2) \approx \sum_{k}C_{12k} \frac{(x_{12}\cdot\mathcal{D}_{z})^J\mathcal{O}_{k,J}(x_1,z)}{(x_{12}^2)^{\frac{2\Delta_\phi-\tau_k}{2}}}+\dots \, \  \ \ \ \  \   \ \  \  \ \  \ \textrm{{\color{blue}Euclidean}}\\
&\phi(x_1)\,\phi(x_2) \approx  \sum_{k}C_{12k} \int_{0}^{1}  [dt]\,\frac{\mathcal{O}_{k,J}(x_1+tx_{21},x_{12}) }{(x_{12}^2)^{\frac{2\Delta_\phi-\tau_k}{2}}} 
+\dots \  \ \ \ \  \  \textrm{{\color{blue}Lorentzian}}\label{eq:lightconeOPEAppendix}
\end{align}
where the $\dots$ represent subleading terms in each expansion, $z$ is a null polarization vector, 
\begin{equation}
[dt] =\frac{\Gamma(\Delta_k+J)}{\Gamma^2(\frac{\Delta_k+J}{2})} (t(1-t))^{\frac{\Delta_k+J}{2}-1} dt \,,
\end{equation}
and $\mathcal{D}_z$ is the so-called Todorov operator \cite{todorov:1976}
\begin{align}
\mathcal{D}_{z} =\left(\frac{d}{2}-1+z\cdot \frac{\partial}{\partial z}\right)\frac{\partial}{\partial z^{\mu}}-\frac{1}{2}z^{\mu}\frac{\partial^2}{\partial z\cdot \partial z} \,. \label{eq:TodorovD}
\end{align} 
The formulae above are key in obtaining the conformal block expansion around  both limits. 
For example, in the four-point function case it is trivial to obtain the lightcone block from (\ref{eq:lightconeOPEAppendix}), with the result
\begin{align}
&\langle \phi(x_1)\dots \phi(x_4) \rangle \approx \sum_{k}\frac{C_{12k}}{(x_{12}^2)^{\frac{2\Delta_\phi-\tau_k}{2}}} \int [dt]\, \langle \mathcal{O}_k(x_1+tx_{21},x_{12})\phi(x_3)\phi(x_4) \rangle
\\
&=\sum_{k}\frac{C_{12k}^2}{(x_{12}^2x_{34}^2)^{\frac{2\Delta_\phi-\tau_k}{2}}} \int \frac{[dt] \,(x_{13}^2x_{24}^2-x_{14}^2x_{23}^2)^{J}}{(x_{23}^2t+(1-t)x_{13}^2)^{\frac{\Delta_k+J}{2}}(x_{24}^2t+(1-t)x_{14}^2)^{\frac{\Delta_k+J}{2}}}\,,
\nonumber
\end{align}
where we have changed variables $t\rightarrow  t/(t+1)$ and $t\rightarrow t  x_{24}^2 /x_{14}^2$. 
The lightcone block for the exchange of an operator ${\cal O}_k$
 is defined by this leading term in the expansion
\begin{equation}
\langle \phi(x_1)\dots \phi(x_4)\rangle \approx   \frac{1}{(x_{12}^2x_{34}^2)^{\Delta_\phi}}\sum_{k} C_{12k}^2  \,\left(\mathcal{G}_{k} (u,v) + \dots \right)\,,
\label{eq:4ptlightconeexp}
\end{equation}
where
\begin{equation}
\mathcal{G}_{k} (u,v)= u^{\tau_k/2}(1-v)^{J_k}\,_2F_1\left(\frac{\Delta_k+J_k}{2},\frac{\Delta_k+J_k}{2},\Delta_k+J_k,1-v\right)  \equiv  u^{\tau_k/2}  g_{k}(v)\,.
\label{eq:LCblock4pt}
\end{equation}
We defined the function $ g_{k}(v)$ for later convenience. Note that the expansion (\ref{eq:4ptlightconeexp}) is merely schematic, since subleading terms in the lightcone limit of a lower twist block can dominate with respect to the lightcone limit of a higher twist block.

\subsection{Lightcone conformal blocks}

\begin{figure}
 	\centering
 	\includegraphics[width=0.8\linewidth]{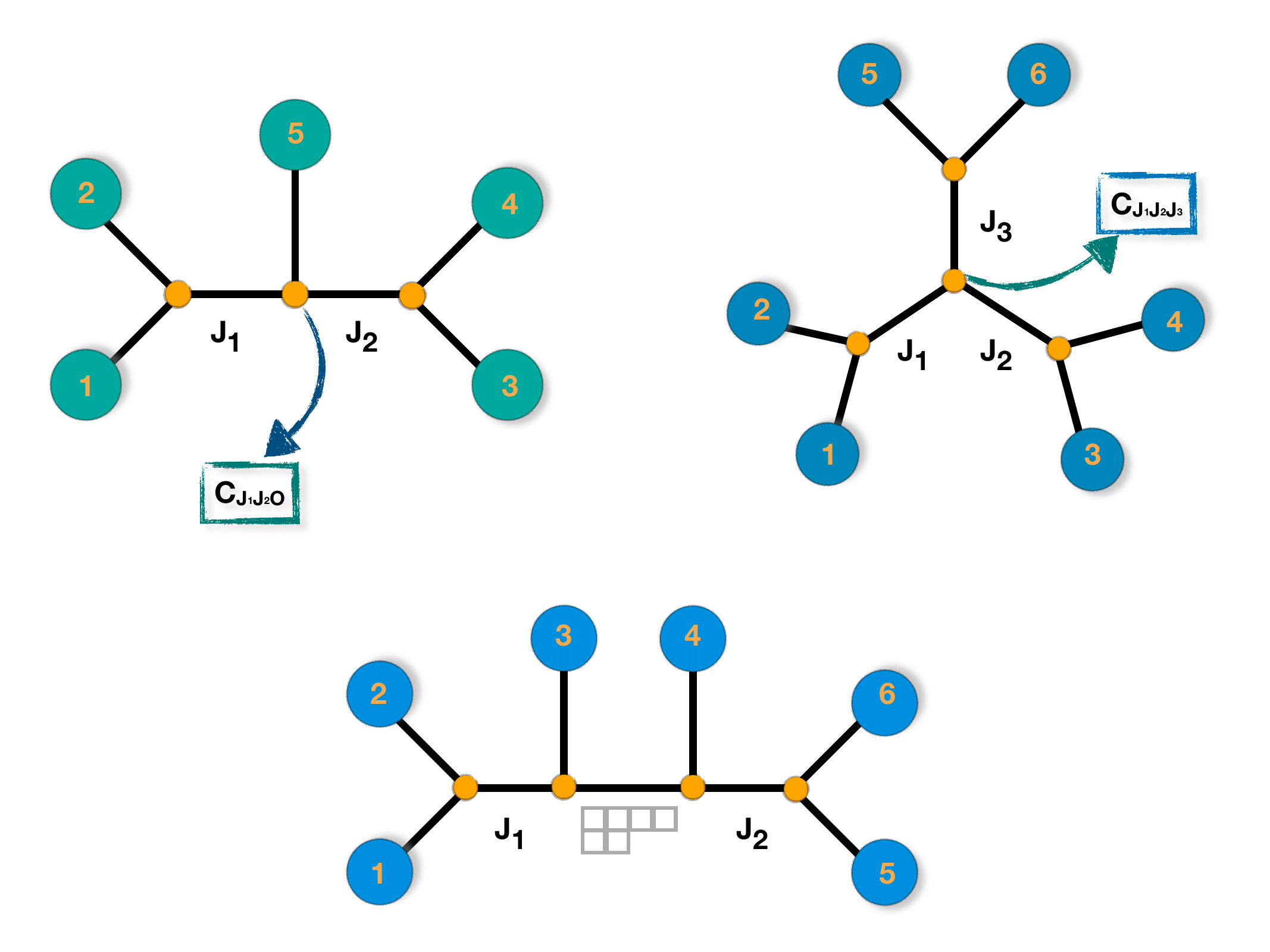}
 	\caption{Schematic representation of the OPE channels for five- and six- point functions. In the top left we have the snowflake decomposition of the five-point function, where we emphasize the OPE coefficient involving two spinning operators. In the top right we have the snowflake decomposition of the six-point function, emphasizing the OPE coefficient of three spinning operators. In the bottom, we depict the comb channel expansion, which may involve mixed-symmetry tensors and which we will not analyze in detail. }
 	\label{fig:OPEsdecompositions}
\end{figure}
Let us start with the lightcone expansion of the five-point conformal block.  Applying twice the OPE limit (\ref{eq:lightconeOPEAppendix}) we obtain 
\begin{equation}
\langle \phi(x_1)\dots \phi(x_5) \rangle \approx \sum_{k_i}    \left( \prod_{i=1}^2  C_{\phi\phi k_i}\int [dt_i]\right) \frac{\langle \mathcal{O}_{k_1}(x_1+t_1x_{21},x_{12})\mathcal{O}_{k_2}(x_3+t_2x_{43},x_{34})\phi(x_5)  \rangle}{(x_{12}^2)^{\frac{2\Delta_{\phi}-\tau_{k_1}}{2}}(x_{34}^2)^{\frac{2\Delta_{\phi}-\tau_{k_2}}{2}}} \, .
\end{equation}
The limits $x_{12}^2\to 0$ and $x_{34}^2\to 0$ correspond to $u_1\to 0 $ and  $u_3\to 0 $, respectively.
The three-point function in the integrand involves the external scalar and two symmetric traceless operators  with arbitrary spin as depicted in the top-left part of figure \ref{fig:OPEsdecompositions}. Our convention for three-point functions of symmetric and traceless operators is \cite{costa:2011}
\begin{align}
\langle \mathcal{O}_{k_1}(x_1,z_1)\dots\mathcal{O}_{k_3}(x_3,z_3) \rangle  =\sum_{\ell_i} \frac{C_{J_1J_2J_3}^{\ell_1\ell_2\ell_3}V_{1,23}^{J_1-\ell_2-\ell_3}V_{2,31}^{J_2-\ell_1-\ell_3}V_{3,12}^{J_3-\ell_1-\ell_2}H_{12}^{\ell_3}H_{13}^{\ell_2}H_{23}^{\ell_1}}{(x_{12}^2)^{\frac{h_1+h_2-h_3}{2}}(x_{13}^2)^{\frac{h_1+h_3-h_2}{2}}(x_{23}^2)^{\frac{h_2+h_3-h_1}{2}}}\,\label{eq:SpinningThreepointFunction},
\end{align}
where we used a null polarization vector $z_i$ to encode the indices of the operators, $h_i=\Delta_i+J_i$ and $V$ and $H$ are defined as
\begin{align}
V_{i,jk} =\frac{ (z_i\cdot x_{ij} ) x_{ik}^2 - (z_i\cdot x_{ik}) x_{ij}^2}{x_{jk}^2} \,, \qquad  H_{ij} =(z_i\cdot x_{ij} )(z_j\cdot x_{ij}) -\frac{x_{ij}^2 (z_i\cdot z_{j} )}{2}\,.
\label{eq:estruturas3pf}
\end{align}
The sum in $\ell_i \in \{ 0,\dots, {\rm min}(J_k)  \}$ counts the possible tensor structures. 
In the five-point case we have a three-point function of a scalar with two operators of spin $J_1$ and $J_2$, therefore the different structures are labelled by $\ell_3\equiv \ell$ and $\ell_1$ and  $\ell_2$ vanish.
After doing simple and straightforward manipulations we arrive at the explicit expression for the lightcone block defined by
\begin{equation}
\langle \phi(x_1)\dots \phi(x_5) \rangle \approx  \frac{1}{(x_{12}^2x_{34}^2)^{\Delta_{\phi}}}\left(\frac{x_{13}^2}{x_{15}^2x_{35}^2}\right)^{\frac{\Delta_{\phi}}{2}}
  \sum_{k_1,k_2,\ell} P_{k_1k_2\ell}
\,\mathcal{G}_{k_1k_2\ell}(u_i) \,,
\label{eq:5ptexpansion}
\end{equation}
where 
\begin{align}
&\mathcal{G}_{k_1k_2\ell}(u_i)= u_1^\frac{\tau_1}{2}u_3^\frac{\tau_2}{2}(1-u_2)^{\ell}u_5^{\frac{\Delta_\phi}{2}} \int [dt_1][dt_2]
\label{eq:5ptlightconeblockdef}\\
&   \frac{\big(1-t_1(1-u_2)u_4-u_2u_4\big)^{J_2-\ell}\big(1-t_2(1-u_2)u_5-u_2u_5\big)^{J_1-\ell}}{ \big(1-(1-u_4)t_2\big)^{\frac{h_2-\tau_1-2\ell+\Delta_\phi}{2}}\big(1-(1-u_5)t_1\big)^{\frac{h_1-\tau_2-2\ell+\Delta_\phi}{2}} \big(1-(1-t_1)(1-t_2)(1-u_2)\big)^{\frac{h_1+h_2-\Delta_\phi}{2}}}\,.
\nonumber
\end{align}
The expansion (\ref{eq:5ptexpansion}) includes a product of three OPE coefficients that we denote  by
\begin{equation}
P_{k_1k_2\ell} =  C_{\phi\phi k_1}C_{\phi\phi k_2}C_{\phi k_1k_2}^{(\ell)}\,.
\label{eq:P5pt}
\end{equation}
Formula (\ref{eq:5ptlightconeblockdef})  is valid as long as one of the exchanged operators is not the identity. In such a case the OPE instead simplifies to
\begin{align}
\phi(x_1) \phi(x_2)   \approx \frac{C_{\phi \phi \mathcal{I}}}{(x_{12}^2)^{\Delta_\phi}}\,\mathcal{I} \,,
\end{align}
which forces the other exchanged operator to be the same as the external one.
When the exchanged operator in the $(12)$ OPE is the identity we have  (in this case there is a single $\ell=0$ structure)
\begin{align}
\mathcal{G}_{ \mathcal{I} \, \phi}(u_i) = \left(\frac{u_3u_5}{u_4}\right)^{\frac{\Delta_\phi}{2}}  \,, 
\label{eq:5ptblockidentitylimit}
\end{align}
on the other hand, when the identity is flowing in the $(34)$ OPE, we have
\begin{align}
\mathcal{G}_{\phi\, \mathcal{I} }(u_i) = u_1^\frac{\Delta_\phi}{2}   \,.
\label{eq:5ptblockidentitylimit2}
\end{align}

For the lightcone expansion of the six-point conformal block we need to apply the OPE limit (\ref{eq:lightconeOPEAppendix})  three times. We will choose the snowflake 
channel as illustrated in the top-right of figure \ref{fig:OPEsdecompositions}. In this choice the exchanged operators are always symmetric traceless tensors of spin $J_i$.
This gives
\begin{align}
\label{eq:6ptexpansion}
&\langle \phi(x_1)\dots \phi(x_6) \rangle \approx \frac{1}{(x_{12}^2x_{34}^2x_{56}^2)^{\Delta_{\phi}}} \sum_{k_i,\ell_i}   P_{k_i\ell_i}\mathcal{G}_{k_i \ell_i}(u_i,U_i)=
\\
 &\sum_{k_i}  \left( \prod_{i=1}^3  C_{\phi\phi k_i}\int [dt_i]\right) 
\frac{\langle \mathcal{O}_{k_1}(x_1+t_1x_{21},x_{12})\mathcal{O}_{k_2}(x_3+t_2x_{43},x_{34})\mathcal{O}_{k_3}(x_5+t_3x_{65},x_{56})  \rangle}{(x_{12}^2)^{\frac{2\Delta_{\phi}-\tau_1}{2}}(x_{34}^2)^{\frac{2\Delta_{\phi}-\tau_2}{2}}(x_{56}^2)^{\frac{2\Delta_{\phi}-\tau_3}{2}}}\,.\nonumber
\end{align}
Using the three-point function conventions (\ref{eq:estruturas3pf}) and defining ${\cal T} =\sum_i \tau_i$,  $L=\sum_i \ell_i$ and $H=\sum_i h_i$
we obtain
\begin{align}
&\mathcal{G}_{k_i \ell_i}(u_i,U_i)\equiv  u_1^{\frac{\tau_1}{2}}u_3^{\frac{\tau_2}{2}}u_5^{\frac{\tau_3}{2}} g_{k_i\ell_i} (u_2,u_4,u_6,U_i) 
\label{eq:6ptlightconeblockdef}
\\
&
=  \prod_{i=1}^3u_{2i-1}^\frac{\tau_i}{2} \int [dt_i]\frac{u_{2i}^{\ell_i}\,\chi_i^{\ell_{1-i}}(1-\chi_i)^{\ell_{2-i}-\tau_{2-i}+{\cal T}/2 }(1-u_{2i})^{J_{i+1}+\ell_{i+1}-L}\mathcal{A}_i^{J_i+\ell_i-L}}{\mathcal{B}_i^{\ell_i-\Delta_i -L+H/2 }} \,,
\nonumber
\end{align}
where we use the notation $\ell_i\equiv \ell_{i+3}$ and\footnote{The reader may have realized that due to the cyclic defining property of the cross-ratios we can for example refer to the even cross-ratios $u_2$, $u_4$, $u_6$ in  the product  as $u_{2(i-1)}$.}
\begin{align}
\mathcal{A}_i &=\frac{1}{(1-u_{2(i-1)})}\bigg[(1-t_{i-1})(1-\chi_{1-i})\big(-1+u_{2(i-1)}-(1-t_{i+1})u_{2(i-1)}\chi_{2-i}+\chi_{3-i}\big)\nonumber\\
&+t_{i-1} u_{2(i+1)}(1-\chi_{3-i})\big(-1+u_{2(i-1)}-(1-t_{i+1})u_{2(i-1)}\chi_{2-i}\big)\bigg]   \,, \\
\mathcal{B}_i &= 1-\chi_{2-i}-t_{1+i}(1-u_{2i}-\chi_{2-i}+(1-t_{i-1})u_{2i}\chi_{1-i})\,,\nonumber
\end{align}
with $\chi_i$ defined as $\chi_i= \frac{U_i-u_{2(2-i)}}{U_i}.$
A nice property of the $\chi$ variables is that the conformal block factorizes in products of three $\,_2F_1$ in the limit $\chi_i\rightarrow 0$. Another nice property is that $\ell_i$ determines the leading power
of $\chi_i$, as can easily be seen in (\ref{eq:6ptlightconeblockdef}).

When one of the exchanged operators is the identity, the remaining two are equal to each other, which leads to the simplified expression
\begin{align}
\mathcal{G}_{k k {\cal I}}(u_i,U_i)= \left(\frac{u_1u_3}{U_2}\right)^\frac{\tau_k}{2}g_{k}(u_2/U_1)\,,
\label{eq:6ptOneIDblock}
\end{align}
where $g_{k}(v)$ contains is the four-point  block as defined in ({\ref{eq:LCblock4pt}}).

\section{Snowflake bootstrap}
\label{sec:Bootstrap}

Let us start by recalling the basic features of the lightcone bootstrap for four-point correlators  \cite{Komargodski:2012ek,Fitzpatrick:2012yx}.
A four-point function of local operators $\phi$ can be decomposed in the $(12)$ or $(23)$ OPE channels
\begin{align}
\frac{1}{(x_{12}^2x_{34}^2)^{\Delta_\phi}}\sum_{\mathcal{O}_{k}} C_{\phi\phi k}^2 G_{k}(u,v) = \frac{1}{(x_{23}^2x_{14}^2)^{\Delta_\phi}}\sum_{k} C_{\phi\phi k}^2 G_{k}(v,u)\,,\label{eq:bootstrapequation}
\end{align}
where  $G_{k}(u,v)$ is the full conformal block in the $(12)$ channel. This bootstrap equation has been used to extract properties of conformal field theories following both analytic  and numerical approaches. 

\begin{figure}
 	\centering
 	\includegraphics[width=0.45\linewidth]{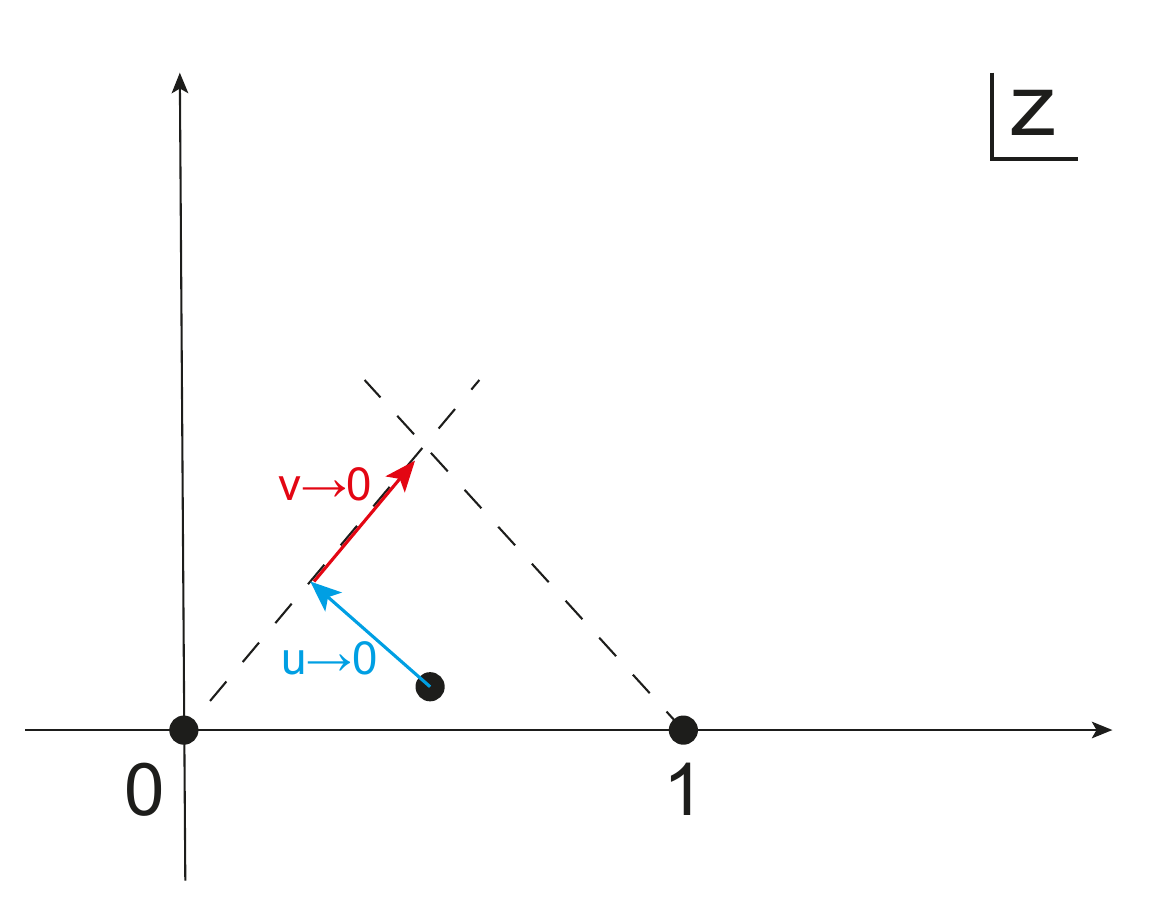}
 	\caption{Schematic representation of the relevant lightcone limit in the $z$-plane. The point $x_2$ first approaches the lightcone of the operator at the origin, as $u\to0$.  Subsequently, it approaches the lightcone of the operator at $x_3=(1,0)$, which corresponds to taking $v\to0$.}
 	\label{fig:doublelc}
 \end{figure}

Low twist  operators dominate in the lightcone $x_{12}^2\rightarrow 0$ limit of the left hand side of the bootstrap equation. Unitary CFTs obey the following bounds for the  twist of operators 
\begin{align}
\tau= 0\quad\textrm{identity}\,,\qquad \tau\equiv\Delta-J \ge \begin{cases}(d-2)/2\, &\textrm{scalar} \\ d-2\, \, &\textrm{spin}\,,
 \end{cases}
\end{align}
and so the leading term on the left hand side of the bootstrap equation is given by
\begin{align}
\frac{1}{(x_{12}^2x_{34}^2)^{\Delta_\phi}}\sum_{k} C_{\phi\phi k}^2 G_{k}(u,v) = \frac{1}{(x_{12}^2x_{34}^2)^{\Delta_\phi}}\left[1+ C_{\phi\phi k_*}^2u^\frac{\tau_{k*}}{2}g_{k_*}(v)+\dots \right],
\end{align}
where we have used that the conformal block behaves as $G_{k}(u,v)\rightarrow \mathcal{G}_k(u,v)  =u^{\frac{\tau}{2}}g_k(v)$ in the $u\rightarrow 0$ limit. 
The assumption is that above the identity there is a unique operator 
 ${\cal O}_{k_*}$ with  leading twist. Next  we take the limit of $x_{23}^2\to 0$, which moves the point $x_2$ to the corner of the  square made by the lightcones of points $1$ and $3$, which can be taken respectively at $0$ and $1$ in the complex $z$-plane, as shown in figure \ref{fig:doublelc}. It is possible to take this second limit, which corresponds to $v$ small, and use the right hand side of (\ref{eq:bootstrapequation}).
 
Each term in the $u\rightarrow0$ limit will diverge at most logarithmically, which apparently contradicts the power law divergence of the left hand side of the equation. The emergence of the power law singularity was addressed in \cite{Komargodski:2012ek,Fitzpatrick:2012yx} and it boils down to the contribution of double-twist  operators $ [\phi\phi]_{0,J} \sim  \phi \Box^0 \partial^J \phi$ whose twist approaches $2\Delta_\phi$ at large spin. The stronger divergence is recovered by performing the infinite sum over spin of these double-twist families. 
In particular, this fixes the density of OPE coefficients for this family of operators at large spin to be\footnote{This differs from some conventions in the literature by a factor of $2^J$ due to our conformal block normalization.}
\begin{equation}
		C_{\phi \phi [\phi\phi]_{0,J}}^2 \sim \frac{8 \sqrt{\pi}}{\Gamma(\Delta_\phi)^2 2^{2\Delta\phi+J}}J^{2\Delta_\phi-3/2}\,,
		\label{eq:OPEs/s/dt}
\end{equation}
which is the behaviour of OPE coefficients in Mean Field Theory.

Additionally, the leading twist operator above the identity in the direct-channel leads to $1/J$ suppressed corrections to the OPE coefficients along with {\em anomalous dimension} type corrections, which means the twist of these families behaves as
\begin{equation}
	\tau_{[\phi\phi]_{0,J}}=2\Delta_\phi+ \frac{k}{J^{\tau^*/2}}\,.
\end{equation}
At this level the large spin expansion is merely asymptotic, and the OPE coefficients and anomalous dimensions cannot be assigned to a single operator of a given spin. However, the large spin expansion actually converges at least down to spin 2, and the OPE coefficients are really associated to a unique operator at each spin, which follows from the fact that the double-twist operators really sit in Regge trajectories that are analytic in spin. All these remarkable facts were established through the Lorentzian inversion formula \cite{Caron-Huot:2017vep}. 
This formula systematizes the large spin perturbation theory/lightcone-bootstrap and essentially supersedes it as a computational tool \cite{Alday:2017zzv,Henriksson:2018myn,Alday:2019clp}. 
 In this work, however, we are interested in higher-point functions which are much richer, and for which a  Lorentzian inversion formula  is presently unavailable. 
 Therefore we must resort to the more pedestrian large spin perturbation theory. It would of course be interesting to develop higher-point Lorentzian inversion formulae and reproduce and extend the results we will derive below.

\subsection{Five-point function}

Let us consider the more complicated case of the five-point function. We now have an exchange of two operators, and their contribution is captured by the block expansion
in a given channel. We consider the $(12)(34)$ and $(23)(45)$ channels for the five-point function $\langle \phi(x_1)\phi(x_2)\phi(x_3)\phi(x_4) \phi(x_5)\rangle$,
\begin{equation}
\frac{(x_{13}^2)^{\frac{\Delta_\phi}{2}}}{(x_{12}^2x_{34}^2)^{\Delta_\phi}(x_{15}^2x_{35}^2)^{\frac{\Delta_{\phi}}{2}}} \sum_{k_1,k_2,\ell}P_{k_1k_2\ell} 
G_{k_1k_2\ell}^{12,34}(u_i)=\frac{(x_{24}^2)^{\frac{\Delta_\phi}{2}}}{(x_{23}^2x_{45}^2)^{\Delta_\phi}(x_{12}^2x_{14}^2)^{\frac{\Delta_{\phi}}{2}}} \sum_{n_1,n_2,\ell}P_{n_1n_2\ell} G_{n_1n_2\ell}^{23,45}(u_i)\,.
\label{eq:5ptCrossing}
\end{equation}
The limit $x_{12}^2,x_{34}^2\rightarrow 0$ is dominated by low twist operators in the $(12)(34)$ channel. The natural candidate to lead this expansion is the identity operator, however it is not possible to have two identities being exchanged at the same time, since that would imply a nonzero three-point functions between two identities and the scalar operator $\phi(x_5)$.  It is however possible to have one identity being exchanged in one OPE and another operator in the other OPE. In this case the conformal blocks simplify considerably and the exchanged operator must be the external one. The block simplifies to a product of a two- and three-point function,  
check (\ref{eq:5ptblockidentitylimit}) and (\ref{eq:5ptblockidentitylimit2}). 
Thus, we conclude that the first terms in the lightcone limit in the channel $(12)(34)$ are given by
\begin{equation}
C_{\phi\phi\phi} {\cal G}_{\mathcal{I} \, \phi}(u_i)+
C_{\phi\phi\phi} {\cal G}_{\phi\, \mathcal{I}}(u_i) =
C_{\phi\phi\phi} \left(
\left(\frac{u_3u_5}{u_4}\right)^\frac{\Delta_\phi}{2}
+
u_1^\frac{\Delta_\phi}{2}\right)
.
\end{equation}
There is possibly another leading term from two exchanges  of  the leading twist operator ${\cal O}_{k_*}$. This term has a lightcone limit in the channel $(12)(34)$ given by
\begin{equation}
C_{\phi\phi k_*}C_{\phi\phi k_*}C_{k_*k_*\phi} {\cal G}_{k_* k_* \ell}(u_i)\,.
\end{equation}
The term that dominates is determined by the rate at which $u_1$ and $u_3$ go to zero and by the twist of $\phi$ and ${\cal O}_{k_*}$. Below we shall address both possibilities.
We may then take the other limits $x_{23}^2,x_{45}^2,x_{15}^2\rightarrow 0$, corresponding to $u_2,u_4,u_5\rightarrow 0$,  which as we shall see, are suitable for the expansion in the 
 $(23)(45)$ channel. The decomposition in this channel takes the form
 \begin{equation}
\left(\frac{u_1u_3^2u_5}{u_2^2u_4^2}\right)^{\Delta_\phi/2} \sum_{n_1,n_2,\ell}P_{n_1n_2\ell}\, {\cal G}_{n_1n_2\ell}^{23,45}(u_i)\,,
\label{eq:5ptCrossedChannel}
\end{equation}
where we collected here the prefactors on both sides of (\ref{eq:5ptCrossing}).
The powers of  $u_2,u_4$ in the  denominator of (\ref{eq:5ptCrossedChannel}) impose constraints on the operators that need to be present in the conformal block decomposition of the channel $(23)(45)$. 

 \subsubsection{Identity in the (12) OPE}
 \label{sec:ID(12)5-point}

Let us understand this in more detail. First consider the term 
\begin{equation}
C_{\phi\phi\phi} {\cal G}_{\mathcal{I} \, \phi}(u_i)= C_{\phi\phi\phi} 
\left(\frac{u_3u_5}{u_4}\right)^\frac{\Delta_\phi}{2}
\,,
\end{equation} 
where the identity is exchanged in the $(12)$ OPE.
The cross-ratios $u_2$ and $u_4$, when taken to be small, control the twist of the exchanged operators in the cross-channel. We can use this to infer what class of operators are contributing in the  cross-channel where the blocks behave as
 \begin{equation}
 	{\cal G}_{n_1n_2\ell}^{23,45}(u_i)= u_2^{\tau_{n_1}/2} u_4^{\tau_{n_2}/2} g_{n_1n_2\ell}(u_1,u_3,u_5)\,.
 \end{equation}
 Combining these behaviours with the prefactor in (\ref{eq:5ptCrossedChannel}) we can conclude that  the operators $n_1$ have a twist that approaches $2\Delta_\phi$, and therefore correspond to the usual leading double- twist  operators. Moreover, in this case the operator $n_2$ must have twist $\Delta_\phi$. This corresponds to the exchange of the external operator itself. Therefore the cross-channel OPE data
is given by 
\begin{equation}
	P_{[\phi\phi]_{0,J},\phi} = C_{\phi \phi [\phi\phi]_{0,J}} C_{\phi \phi \phi} C_{\phi \phi [\phi\phi]_{0,J}}\,,
\end{equation}
from which we can see that the single-trace OPE coefficient cancels on both sides of the crossing equation, and we are left with data that is known from the four-point bootstrap, namely scalar/scalar/double-twist OPE coefficients.

Actually this case reduces to the crossing of the four-point function of $\phi$ and its descendants.  Firstly, in the direct-channel, since the five-point function factorizes into a product of 2 and 3-pt functions, we can use the (45) OPE into the exchanged scalar operator $\phi$, which acts on the MFT 4-pt function of $\phi$ at points $1235$. Secondly, in the cross-channel the (45) OPE reduces the five-point block into an action on the four-point block with external $\phi$ at points $1523$ and double-twist exchange. This shows the problem reduces to that of the  four-point function.

Nevertheless it is instructive to check this result explicitly using the lightcone blocks in (\ref{eq:5ptlightconeblockdef}) to describe the cross-channel contributions. In this case $J_2=\ell=0$ and $\Delta_2=\Delta_\phi$. Additionally for large spin $J_1$ the dimension of the exchanged operator approaches the double-twist value 
$\Delta_1=2\Delta_\phi + J_1$. This significantly simplifies the expression (\ref{eq:5ptlightconeblockdef}) for the blocks. In practice, it is useful to expand the integrand using the binomial theorem and performing the $t_i$ integrals, which leads to a representation in terms of an infinite sum of hypergeometric functions.
In fact, the sum is dominated by the region $u_1 \sim J_1^{-2}$, similarly to the four-point case. This allows one to simplify the hypergeometric functions into Bessel functions, so the large spin limit
of the lightcone block reads
\begin{equation}
{\cal G}_{[\phi\phi]_{0,J_1} \phi }^{23,45}(u_i) \approx \sum_{n=0}^\infty \frac{ J_1^{n+\frac{1}{2}} \Gamma \Big(\frac{\Delta _{\phi}+1}{2} \Big) \Gamma \Big(\frac{2n+\Delta _{\phi }}{2}\Big)
u_1^{\frac{\Delta _{\phi+n}}{2}} u_2^{\Delta _{\phi }}
    (1-u_3)^n u_4^{\frac{\Delta _{\phi }}{2}} K_{n}\left(2 J_1 \sqrt{u_1}\right)}
   {2^{1-3 \Delta _{\phi }-J_1} \pi  \Gamma (n+1) \Gamma \big(n+\Delta _{\phi }\big)}\,.
\end{equation}
Imposing the well-known large spin asymptotics of the scalar/scalar/double-twist OPE coefficients (\ref{eq:OPEs/s/dt}), one can do the sum over $J_1$ by approximating it as an integral. This reproduces the correct power of 
$u_1$ at fixed $n$. The correct power of $u_3$ is then recovered by doing the infinite sum over $n$.

We remark that one can then consider the related contribution where we swap the exchanged operators in the cross-channel, meaning we have $\mathcal{O}_{n_1}= \phi$ and $\mathcal{O}_{n_2}=[\phi \phi]_{0,J_2}$. This obviously corresponds to a factorized correlator in a different channel which is subleading in the lightcone limit here considered.

 \subsubsection{Identity in the (34) OPE}

On the other hand, when we exchange the identity in the $(34)$ OPE, the direct-channel contribution is
\begin{equation}
C_{\phi\phi\phi}  \,u_1^\frac{\Delta_\phi}{2}\,.
\end{equation}
Thus, since the leading powers of $u_2$ and $u_4$  in the cross-channel expression (\ref{eq:5ptCrossedChannel}) 
are the same, the operators that are exchanged in the cross-channel will both have the double-twist value $2\Delta_\phi$.
This allows us to probe the double-twist/double-twist/scalar OPE coefficient on the cross-channel
\begin{equation}
P_{[\phi\phi]_{0,J_1}[\phi\phi]_{0,J_2} \ell} = C_{\phi \phi [\phi\phi]_{0,J_1}} C_{\phi \phi[\phi\phi]_{0,J_2}} C_{\phi [\phi\phi]_{0,J_1} [\phi\phi]_{0,J_2}}^{(\ell)}\,.
\end{equation}
It is important to notice that the double-twist/double-twist/scalar OPE coefficient depends on the additional quantum number $\ell$, which encodes the tensor structure associated to  spin-spin-scalar three-point functions.

Since the scalar/scalar/double-twist coefficients are fixed from the four-point analysis, matching to the direct-channel we immediately discover the remarkable non-perturbative relation
\begin{equation}
	C^{(\ell)}_{\phi [\phi\phi]_{0,J_1} [\phi\phi]_{0,J_2}} \propto C_{\phi\phi\phi}\,,
\end{equation}
which would be expected in a perturbative theory.
With a more careful analysis, we will now fix the large spin asymptotics of this OPE coefficient, along with its $\ell$ dependence. 

We need to reproduce the power law behaviour in the variables $u_1$, $u_3$ and $u_5$, which will emerge from the infinite sum over $J_1$, $J_2$ and $\ell$ in the cross-channel. More specifically, we consider the limit 
$J_1,J_2 \to \infty$ with $u_1 J_1^2$ and $u_5 J_2^2$ fixed.  It is possible to approximate the lightcone block in this regime by approximating the integrand in (\ref{eq:5ptlightconeblockdef}), so that one finds 
integral representations  of two Bessel functions,\footnote{This procedure deserves a word of caution. Strictly speaking we should first take the limit of $u_1,u_3\to0$, keeping large spin
contributions, and only then take $u_2,u_4\to0$. In practice, since we use the lightcone block expansion (\ref{eq:5ptlightconeblockdef}) in the cross-channel, we are swapping the order of limits. This is justified  a posteriori since the asymptotics of OPE coefficients at large spin that we obtain match the examples studied in section \ref{sec:examples}.}
 \begin{align}
	{\cal G}^{23,45}_{[\phi\phi]_{0,J_1}[\phi\phi]_{0,J_2} \ell}   (u_i)
	&\approx \frac{2^{4 \Delta_\phi +J_1+J_2}}{\pi} J_1^{1/2} J_2^{1/2} u_2^{\Delta_\phi } u_4^{\Delta_\phi } (1-u_3)^{\ell } \nonumber\\
		&u_1^{\frac{1}{4} (3 \Delta_\phi +2 \ell )} u_5^{\frac{1}{4} (\Delta_\phi +2 \ell )} K_{\ell +\frac{\Delta_\phi }{2}}\left(2 J_1
		u_1^{1/2}\right) K_{\ell +\frac{\Delta_\phi }{2}}\left(2J_2
		u_5^{1/2}\right).
\end{align} 
It is not hard to see that for consistency with the $u_3\to 0$ limit the power law behavior in $u_1,u_5$ has to be reproduced term by term in the sum over $\ell$. This leads to the ansatz
\begin{equation}
	P_{[\phi\phi]_{0,J_1}[\phi\phi]_{0,J_2} \ell} \approx  C_{\phi\phi\phi}\,b_\ell  \, 2^{-J_1-J_2} J_1^{\ell +3(\Delta_\phi-1)/2} J_2^{\ell +3(\Delta_\phi-1)/2} \,,
	\label{eq:OPEid(34)5-pt}
\end{equation}
which, upon performing the integrals over $J_1$ and $J_2$, reproduces the power law behavior in $u_1$ and $u_5$. Since $\ell\in\{0,\dots,\min(J_1,J_2)\}$, this leaves us with an infinite sum over $\ell$ to perform, which will recover the power law behavior in $u_3$. In particular, we need to zoom in on the $\ell \to \infty$ region, with $u_3$ approaching zero such that $u_3 \ell$ is kept fixed. In this limit, we can use the approximation 
$(1-u_3)^\ell \approx  e^{- u_3 \ell}$.
Then, we can take the asymptotic large $\ell$ behaviour of the coefficient $b_\ell$ to be~\footnote{The same result could be obtained by explicitly performing the sum over $\ell$ assuming $b_\ell \propto \frac{1}{\ell! \Gamma\left(\ell+\Delta_\phi\right)}$. However, this cannot be used to determine the form of the coefficients at finite $\ell$ since the leading singularity in $u_3\to 0$ only determines the asymptotic behaviour at $\ell\to \infty$. Remarkably this turns out to be the exact form of the coefficients in the disconnected correlator in section \ref{disconnected5pt}. A similar situation also occurs for the six-point case.}
\begin{equation}
	b_\ell \approx\frac{ \Delta_\phi \Gamma\left(\frac{1+\Delta_\phi}{2}\right)}{2^{3 \Delta_\phi-3} \sqrt{\pi}\, \Gamma(\Delta_\phi)^2\Gamma\left(1+\frac{\Delta_\phi}{2}\right)}\,  \ell^{-2\ell}e^{2\ell} \ell^{-\Delta_\phi} \,.
\end{equation}
We can then approximate the sum over $\ell$ by an integral, which gives the correct power law behaviour in $u_3$ and finally reproduces the identity contribution in the direct-channel.

\begin{figure}
 	\centering
 	\includegraphics[width=0.75\linewidth]{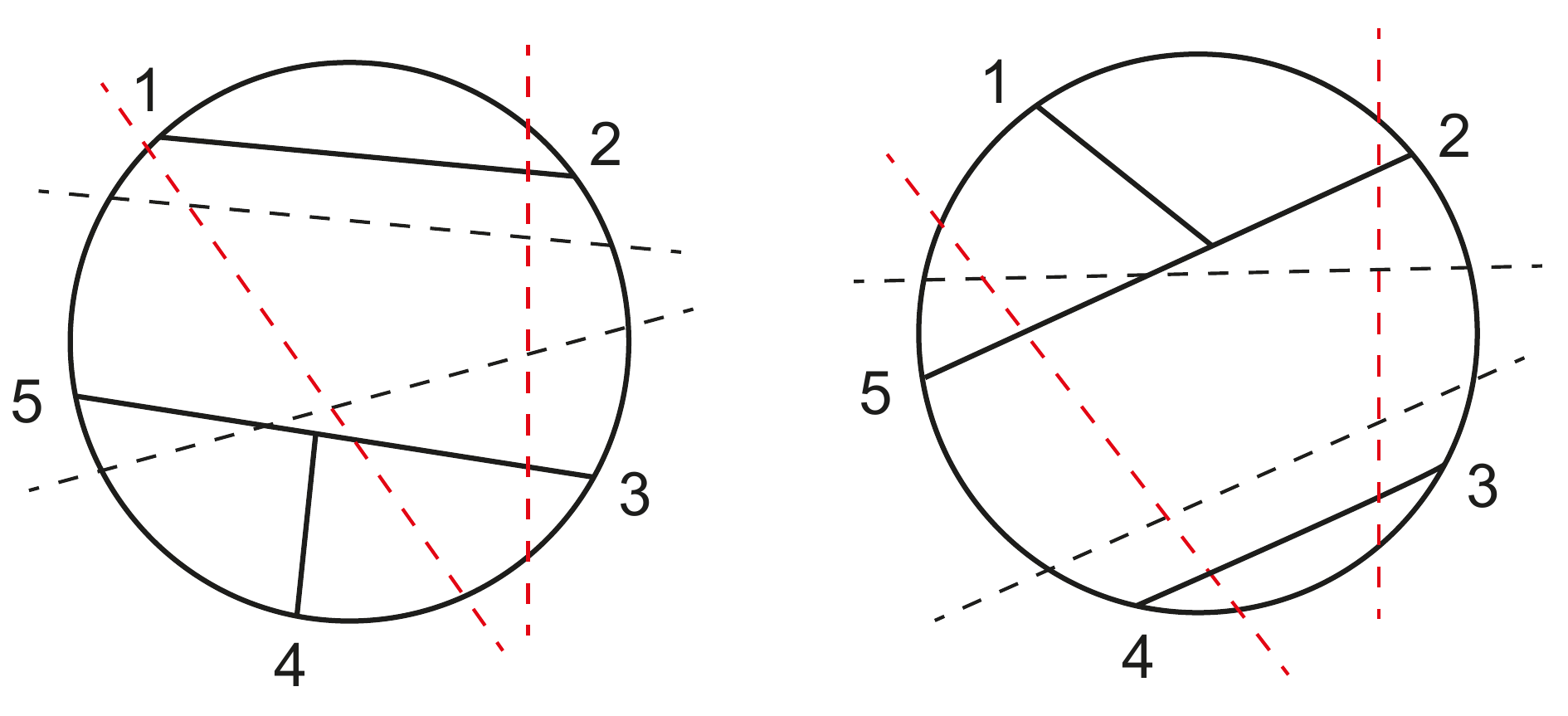}
 	\caption{Witten diagrams corresponding to the leading order five-point function in a large N theory. 
	The black and red dashed lines correspond to the unitarity cuts in the direct and crossed OPE channels, allowing us to infer what the exchanged operators are.}
 	\label{fig:fiveptcuts}
 \end{figure}

Both leading terms with an identity exchange are understood as a five-point function which factorizes into a product of a two-  and three-point functions. A simple example of CFTs expected to present this behaviour are holographic theories with cubic couplings. We can draw bulk Witten diagrams and look at their unitarity cuts to infer the exchanged operators in the corresponding channel. This is presented  in figure \ref{fig:fiveptcuts}.
Clearly, this picture is consistent with the results obtained from the lightcone limit analysis.

\subsubsection{Two non-trivial exchanges}

The case of two non-trivial exchanges is more subtle. When the exchanged operators are identical to the external ones, the lightcone limit of the block  in the channel $(12)(34)$ is given by
\begin{equation}
C_{\phi\phi\phi}^3(u_1u_3u_5)^\frac{\Delta_\phi}{2}\frac{\Gamma(\Delta_\phi)^2}{\Gamma(\frac{\Delta_\phi}{2})^4}\left(\zeta_2+\ln u_4\ln u_5+ 2S_{\frac{\Delta_\phi-2}{2}}(\ln u_4 +\ln u_5 )+4S_{\frac{\Delta_\phi-2}{2}}^2-S_{\frac{\Delta_\phi-2}{2}}^{(2)}+\dots\right),
\end{equation}
where  $S^{(n)}_\alpha$ denotes the degree-$n$  harmonic number and  the dots represent subleading terms in $u_2$, $u_4$ and $u_5$.
 The powers of $u_2$ and $u_4$ indicate that the exchanged operators in the cross-channel should once again be of double-twist type. 
 However, since the powers of $u_5$ are the same for both block expansions in the small $u_5$ limit, one cannot employ the usual argument which ensures that operators with large spin $J_2$ dominate the cross-channel. This means that the information in this OPE is not universal. The leading power of $u$ is a constant, which can be achieved block by block in the cross-channel, and therefore the usual argument for the necessity of large spin double-twist operators is not valid.

One can instead study the case where the two exchanged scalar operators $\mathcal{O}_{k_*}$ are different from the external one, but identical among themselves. 
\begin{equation}
	{\cal G}_{k_* k_*}^{12,34}(u_i) \approx a_{\Delta_*\Delta_\phi} (u_1 u_3 u_5)^{\Delta_* /2} u_4^{\frac{\Delta_*-\Delta_\phi}{2}} \,,
	\label{eq:direct2equal}
\end{equation}
with 
\begin{equation}
	a_{\Delta_*\Delta_\phi}=\frac{\pi  4^{\Delta ^*-1} \Gamma \big(\frac{\Delta_*+1}{2}\big){}^2 \csc
		^2\big(\pi  (\frac{\Delta_*-\Delta _{\phi }}{2}) \big)}{\Gamma
		\big(\frac{\Delta_*-\Delta _{\phi }}{2}+1\big)^2 \,\Gamma
		\big(\frac{\Delta _{\phi }}{2}\big)^2}\,.
\end{equation}
When $\Delta_*<\Delta_\phi$ this is the leading term. On the other hand, for $\Delta_* \geq \Delta_\phi$ the leading powers are instead  integers and lead to the same limitation discussed above.
Nevertheless, the term (\ref{eq:direct2equal}) is still present and can also be bootstrapped. 

Notably, the power of $u_4$ will change the nature of the exchanged operators in the (45) OPE. 
In particular, we now have that the operator must have dimension asymptoting to $ \Delta_* + \Delta_\phi + J_2$. Thus we 
 prove the existence of the double-twist operators 
 $[\phi {\cal O}_*]_{0,J_2}$ 
  built out of the external $\phi$ and the internal $\mathcal{O}_*$.
  We see an asymmetry between the exchanges in the cross-channel, since the operators in the (23) channel are 
  still the double-twist composites $[\phi \phi]_{0,J_1}$. 
  This is similar to the case of identity exchange in the (12) channel which also leads to an asymmetry in the cross-channel exchanges. In particular, swapping the cross-channel exchanges in the (23) and (45) OPEs leads to a subleading contribution in the direct-channel.

The calculation in the cross-channel is similar to that of the previous subsection. Both families of double-twist operators must be in the large spin regime, which gives the following approximation for the cross-channel conformal block
\begin{align}
	{\cal G}^{23,45}_{[\phi\phi]_{0,J_1}[\phi\mathcal{O}_*]_{0,J_2} \ell}  (u_i)&
	\approx\frac{2^{3 \Delta_\phi+\Delta_* +J_1+J_2}}{\pi} J_1^{1/2} J_2^{1/2} u_2^{\Delta_\phi } u_4^{(\Delta_
		\phi+\Delta_*)/2 } (1-u_3)^{\ell } \nonumber\\&
	u_1^{\frac{1}{4} (2 \Delta_\phi+\Delta_* +2 \ell )} u_5^{\frac{1}{4} (\Delta_\phi +2 \ell )} K_{\ell +\frac{\Delta_* }{2}}\left(2 J_1
	u_1^{1/2}\right) K_{\ell +\frac{\Delta_\phi }{2}}\left(2J_2
	u_5^{1/2}\right)\,.
\end{align}
 Once again the sum over large spins $J_1$ and  $J_2$ must be done for fixed $\ell$ and we then sum over $\ell$. The correct asymptotics for the OPE coefficients in this case is given by
\begin{equation}
	P_{[\phi\phi]_{0,J_1}[\phi\mathcal
		{O}^*]_{0,J_2} \ell}
		\approx 
		q_{\Delta_*\Delta_\phi} 2^{-J_1-J_2} J_1^{\frac{ 4\Delta_\phi-3+2\ell - \Delta_* }{2}} J_2^{\frac{3\Delta_\phi-3+2\ell - 2\Delta_* }{2}} \ell^{-2 \ell} e^{2\ell} \ell^{-\Delta_\phi}\,,
\label{eq:OPE5ptlt}	
\end{equation}
where
\begin{equation}
	q_{\Delta_*\Delta_\phi} = P_{\mathcal{O}_*\mathcal{O}_*} a_{\Delta_*,\Delta_\phi} \frac{2^{5-3\Delta_\phi-\Delta_*}}{\Gamma(\frac{\Delta_\phi-\Delta_*}{2})\Gamma(\Delta_\phi -\frac{\Delta_*}{2})^2}\,.
\end{equation}
The factor of $ P_{\mathcal{O}_*\mathcal{O}_*}   = C^2_{\phi\phi \mathcal{O}_*} C_{\phi\mathcal{O}_*\mathcal{O}_*}$ is needed to match the direct-channel.

\subsubsection{Stress-tensor exchange}
In a general CFT, the leading twist operators are usually scalars of scaling dimension less than $d-2$ or the stress tensor which has dimension $d$ and spin 2, and therefore twist $d-2$. A spin 1 conserved current also has twist $d-2$ but, since we are studying the OPE of identical scalars, only even spin operators can be exchanged. Thus, we are only left to consider  the case of the stress tensor\footnote{Higher spin conserved currents also have twist $d-2$ but they only exist in free theories and we therefore ignore them.}.

In this case, the direct-channel contribution has three terms associated to the tensor structures with $\ell=0,1,2$. In the cyclic lightcone limit, it turns out that the powerlaw behavior in $u_4\to 0$ is suppressed by $\ell$ and therefore the tensor structure with $\ell=0$ dominates. The block behaves very similarly to the scalar case, with the role of $\Delta_*$ being played by the twist of the stress tensor $d-2$, up to some extra prefactors. Concretely, 
 the direct-channel block contains the following term in the lightcone expansion
\begin{equation}
	{\cal G}_{TT\,\ell=0} \approx	a_{T,\Delta_\phi} (u_1 u_3 u_5)^{(d-2)/2} u_4^{\frac{d-2-\Delta_\phi}{2}} \,,
\label{eq:stresstensor5pt}	
\end{equation}
with
\begin{equation}
	a_{T,\Delta_\phi}=\frac{\pi  4^{d-1} \Gamma \left(\frac{d+3}{2}\right)^2 \sec ^2\left(\pi\frac{\Delta _{\phi }+3-d}{2} \right)}{\Gamma^2 \left(\frac{\Delta _{\phi
			}+4}{2}\right) \Gamma^2 \left(\frac{d-\Delta _{\phi }}{2} \right)}\,.
\end{equation}
In the block expansion this term will come multiplied by the product of OPE coefficients $P_{TT\,\ell=0} $.
Once again there are terms where the powers of $u_4$ and $u_5$ are constant and cannot be reproduced by large spin double twist families in the cross-channel.
The term in (\ref{eq:stresstensor5pt}) is the leading one for $d-2-\Delta_\phi<0$, but it remains in the expansion otherwise, so it can be bootstrapped. 
The physics in the cross-channel is very similar to the scalar case as well. The small $u_2$ and $u_4$ behavior is matched by operators of the form $[\phi\phi]_{0,J_1}$ in the (23) OPE and $[\phi T]_{0,J_2}$ in the (45) OPE,
with twists asymptoting to $2\Delta_\phi$ and $d-2+\Delta_\phi$ at large $J_1$ and $J_2$, respectively. The large spin limit is needed to obtain the right power law behavior in $u_1$ and $u_5$, and finally the large $\ell$ limit reproduces the small $u_3$ behavior. The cross-channel blocks and OPE coefficients are the same as in the scalar case with the replacement $\Delta_* \to d-2$, up to the different prefactor which is fixed by the direct-channel block.
More concretely, the cross-channel block in the large spin limit becomes
\begin{align}
{\cal G}^{23,45}_{[\phi\phi]_{0,J_1}[\phi T]_{0,J_2} \ell}  &\approx 
\frac{2^{3 \Delta_\phi+d-2 +J_1+J_2}}{\pi} J_1^{1/2} J_2^{1/2} u_2^{\Delta_\phi } u_4^{(\Delta_\phi+d-2)/2 } (1- u_3)^{\ell} \nonumber\\&
u_1^{\frac{1}{4} (2 \Delta_\phi+d-2 +2 \ell )} u_5^{\frac{1}{4} (\Delta_\phi +2 \ell )} K_{\ell +\frac{d-2}{2}}\left(2 J_1 u_1^{1/2}\right) K_{\ell +\frac{\Delta_\phi }{2}}\left(2J_2 u_5^{1/2}\right)\,,
\end{align}
and the OPE coefficients
\begin{equation}
P_{[\phi\phi]_{0,J_1} [\phi T]_{0,J_2} \ell} 
	\approx q_{T \Delta_\phi} 2^{-J_1-J_2} J_1^{\frac{1}{2}(-1+2\ell -d + 4\Delta_\phi)} J_2^{\frac{1}{2}(1+2\ell - 2d + 3\Delta_\phi)} \ell^{-2 \ell} e^{2\ell} \ell^{-\Delta_\phi}\,,
\end{equation}
where
\begin{equation}
q_{T \Delta_\phi} = P_{TT \ell=0}\, a_{T \Delta_\phi} \frac{2^{7-3\Delta_\phi-d}}{\Gamma\Big(\frac{\Delta_\phi-d+2}{2}\Big)\Gamma\Big(\Delta_\phi -\frac{d-2}{2}\Big)^2}\,.
\end{equation}
\subsection{Six-point function -- snowflake}
\label{sec:6-pt}
\begin{figure}
 	\centering
 	\includegraphics[width=0.8\linewidth]{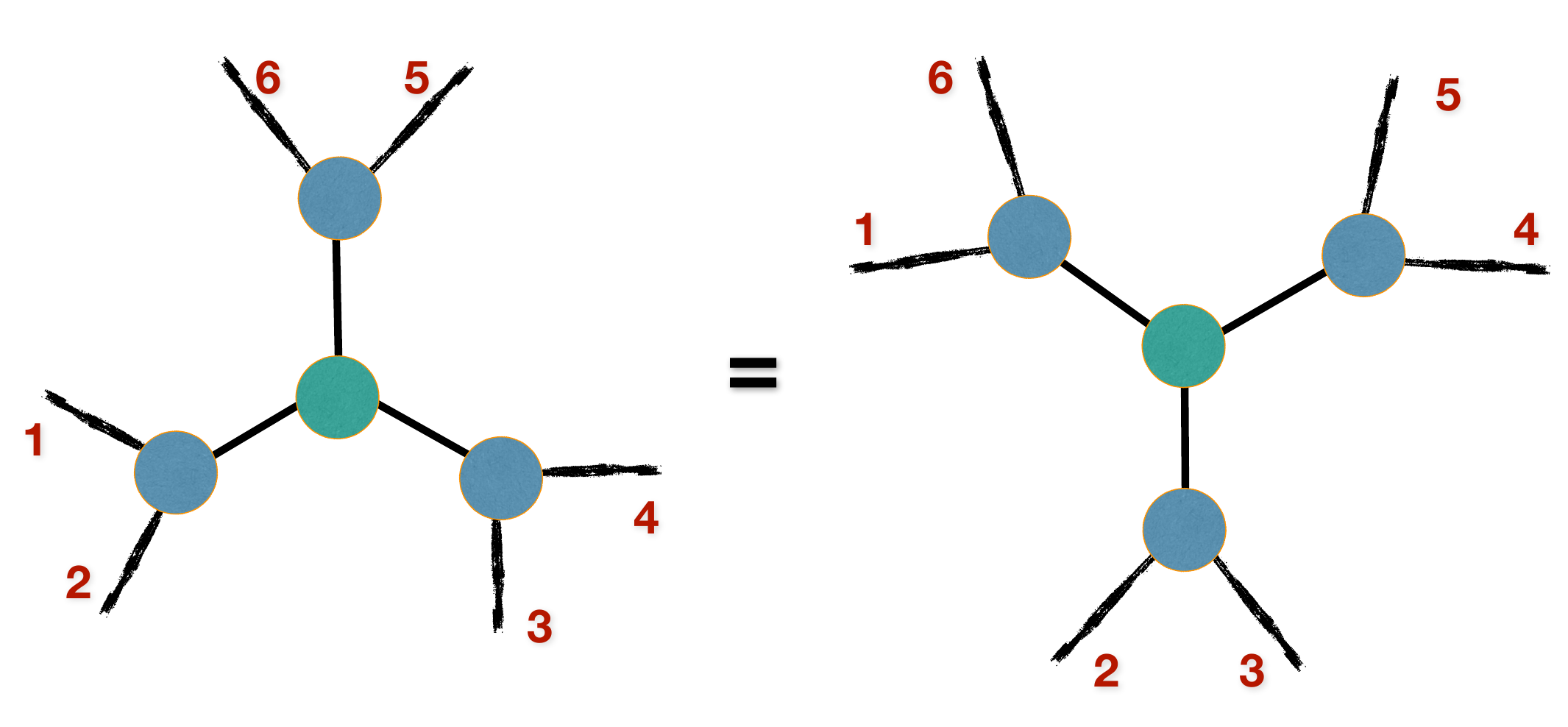}
 	\caption{A schematic form of the six-point snowflake bootstrap equation. The left hand side represents the $(12)(34)(56)$ direct-channel expansion while the right hand side represents the $(23)(45)(61)$ cross-channel.}
 	\label{fig:snowflake6pt}
 \end{figure}

The six-point function is a richer object as it admits two very different OPE decompositions that are usually denoted by snowflake and comb. One distinction between them is that in the snowflake decomposition we do three OPEs in nonconsecutive  pairs of points and therefore all OPEs involve two external scalars. Therefore there will be  an OPE coefficient between three symmetric traceless operators of arbitrary spin, as can be seen in the top-right of figure \ref{fig:OPEsdecompositions}. 
On the other hand, in  the comb channel the OPE involves consecutive pairs of operators. Thus, after performing the OPE between two external scalars,  the resulting symmetric traceless operator will be fused with another external scalar and can
produce a mixed symmetry tensor operator, which in the mean field theory limit should correspond to a triple-twist operator. The bottom part of figure \ref{fig:OPEsdecompositions} illustrates this structure. In this paper we use the lightcone OPE between scalars (\ref{eq:lightconeOPEAppendix}) and therefore limit our analysis to the snowflake channel, whose bootstrap equation we depict in figure \ref{fig:snowflake6pt}.

We start by considering the block expansion in the direct $(12)(34)(56)$ channel
\begin{equation}
\langle \phi(x_1)\dots \phi(x_6)\rangle =\frac{1}{(x_{12}^2x_{34}^2x_{56}^2)^{\Delta_\phi}}  \sum_{k_i, \ell_i}P_{k_i \ell_i} G_{k_i \ell_i}^{12,34,56}(u_i,U_i)\,.
\end{equation}
and take the  lightcone limits  $x_{12}^2\to0$, $x_{34}^2\to0$, $x_{56}^2\to 0$, which correspond to $u_1\to0$,   $u_3\to 0$,   $u_5\to 0$.
The leading contributions in this limit come from the exchange of three identities, one identity and two leading twists or three leading twists. For now we take the 
leading twist to be a scalar, so that 
\begin{align}
\langle \mathcal{O}(x_1)\dots \mathcal{O}(x_6)\rangle& \approx \frac{1}{(x_{12}^2x_{34}^2x_{56}^2)^{\Delta_\phi}}
\Big[ P_{{\cal I}{\cal I}{\cal I}} {\cal G}_{{\cal I}{\cal I}{\cal I}}(u_i,U_i)  +  \Big( P_{{\cal I}k_*k_*} {\cal G}_{{\cal I}k_*k_*}(u_i,U_i) +{\rm perm}\Big)  
\nonumber\\
&\qquad\qquad\qquad\qquad
+    P_{k_*k_*k_*} {\cal G}_{k_*k_*k_*}(u_i,U_i)\Big]=
\nonumber
\\
&=\frac{1}{(x_{12}^2x_{34}^2x_{56}^2)^{\Delta_\phi}} \Bigg[1+ \bigg(C_{\phi\phi k_*}^2\Big(\frac{u_1u_3}{U_2}\Big)^\frac{\tau_{k_*}}{2}g_{k_*}(u_2/U_1) +\textrm{perm}\bigg)
\label{eq:confblockdecompdirectchannel}\\
&\qquad\qquad\qquad\qquad
+C_{\phi\phi k_*}^3C_{k_*k_*k_*}(u_1u_3u_5)^{\frac{\tau_{k_*}}{2}}g_{k_*k_*k_*}(u_{2i},U_i)\Bigg]\,,
\nonumber
\end{align}
where
 $\Delta_{*}$ is the dimension of the leading twist operator ${\cal O}_{k_*}$ and the functions
  $g_{k_*}$ and $g_{k_*k_*k_*}$  are defined from the 
 four- and six-point  lightcone blocks in ({\ref{eq:LCblock4pt}})
 and (\ref{eq:6ptlightconeblockdef}), respectively.
  Then we take  the three distances  $x_{23}^2$, $x_{45}^2$ and $x_{16}^2$ to zero, or in cross-ratios $u_{2i}\rightarrow 0$, 
  which will be appropriate  to study the OPE decomposition in the crossed  channel $(23)(45)(16)$ in the lightcone limit. The four-point conformal block $g_{k_*}$ simplifies considerably in this limit
\begin{align}
g_{k_*}(u_i/U_j) \approx -\frac{\Gamma(\Delta_*+J_*)}{\Gamma^2(\frac{\Delta_*+J_*}{2})}\left(S_{\frac{\Delta_*+J_*-2}{2}}+\ln(u_i/U_j)\right)+\dots\,,
\label{eq:4ptSmallv}
\end{align}
where the $\dots$ represent subleading terms in $u_i/U_j$. However, after taking $u_{2i}\rightarrow 0$ the function  $g_{k_*k_*k_*}(u_{2i},U_i)$ of the 
 six-point conformal lightcone block is still a nontrivial function of the cross-ratios $U_i$, so we take one further limit $x_{24}^2,x_{26}^2,x_{46}^2\rightarrow 0 $, or equivalently $U_i\rightarrow 0$, which we refer to as the origin limit \cite{Bercini:2021jti}.
 Let us remark that we do this just to make the problem technically simpler. With this extra limit one gets
\begin{align}
g_{k_*k_*k_*}(u_{2i},U_i) & \approx -\frac{\Gamma^3(\Delta_*)}{\Gamma^6(\frac{\Delta_*}{2})}\bigg[\frac{\prod_{i}\ln U_i}{3}+2S_{\frac{\Delta_*-2}{2}}\ln U_1\ln U_2+\left(4S_{\frac{\Delta_*-2}{2}}^2-S_{\frac{\Delta_*-2}{2}}^{(2)}+\zeta_2\right)\ln U_1\nonumber\\
&+\frac{2}{3}S_{\frac{\Delta_*-2}{2}}\bigg(4S_{\frac{\Delta_*-2}{2}}^2-3S_{\frac{\Delta_*-2}{2}}^{(2)}+3\zeta_2\bigg) +\dots   \bigg]+\textrm{perm}\,,\label{eq:blockoriginlimitscalar}
\end{align}  
where the $\dots$ represent subleading terms. We give the derivation os this result in appendix \ref{app:Blocks}. Notice that up to this order the correlator  is polynomial of degree three in the logarithm of the cross-ratios, which contrasts with the  behavior in a planar gauge theory\cite{Vieira:2020xfx}.

\subsubsection{Exchange of three identities}

\begin{figure}
	\centering
	\includegraphics[width=0.35\linewidth]{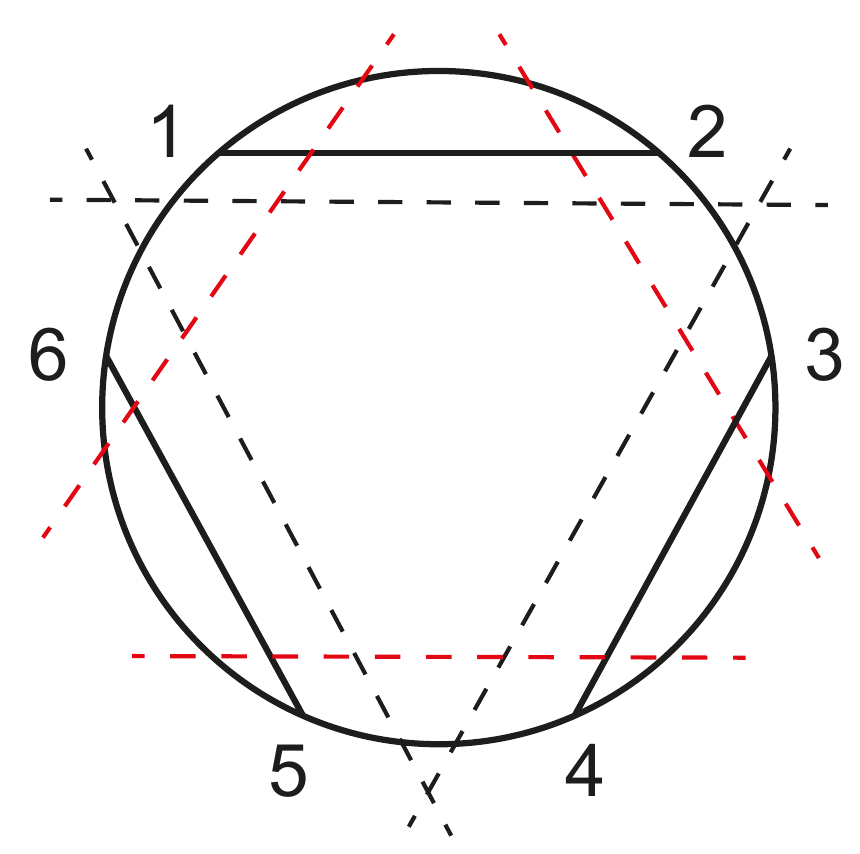}
	\caption{Witten diagrams corresponding to the leading order six-point function in a large N theory. 
		The black and red dashed lines correspond to the unitarity cuts in the direct and crossed OPE channels, allowing us to read-off the exchanged identity and double-twist operators, respectively.}
	\label{fig:witten6pt}
\end{figure}

Given the crossing equation 
\begin{align}
\sum_{k_i,\ell_i} P_{k_i\ell_i} G_{k_i \ell_i}^{12,34,56}(u_i,U_i)&=\prod_{i=1}^3\left(\frac{u_{2i-1}}{u_{2i}}\right)^{\Delta_{\phi}} \sum_{k_i,\ell_i}P_{k_i\ell_i} G_{k_i\ell_i}^{23,45,16}(u_i,U_i)\,,
\label{eq:conformalblockdecomp6pt}
\end{align}
the limit taken above should be compatible with the   cross-channel decompositions in the channel $(23)(45)(16)$. 
As we just described, the left hand side of this equation starts with a one and then has subleading corrections in the cross-ratios $u_{\textrm{odd}}\rightarrow 0$, while on the right hand side there is an aparent power law divergence in $u_{\textrm{even}}$ in the prefactor. This implies that the cross-channel decomposition involves
operators with dimension approximately equal to  $ 2\Delta_\phi+J$ that cancel the prefactor $u_{2i}^{\Delta_\phi}$ in the denominator. Each individual conformal block in the $(23)(45)(16)$  channel is regular in the cross-ratios $u_{\textrm{odd}}$ as they approach zero, which is 
not enough to cancel the prefactor  $u_{2i-1}^{\Delta_\phi}$   and recover the identity contribution of the direct-channel.\footnote{This 
behavior is similar to that of  scalar exchange  in  the direct-channel  (\ref{eq:blockoriginlimitscalar})  and is given in appendix \ref{app:Blocks} for general spin.}
The solution is similar to that of the four- and five-point correlators in the sense that the identity is recovered from the infinite sum of double-twist  operators with large spin. This can also be intuitively understood by looking at the "unitarity cuts" of a disconnected Witten diagram as in figure \ref{fig:witten6pt}.

 We will now choose the kinematics  where both $u_{\rm odd}$ and  $U_i$ are sent to zero with the same rate $J^{-2}$, with $\ell_i$ fixed. 
 This is not the choice we did in the direct-channel above, but we will recover its kinematics by sending $u_{\rm odd}/U_i\to 0$ afterwards.
 The conformal block simplifies considerably in this limit and is given by a product of three Bessel functions
\begin{align}
{\cal G}_{k_i\ell_i}^{23,45,16}\approx  \prod_{i=1}^3\frac{2^{J_i+\tau_i}J_i^{\frac{1}{2}}}{\pi^{\frac{1}{2}}} \,u_{2i}^{\frac{\tau_{i}}{2}} \chi_{i}^{\ell_i}K_{\frac{2\ell_{i-1}-2\ell_{i+1}+\tau_{i+1}-\tau_{i-1}}{2}}\left(2J_i\sqrt{U_{2i-1}}\right)U_{2i-1}^{\frac{2\ell_{i-1}+2\ell_{i+1}+\tau_{i-1}-\tau_{i+1}}{2}}\,,
\label{eq:largespinblock6pt}
\end{align}
where we can see that the parameter $\ell_{i}$ controls the cross-ratio $\chi_{i+1}=1-u_{2i-1}/U_{2i-1}$. 
The direct-channel limit that we took above 
can be recovered in the cross-channel by studying the limit where $\chi_i$ approaches $1$, which in turn is controlled by the large $\ell_i$ region. \footnote{We stress that we made the choice of considering the limit $U_i \to 0$ to simplify the expression for the block. Alternatively, one could mimic the approach of \cite{Bercini:2021jti} and keep these cross-ratios finite. We emphasize however that our choice of taking the origin limit respects an order: $U_i \to 0$ only after $u_i \to 0$. The latter limit is dominated by large $J_i$ and large $\ell_i$, whereas the subsequent $U_i \to 0$ imposes $J_i\gg \ell_i \gg 1$.}
We can now use (\ref{eq:largespinblock6pt}) in the crossing equation (\ref{eq:conformalblockdecomp6pt})
to reproduce the identity exchange of the direct-channel
\begin{align}
1\approx \frac{1}{8} \left(\prod_{n=1}^3\left(\frac{u_{2n-1}}{u_{2n}}\right)^{\Delta_{\phi}} \int dJ_n d\ell_n \right)P_{k_i\ell_i}{\cal G}_{k_i\ell_i}^{23,45,16}(u_i,U_i)\,,
\label{eq:bootstrapequation6pt2}
\end{align}
where we  transformed the sums in $k_i,\ell_i$  in the crossing equation
to integrals in $J_n,\ell_n$ (including  a factor of $1/2$ because we are only summing over even spins). 
We can assume that the product of OPE coefficients $P_{k_i\ell_i}$ has the large $J_i$  power law behavior
\begin{equation}
P_{k_i\ell_i} \approx  C \prod_{n=1}^32^{-J_n}J_n^{a_n}f_n(\ell_n)\,.
\label{eq:P6pt3indentities}
\end{equation}
Integrating over $J_i$ we obtain
\begin{align}
1\approx \prod_{i=1}^3 \prod_{\epsilon=\pm}\int d\ell_i f(\ell_i)\,\frac{2^{2\Delta_{\phi}}u_{2i-1}^{\Delta_{\phi}}\chi^{\ell_{i}}_i}{\pi^{\frac{1}{2}}}\,
\Gamma\bigg(\frac{3+2a_i+2\epsilon(\ell_{i+1}-\ell_{i-1})}{4}\bigg) 
\,U_{2i-1}^{\frac{2(\ell_{i-1}+\ell_{i+1})-2a_{i}-3}{4}}\,,
\label{eq:rhsconfblockdecomp6pt} 
\end{align}
where we  used that $\tau_i = 2\Delta_{\phi}$ to leading order in large $J_i$. Then we consider the limit where $u_{\rm odd}/U_i \to 0$.
Remember that we need a power law divergence  in $u_{\rm odd}$ to kill the prefactor in (\ref{eq:bootstrapequation6pt2}) and, as expected, this is generated by the tail of the sum in $\ell_i$.  
In this regime we can replace $\chi_i^{\ell_i}$ by $exp(-\ell_i  u_{2i-3} / U_{2i-3})$, where we are keeping  fixed the argument of the exponential in the limit. 
The powers of $U_i$ cannot depend on $\ell_i$ otherwise this would give rise to a non-trivial in behavior  $U_i$, which is not consistent with the left-hand side of (\ref{eq:bootstrapequation6pt2}), so we conclude that 
\begin{align}
a_{i}=r+\left(\sum_{j}\ell_j\right) -\ell_i\,,
\end{align}
with $r$ a constant that does not depend on $\ell_i$.  We can, at this point, take the large $\ell_i$ behavior of the $\Gamma$ functions
in (\ref{eq:rhsconfblockdecomp6pt}). The $\ell_i$ behavior of the expression suggests that for large $\ell_i$ the function $f(\ell_i)$ has  the following form
\begin{align}
f_i(\ell_i) \approx e^{2\ell_i}\ell_i^{g-2\ell_i}\,,
\label{eq:threepointlargel}
\end{align}
with $g$ and $c$ constants. Putting everything together and after doing the $\ell_i$ integration we obtain
\begin{align}
1\approx C\, 2^{6\Delta_{\phi}}\Gamma^2\left(\frac{3}{2}+g+r\right)  \prod_{i=1}^3 u_{2i-1}^{\Delta_{\phi}-\frac{3}{2}-g-r}U_{i}^{\frac{3}{4}+g+\frac{r}{2}}\,,
\end{align}
which fixes both $r,g$ and $c$ to be
\begin{align}
r=\frac{4\Delta_{\phi}-3}{2}\,, \qquad
g=-\Delta_{\phi}\,, \qquad
C=\frac{1}{2^{6\Delta_{\phi}}\Gamma^3\left(\Delta_{\phi}\right)}\,.
\end{align}
This fixes the asymptotic form of $P_{k_i\ell_i}$ proposed in (\ref{eq:P6pt3indentities}).

\subsubsection{Exchange of one identity and two leading twist operators}

So far we have only reproduced the contribution of the identity in the direct-channel OPE decomposition (\ref{eq:confblockdecompdirectchannel}). 
As we have seen subleading contributions 
depend non trivially on the cross-ratios, even in the limit where all $u_{i}$ approach zero, cf. (\ref{eq:4ptSmallv}) and (\ref{eq:blockoriginlimitscalar}). One key difference is that we will have to generate $\log$s of the cross-ratios from the cross-channel OPE decomposition. 
Some of these $\log$s are generated by allowing a correction to the dimension of the double-twist operators of the form
\begin{align}
\tau_i = 2\Delta_{\phi} +\frac{k}{J^{a}_i}\,.
\end{align}
The conformal block, in the large spin limit, depends on the twist of the exchanged operator in an explicit way as can be seen in (\ref{eq:largespinblock6pt}). It is easy to perturb the previous computation,
done to reproduce the contribution of the identity with the cross-channel double-twist exchange, and include the correction to the dimension of these operators.
First we expand (\ref{eq:largespinblock6pt}) at large $J_i$ and keep the first subleading term in the series. Then, performing the integrals in $J_i$ and $\ell_i$ we obtain the following
correction to the contribution of the  leading twist operators  exchange 
\begin{align}
 k \frac{\Gamma^2\left(\frac{2\Delta_\phi-\tau_*}{2}\right)}{\Gamma^2(\Delta_\phi)}\sum_{j}\bigg[\ln \frac{u_{2j}u_{2j+3}U_{2j+1}^\frac{1}{2}}{(u_{2j-1}u_{2j+1}U_{2j-1}^3)^{\frac{1}{2}}}-(S_{\Delta_{\phi}}-S_{\Delta_{\frac{2\phi-a}{2}}})\bigg]\left(\frac{u_{2j-1}u_{2j+1}}{U_{2j+1}}\right)^{\frac{a}{2}}.
\end{align}
This term has the correct power law behavior coming from the direct-channel contribution of the identity and two leading twist operators, cf. 
(\ref{eq:confblockdecompdirectchannel}) or (\ref{eq:6ptOneIDblock}). 
This fixes $a=\tau_*$, in agreement with  the four-point function calculation.
Moreover, it contains some of the $\log$s coming from the four-point block  function 
$g_{k_*}$, but it also has some unexpected $\log$ terms. It is precisely these terms that will allow us to fix the correction to the OPE coefficient between
three double-twist operators
\begin{align}
P_{k_i\ell_i} = P_{k_i\ell_i}^{MFT} \left(1+\sum_{j}\frac{\sum_{k} \left(c_{j,k}\ln J_k +b_{j,k}\ln \ell_k\right)+v_{j}}{J_j^{\tau_*}}+\dots\right)\,,
\label{eq:OPEsixpt1identity}
\end{align}
where $c_{i,j}$, $b_{i,j}$ and $v_i$ are coefficients that we will fix.
Upon inserting this in the cross-channel conformal block decomposition, and integrating over $J_i$ and $\ell_i$, we obtain 
\begin{align}
 \sum_{j}\bigg[\ln   \left( \prod_{i}u_{2i-1}^{-b_{j,i+1}-\frac{c_{j,i}+c_{j,i-1}}{2}}U_{2i-1}^{b_{j,i+1}+\frac{c_{j,i-1}}{2}}\right)
 -\frac{2v_j}{k}- \left(S_{\Delta_{\phi}}-S_{\Delta_{\frac{2\phi-\tau_*}{2}}}\right)
 \bigg]\left(\frac{u_{2j-1}u_{2j+1}}{\tilde{U}_{j+1}}\right)^{\frac{\tau_*}{2}}. 
\end{align}
The correct $\log$ behavior  imposes that
\begin{align}
&b_{i,i}=0, \,b_{i,i+1}=b_{i,i+2}= \frac{k}{2} \,, \quad  
c_{i,i}=0,\, 
c_{i,i+1} =c_{i,i+2}= -\frac{k}{2}\,, \,\, v_{1} = k S_{\frac{\tau+2J}{2}}\nonumber\\
&k= -\frac{C_{\phi\phi \tau*}^2\Gamma^2(\Delta_{\phi})\Gamma(2J+\tau*)}{2^{2J*-1}\Gamma^2(\frac{2\Delta_{\phi}-\tau*}{2})\Gamma^2(\frac{2J+\tau*}{2})}\,.
\end{align}

Thus, we see that we can reproduce exchanges in the direct-channel that include at least one identity by 
 taking into account the contribution of large spin double-twist operators in the cross-channel. 
 Moreover this procedure fixes the dimension and OPE coefficients of these operators at large spin. 
 The formula for the OPE coefficients is one of the main results of this paper.
 
\subsubsection{Exchange of three leading twist operators}

Before analysing the contribution of the exchange of three leading twist operators in the direct-channel, let us 
 see what is the effect of dressing the large spin double-twist contribution in the cross-channel by a term of the form
 $\prod_{i=1}^3J_i^{q_i}\ell_i^{r_i}$. This can be used, for example, to check what is the cross-ratio dependence  of the corrections to the double-twist exchange in the cross-channel 
 at large spin
\begin{align}
\prod_{i=1}^3\left(\frac{u_{2i-1}}{u_{2i}}\right)^{\Delta_{\phi}} \int dJ_i d\ell_i P_{J_i,\ell_i}^{\textrm{tree}}\bigg[\prod_{j=1}^3J_j^{q_j}\ell_i^{r_j}\bigg]G_{k_i\ell_i}^{23,45,16}(u_i,U_i) \propto  \prod_{j=1}^3 \frac{U_{2j-1}^{\frac{q_{j-1}+2r_{j+1}}{2}}}{u_{2j-1}^{\frac{q_{j}+q_{j-1}}{2}+r_{j+1}}}\,.
\label{eq:toymodelforcorrections}
\end{align}
It follows  that  multiple corrections to the dimension of operators exchanged in the OPEs  $(23)(45)$ and $(23)(45)(16)$, where $r_i=0$ and two or three
nonvanishing  exponents $q_i$ equal $-\tau_*$,
have, respectively, terms of the form
\begin{align}
&\left(\frac{u_{1}u_{5}}{U_2 U_3}\right)^{\frac{\tau_*}{2}}u_{3}^{\tau_*}\bigg[\ln u_{2}\ln u_{4} +\dots\bigg]\,,
 \ \ \ \frac{(u_{1}u_3u_5)^{\tau_*}}{(U_1U_2U_3)^{\frac{\tau_*}{2}}}\bigg[\ln u_{2}\ln u_{4}\ln u_{6} +\dots\bigg]\,,
\end{align}
where the $\dots$ stand for the contribution of $\log$ terms in other cross-ratios that are not important for the present discussion. 
One important feature of these two results is that at least one power of $u_{\textrm{odd}}$ is given by $\tau_*$. This can be thought as coming from the direct-channel contribution of a family of operators whose twist asymptotes to $2\tau_*$. Another curious feature is that there is necessarily a dependence on  $\ln u_{\textrm{even}}$ that cannot be generated by the contribution of a single conformal block, as we can see from (\ref{eq:blockoriginlimitscalar}). This suggests that this term comes from the contribution
in the direct-channel of an infinite family of operators with twist  $2\tau_*$. This behavior was already observed in \cite{Simmons-Duffin:2016wlq} for the case of the four-point function from the existence 
of $\log^2v$ terms. 

Now we are ready to reproduce the last term in (\ref{eq:confblockdecompdirectchannel}) from the cross-channel decomposition.
Since the direct-channel contribution (\ref{eq:blockoriginlimitscalar}) does not have any $\ln u_{\textrm{even}}$ we conclude from the analysis of the previous paragraph
that this term  does not come from the correction of the dimension of  double-twist operators.
Therefore it must come solely from the correction to the OPE coefficient, which we propose to have the  form
\begin{align}
P_{J_i,\ell_i} = P_{J_i,\ell_i}^{tree}\left(1+\sum_{j}\frac{\sum_{k} \left(c_{j,k}\ln J_k +b_{j,k}\ln \ell_k \right)+v_{j}}{J_j^{\tau_*}}+\frac{p(\ln J_j,\ln \ell_j)}{\prod_{j}J_j^{{\tau_*}}\ell_j^{-\frac{\tau_*}{2}}}+\dots \right). 
\label{eq:OPE6p3leadtwist}
\end{align}
where the $c_{i,j}$, $b_{i,j}$ and $v_i$ were already fixed in the previous section and
$p(\ln J_j,\ln \ell_j)$ is a polynomial function of the third degree\footnote{This ansatz is justified because 
 the scalar conformal block is a polynomial of degree $3$ in $\log$ of cross-ratios}
\begin{align}
\!\!\!\!\!\!&p(\ln J_j,\ln \ell_j) = c_1 - c_2\ln \frac{J_3^2}{\ell_1\ell_2}\ln \frac{J_2^2}{\ell_1\ell_3}\ln \frac{J_1^2}{\ell_2\ell_3}+c_3\ln \frac{J_1J_2J_3}{\ell_1\ell_2\ell_3}+2c_4\bigg[\ln J_1\ln \left(\frac{J_2J_3}{\ell_1}\right)^2\frac{1}{\ell_2\ell_3}\nonumber\\
\!\!\!\!\!\!&+\ln J_2\ln \frac{J_3^2}{\ell_2^2\ell_1\ell_3}-\ln J_3\ln\ell_3^2\ell_2\ell_1+\frac{3(\ln \ell_1\ln \ell_2\ell_3+\ln \ell_2\ln\ell_3)}{2}+\frac{\ln^2\ell_1+\ln^2\ell_2+\ln^2\ell_3}{2}\bigg].
\end{align}
This polynomial generates the terms
\begin{align}
&\frac{(\prod_{i}u_i)^{\frac{\tau_*}{2}}\Gamma^3\Big(\frac{2\Delta_\phi-\tau_*}{2}\Big)}{\Gamma^3(\Delta_{\phi})}\bigg[8c_1+ c_2\ln U_1\ln U_2\ln U_3-4c_3\ln U_1U_2U_3+2c_4\sum_{i<j}\ln U_i\ln U_j \bigg],
\end{align}
upon integration in $J_i$ and $\ell_i$. A simple comparison with (\ref{eq:blockoriginlimitscalar}) fixes the values of $c_i$ to be
\begin{align}
&c_2= P_{k_*k_*k_*} \frac{\Gamma(\Delta_*)\Gamma^3(\Delta_{\phi})}{\Gamma^2(\frac{\Delta_*}{2})\Gamma^3\Big(\frac{2\Delta_\phi-\Delta_*}{2}\Big)}\,,\qquad c_3=   \frac{1}{4} \left( S_{\frac{\Delta_*-2}{2}}^{(2)} - 4S_{\frac{\Delta_*-2}{2}}^2-\zeta_2 \right) c_2  \,,
\nonumber\\
&c_1=   \frac{1}{4}  \, S_{\frac{\Delta_*-2}{2}}\left(4S_{\frac{\Delta_*-2}{2}}^2-3S_{\frac{\Delta_*-2}{2}}^{(2)}+3\zeta_2\right) c_2 \,,\qquad
c_4=  S_{\frac{\Delta_*-2}{2}} \,c_2\,.
\end{align}
for a scalar leading twist operator and 
\begin{align}
&c_2=\Gamma^3(\Delta_{\phi})\frac{P_{000} \,\mathbb{B}_{000}^{(3)}}{\Gamma^3(\frac{2\Delta_{\phi}-\tau*}{2})},\, c_4=\Gamma^3(\Delta_{\phi})\frac{P_{000} \,\mathbb{B}_{000}^{(2)}}{\Gamma^3(\frac{2\Delta_{\phi}-\tau*}{2})},\, c_1=\Gamma^3(\Delta_{\phi})\frac{P_{000} \,\mathbb{B}_{000}^{(0)}+3P_{001} \,\mathbb{B}_{001}^{(0)}+3P_{002}\mathbb{B}_{002}^{(0)}}{\Gamma^3(\frac{2\Delta_{\phi}-\tau*}{2})}\,,\nonumber\\
&c_3=2\Gamma^3(\Delta_{\phi})\frac{P_{000} \,\mathbb{B}_{000}^{(1)}+P_{001} \,\mathbb{B}_{001}^{(1)}+P_{002}\mathbb{B}_{002}^{(1)}}{\Gamma^3(\frac{2\Delta_{\phi}-\tau*}{2})}\,,
\end{align}
for the exchange a stress tensor, where we used the block for stress-tensor exchange derived in appendix \ref{app:HyperInt} and wrote $P_{\ell_1 \ell_2 \ell_3} \equiv P_{TTT\ell_1 \ell_2 \ell_3}$. We emphasize the absence of the OPE coefficients associated with the structures where two or three of the $\ell_i$'s are equal to 1. This happens since such structures are subleading in the $U_i \to 0$ limit. The constants $\mathbb{B}_{\ell_1\ell_2\ell_3}^{(m)}$ are the coefficients multiplying the degree-$m$ polynomial of $\ln U_i$ in the block associated to the tensor structure labeled by $\ell_1,\,\ell_2$ and $\ell_3$. These coefficients can be read off from equation  (\ref{eq:blockstresstensorWL}) in appendix \ref{app:HyperInt}. We remark that, as is well known, the OPE coefficients of the stress tensor are not all independent and in fact satisfy
\begin{align}
	P_{011}&= -2\,\frac{8 (P_{000}+P_{001})+ d(d+2)P_{002}}{(d+4)(d-2)}\,, \\ P_{111}&= \frac{32(2+d)P_{000} +8d(6+d)P_{001} -4d(d^2-20)P_{002}}{(d-2)^2(d+2)(d+4)}\,, \nonumber
\end{align}
since its correlation functions satisfy conservation equations \cite{costa:2011}. This means that the different OPE coefficients associated to the $\ell_i$ tensor structures are related to a set of three independent numbers.

\begin{figure}
	\centering
	\includegraphics[width=0.49\linewidth]{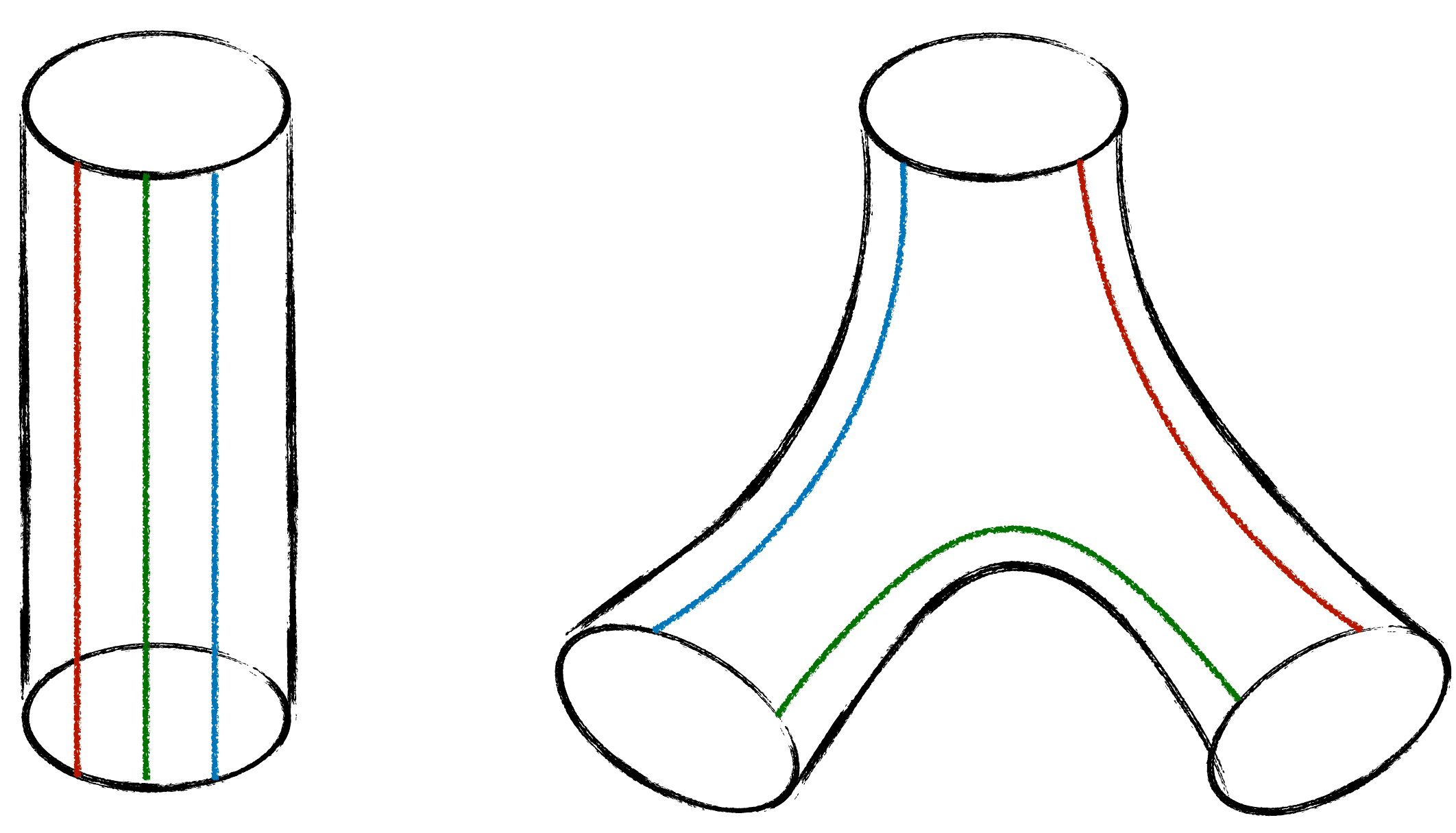}
	\caption{Schematic representation of the gravitational processes dual to the six-point comb channel on the left and to the six-point snowflake channel on the right. In the comb case, three particles come from the infinite past, interact weakly and continue towards future infinity. In the snowflake case, the blue and red particles come from the past infinity of two different time coordinates, say $t_1$ and $t_2$, respectively. The blue one travels to future infinity along $t_1$ and the red one along $t_2$. A third, green particle comes from past infinity in the $t_1$ direction and moves towards past infinity in $t_2$. The process can also be interpreted in other similar ways by permuting the role of the OPEs.}
	\label{fig:snowflakequantization}
\end{figure}

We end this section with a speculative holographic interpretation of our bootstrap results which can be skipped by the more orthodox readers. In a four-point function, radial quantization allows us to visualize a weak gravitational process in AdS where two particles with large relative angular momentum come from the infinite past, interact, and continue towards the infinite future. This picture can be generalized for the six-point function in the comb channel, which instead corresponds to a three-body gravitational interaction. However, in the snowflake OPE that we analyzed, one cannot assign a single time coordinate which leads to the cylinder picture. Instead, this channel corresponds to a gravitational process where the asymptotic states are defined with respect to distinct time coordinates\footnote{We thank Pedro Vieira for discussions on this point.}, where the underlying geometry is instead a "pair of pants". The physical process is more easily understood by inspecting figure \ref{fig:snowflakequantization}.

\section{Examples}
\label{sec:examples}

Consistency conditions of the bootstrap equations for higher-point functions impose constraints on the behaviour of three point functions of spinning operators as we have seen in the previous sections. The goal of this section is to extract OPE coefficients of spinning operators by performing an explicit conformal block decomposition of the generalized free field theory correlator, as well as theories with cubic couplings, 
and confirm some of our previous results. 

\subsection{Generalized free theory}

The six-point function of operators $\phi$ in a generalized free field theory is given by 
\begin{align}
\langle \prod_{i=1}^6\phi(x_i) \rangle_{\textrm{MFT}} =  \sum_{perm} \langle \phi(x_1)\phi(x_2)\rangle\langle\phi(x_3)\phi(x_4)\rangle\langle\phi(x_5)\phi(x_6)\rangle =
\sum_{perm}  \frac{1}{(x_{12}^2x_{34}^2x_{56}^2)^{\Delta_\phi}}\,,
\label{eq:6ptMFT}
\end{align}
where we should sum over all permutations of  operator positions. We can extract a prefactor $(x_{12}^2x_{34}^2x_{56}^2)^{\Delta_\phi}$ to write  everything just in terms of cross-ratios, 
\begin{align}
&(x_{12}^2x_{34}^2x_{56}^2)^{\Delta_\phi} \langle \prod_{i=1}^6\phi(x_i) \rangle_{\textrm{MFT}}   = 1 + (u_1u_3u_5)^{\Delta_{\phi}} \left(1+(u_2u_4u_6)^{-\Delta_{\phi}}+\sum_{i=1}^3U_i^{-\Delta_{\phi}}\right)
\nonumber\\
&+ \sum_{i=1}^3\left[\left(\frac{u_{2i+1}u_{2i+3}}{U_{2i-1}}\right)^{\Delta_{\phi}} + \left(\frac{u_{2i-1}u_{2i+1}u_{2i+3}}{u_{2i+2}U_{2i-1}}\right)^{\Delta_{\phi}} +\left(\frac{u_{2i+1}u_{2i+3}U_{2i+1}}{u_{2i+2}U_{2i-1}}\right)^{\Delta_{\phi}} \right].
\label{eq:GeneralizedFreeCross}
\end{align}
The prefactor we have extracted is appropriate to analyze the OPE limit in the channel $(12)(34)(56)$. The first term in (\ref{eq:GeneralizedFreeCross}) corresponds to the exchange of three identity operators and the others can contain one identity and two double-twist operators, or three double-twist operators. A systematic analysis of the operators that are exchanged in the OPE in these three channels can be done using the six-point conformal blocks~\cite{Vieira:2020xfx} or the Casimir differential operator together with the boundary condition of the block in the lightcone limit~\cite{Bercini:2021jti}. We obtained for the OPE of three leading double-twist operators, which can not be extracted from the four-point function of $\phi$, the result
\begin{align}
P_{J_i\ell_i}= \prod_{i=1}^3 \frac{\left(J_i+\ell_i-\sum_j \ell_j+1\right)_{(\sum_j \ell_j)-\ell_i}(\Delta_{\phi})_{\frac{J_i}{2}}  (\Delta_{\phi})_{J_i}}{2^{\ell_i-1}J_i!\,\ell_i!\,(\Delta_{\phi})_{\ell_i}\left(\frac{J_i+2\Delta_{\phi}-1}{2}\right)_{\frac{J_i}{2}}}\,.
\end{align}
By taking first the large $J_i$ and then the large $\ell_i$ limit we recover the asymptotic behavior (\ref{eq:P6pt3indentities}) derived from the lightcone bootstrap in the previous section.

Note that for a free theory with $\Delta_\phi=(d-2)/2$ this is the full set of OPE data that can be extracted from this correlator. In a generalized free theory there are subleading double-twist operators $\phi \Box^n \partial^J \phi$ whose OPE coefficients could be extracted. 

\subsection{$\phi^3$ theory in $d=6-\epsilon$}
We now consider turning on a cubic coupling which will allow us to further test our predictions involving, for example, the five-point function which vanishes for mean field theory.
The five-point function in $\phi^3$ theory is given by\footnote{This result can be obtained easily with the method of skeleton expansions as presented in \cite{Goncalves:2018nlv}. It would be interesting to do conformal block decomposition for five- and six-point correlators in $\phi^3$ and see how the respective spinning OPE coefficients compare with the ones in $\mathcal{N}=4$ SYM \cite{Bercini:2021jti}. }
\begin{align}
\langle \prod_{i=1}^5\phi(x_i) \rangle = \sum_{perm}\langle\phi(x_1)\phi(x_2) \rangle \langle\phi(x_3)\phi(x_4)\phi(x_5) \rangle  
+ \langle\prod_{i=1}^5\phi(x_i)\rangle\Big|_{\textrm{conn}} \,.
\end{align}
This correlation function only has odd powers 
of $\epsilon$ as  can be seen by drawing a few Feynman diagrams or from  the strucutre of perturbation theory around the $\mathbb{Z}_2$ symmetric   free theory. 
The leading term is a factorized correlator given by a product of a two-point function and a three-point function. The two-point function starts at the free theory order, but the three-point functions starts at order $\epsilon$, with a tree level contact diagram.
The connected contribution starts at order $\epsilon^3$ and coexists with corrections to the factorized correlator.
To leading order in the $\epsilon$ expansion the connected contribution is given by 
\begin{equation}
 \langle \phi(x_1)\dots \phi(x_5) \rangle\Big|_{\textrm{conn}}  = 
 \sum_{perm} \frac{\big(C^{(1)}_{\phi\phi\phi}\big)^3}{x_{12}^2x_{34}^2}\int \frac{d^6x_0}{x_{10}^2x_{20}^2x_{30}^2x_{40}^2(x_{50}^2)^2}\,.
\end{equation}
This six-dimensional integral is proportional to a $D$-function $D_{11112}$ which we analyze in Appendix \ref{app:Dfuncs}.

\subsubsection{Disconnected contribution to the five-point function}
\label{disconnected5pt}

Let us write the block decomposition as
\begin{equation}
\langle \phi(x_1)\dots \phi(x_5)\rangle^{(1)}= \frac{x_{13}^2}{x_{12}^4x_{34}^4x_{15}^2x_{35}^2} \sum_{k_1,k_2,\ell} P^{(1)}_{k_1k_2\ell}\, G_{k_1k_2\ell}^{(12)(34)}(u_i)\,,
\end{equation}
where the superscript  $(1)$ indicates the order in the $\epsilon$ expansion.
We used that $\Delta_\phi =2 + O(\epsilon)$ and that   $P_{k_1k_2\ell}$ starts at order $\epsilon$. 
Our goal is to derive the spectrum and OPE coefficients of the operators exchanged in the $(12)(34)$  channel for the leading disconnected contribution that is given by
\begin{align}
&\langle \prod_{i=1}^5\phi(x_i) \rangle^{(1)}= \frac{C_{\phi\phi\phi}^{(1)}\,x_{13}^2}{x_{12}^4x_{34}^4x_{15}^2x_{35}^2}\left( u_1^{\frac{\Delta_\phi}{2}}+\left( \frac{u_3 u_5}{u_4}\right)^{\frac{\Delta_\phi}{2}} + \left( \frac{u_1 u_3}{u_2 u_4^2u_5}\right)^{\frac{\Delta_\phi}{2}}\bigg[\big(u_1 u_4^2\big)^{\frac{\Delta_\phi}{2}}+\big(u_3 u_5^2\big)^{\frac{\Delta_\phi}{2}}+\right.  \nonumber\\
&\left.\big(u_2 u_4^2 u_5^2\big)^{\frac{\Delta_\phi}{2}} \Big(u_1^{\frac{\Delta_\phi}{2}}+ u_3^{\frac{\Delta_\phi}{2}}\Big)\bigg] + \left( \frac{u_1^2 u_3^2}{u_2^2 u_4}\right)^{\frac{\Delta_\phi}{2}}\left[ 1+(u_2 u_4 u_5)^{\frac{\Delta_\phi}{2}} + u_2^{\Delta_\phi}\Big(u_4^{\frac{\Delta_\phi}{2}}+u_5^{\frac{\Delta_\phi}{2}} \Big) \right] \right)\,.
\end{align}
To obtain the block decomposition we use two independent methods which  serves as a  cross-check of the calculation. Firstly we consider the Euclidean expansion of the five-point block discussed in Appendix E of \cite{Goncalves:2019znr}, and match it to the small $u_1$ and $u_3$  expansion of the correlator. Using this we can obtain as many OPE coefficients as we desire. We can then conjecture a general form for arbitrary $J_1,J_2$ and $\ell$, which we subsequently test by comparing to the explicit higher order results. Alternatively, we can use a generalization of the technique of \cite{Gliozzi:2017hni} to higher-point correlators~\cite{Bercini:2021jti}. We act with the Casimir differential operators on the correlator in terms of its small $u_1,u_3$ expansion. Since the conformal blocks are eigenfunctions of the Casimir operator, we can fix the OPE coefficients order by order in $u_1,u_3$ by acting recursively with the differential operators. Again, we can do this to arbitrarily high order, guess the general form of the coefficients and check it to even higher order.

We find that depending on which pair of operators form the two-point function we have  different sets of operators being exchanged. 
When the two-point function is between points $x_1$ and $x_2$, we have the identity in the (12) OPE and $\phi$ in the (34) OPE.
The product of OPE coefficients is simply given by $P_{{\cal I}\phi}^{(1)}= C_{\phi \phi \phi}^{(1)}$. Similarly, when 
the two-point function is between points $x_3$ and $x_4$, we have $P_{\phi{\cal I}}^{(1)}= C_{\phi \phi \phi}^{(1)}$.
When the two-point function is between points $x_1$ and $x_5$, or between $x_2$ and $x_5$, 
the result is  less trivial since it leads to an expansion with an infinite number of operators.
Adding up these two contributions,  we find in the (12) OPE the double-twist operators $ [\phi \phi]_{0,J}$, with 
 dimension $4+J$ and (even) spin $J$, along with the operator $\phi$ in the (34) OPE. 
 In this case we obtain $P_{[\phi \phi]_{0,J}\phi}^{(1)}= C_{\phi \phi \phi}^{(1)}C^2_{\phi\phi [\phi \phi]_{0,J}}$, where
		\begin{equation}
		C^2_{\phi\phi [\phi \phi]_{0,J}}= \frac{2^{J+1} \Gamma (J+2)^2 \Gamma
			(J+3)}{\Gamma (J+1) \Gamma (2 J+3)}\,, 
		\end{equation} 
which is the usual formula for the OPE coefficients of two scalar operators and a leading double-twist operator, which holds in MFT with $\Delta_\phi=2$. 
We  may also consider  the factorised correlator with generic $\Delta_\phi$.\footnote{For example studying $\phi^3$ theory in AdS with a massive scalar such that  $m^2=\Delta_\phi(\Delta_\phi-d)$.} In this case  we have several infinite towers of subleading twist operators with dimension $2\Delta_\phi +2n +J$ and spin $J$. We checked that  the  OPE coefficients are again given by the four-point  MFT result. This can be easily understood by using the convergent OPE in the $(34)$ channel, as discussed in section \ref{sec:ID(12)5-point}.
A similar story holds when the  two-point function is between points $x_3$ and $x_5$, or between $x_4$ and $x_5$,

Finally we can have a two-point function between $x_1$ and $x_3$, $x_1$ and $x_4$, $x_2$ and $x_3$, and  $x_2$ and $x_4$,
 which are the most non-trivial and interesting cases. Together they admit an expansion in terms of blocks where the exchanged operators are $ [\phi \phi]_{0,J_1}$  in the 
 (12) OPE and  $ [\phi \phi]_{0,J_2}$ in the (34) OPE. Thus we access OPE coefficients with one scalar and two spinning operators, which have the extra quantum number $\ell$. It is not hard to propose the formula for the OPE coefficients in the case $\ell=0$, where the dependence in $J_1$ and $J_2$ turns out to factorize due to the nature of the tensor structure of $\ell=0$. We find, for generic $\Delta_\phi$,
		\begin{equation}
		P_{[\phi \phi]_{0,J_1}[\phi \phi]_{0,J_2} \ell=0}^{(1)} =\pi  2^{6-4 \Delta _{\phi }}
		 \prod_{i=1}^2\frac{ 2^{-J_i} \Gamma \left(J_i+\frac{\Delta _{\phi
				}}{2}\right) \Gamma \left(J_i+2 \Delta
				_{\phi }-1\right)}
				{\Gamma \left(J_i+1\right)
				 \Gamma \left(\frac{\Delta _{\phi }}{2}\right) \Gamma
				\left(\Delta _{\phi }\right) \Gamma \left(J_i+\Delta _{\phi }-\frac{1}{2}\right)
				}\,,
			\end{equation}	
			which for the $\Delta_\phi=2$ case drastically simplifies to
			\begin{equation}
			P_{[\phi \phi]_{0,J_1}[\phi \phi]_{0,J_2} \ell=0}^{(1)} = \frac{\pi  2^{-J_1-J_2-2} \Gamma \big(J_1+3 \big) \Gamma \big(J_2+3\big)}
			{\Gamma	\big(J_1+\frac{3}{2}\big) \Gamma \big(J_2+\frac{3}{2}\big)}\,.
			\end{equation}
			For higher $\ell$ we find that the $J_1$ and $J_2$ dependence no longer factorizes. Instead, for $\Delta_\phi=2$ 
			we find that the ratio $P^{(1)}_{[\phi \phi]_{0,J_1}[\phi \phi]_{0,J_2} \ell} /	P^{(1)}_{[\phi \phi]_{0,J_1}[\phi \phi]_{0,J_2} \ell=0}$ is given by a symmetric polynomial in $J_1$ and $J_2$, with maximum degree $2\ell$ in both variables combined and maximum degree $\ell$ in each variable separately. For example, the first few polynomials are given by
			\begin{align}
			\frac{P_{\ell=1}}{P_{\ell=0}} &= \frac{1}{2}\Big(3+(J_1+J_2)+ J_1J_2\Big) \,,
			\nonumber\\
			\frac{P_{\ell=2}}{P_{\ell=0}} &= \frac{1}{12} \Big(J_2^2 J_1^2+J_2 J_1^2+J_2^2 J_1+7 J_2 J_1+6(J_1+J_2)+18\Big) ,
		        \\
			\frac{P_{\ell=3}}{P_{\ell=0}} &=\frac{1}{144} \Big(J_2^3 J_1^3-(J_2 J_1^3+J_2^3 J_1)+12 J_2^2 J_1^2
			\nonumber \\
			&+12(J_2 J_1^2+J_2^2
			J_1) +85 J_2 J_1+72( J_1+ J_2)+216\Big),
			\nonumber
			\end{align}
where here we used the shorthand notation $P_{\ell=i}\equiv P^{(1)}_{[\phi \phi]_{0,J_1}[\phi \phi]_{0,J_2}\ell=i}$.
We can easily write down these polynomials to a very high order.\footnote{We can also write down a few of them for general $\Delta_\phi$. In this case there is also a simple additional denominator.} Unfortunately we did not find a closed form at arbitrary $\ell$. Nevertheless, 
we could perform the simpler task of finding the large $J_1,J_2$ at fixed $\ell$ behavior, which in fact we were able to do for generic $\Delta_\phi$. We found that 
			\begin{equation}
			\frac{P^{(1)}_{[\phi \phi]_{0,J_1}[\phi \phi]_{0,J_2}\ell}}{P^{(1)}_{[\phi \phi]_{0,J_1}[\phi \phi]_{0,J_2} \ell=0} }\approx \frac{(J_1 J_2)^\ell}{\Gamma(\ell+1)(\Delta_\phi)_\ell}\,,
			\end{equation}
Combining this result with the large spin behavior of the $\ell=0$ OPE coefficient, and then taking the large $\ell$ limit, we find a perfect match with  formula (\ref{eq:OPEid(34)5-pt}) obtained using the lightcone bootstrap!
	
\subsubsection{Comments on the six-point function}
The six-point function of a scalar $\phi$ in the $\epsilon$ expansion is given by 
\begin{align}
\langle \prod_{i=1}^6\phi(x_i) \rangle &= \sum_{perm} \langle\phi(x_1)\phi(x_2) \rangle \langle\phi(x_3)\phi(x_4) \rangle \langle\phi(x_5)\phi(x_6) \rangle  + 
\sum_{perm}\langle\phi(x_1)\phi(x_2) \rangle\langle \prod_{i=3}^6\phi(x_i) \rangle\big|_{\textrm{conn}} 
\nonumber\\
&+ \sum_{perm} \langle \phi(x_1)\phi(x_2)\phi(x_3)\rangle\langle \phi(x_4) \phi(x_5)\phi(x_6)  \rangle
+\langle \prod_{i=1}^6\phi(x_i) \rangle\big|_{\textrm{conn}} \,. 
\end{align}
The leading term is given by the mean field theory discussed above (with $\Delta_\phi=2+ O(\epsilon)$) and is of order $\epsilon^0$. The partialy factorized terms (two-point function times four-point function and three-point function times another three-point function) begin at order $\epsilon^2$. These have subsequent corrections of order $\epsilon^4$, which is the order at which the connected contributions begin.  
At leading order the latter  is given by 
\begin{equation}
\langle \prod_{i=1}^6\phi(x_i) \rangle\big|_{\textrm{conn}}  =
C_{\phi\phi\phi}^4 \left(
\int \frac{d^6x_0}{x_{12}^2 x_{34}^2 x_{56}^2\prod_{i=1}^6x_{i0}^2}+\int\frac{d^6x_7d^6x_8}{x_{12}^2x_{17}^2x_{27}^2(x_{37}^2)^2x_{47}^2x_{48}^2(x_{58}^2)^2(x_{68}^2)^2x_{78}^2} \right)+\textrm{perm},
\nonumber
\end{equation}
where  the first  integral is the same as the  six-point $D$-function $D_{111111}$, which we analyze in Appendix \ref{app:Dfuncs}. 
It would be nice to systematically study all these corrections and to match the asymptotics of the OPE coefficients with the 
lightcone bootstrap results presented in section  (\ref{sec:6-pt}).

\section{Discussion}
\label{sec:Discussion}

We have shown how to use the lightcone bootstrap for five- and six-point functions to determine the large spin behaviour of some new OPE coefficients. For the
five-point function, in the case of a direct-channel identity exchange we determined the large $J_1,J_2$ and $\ell$ behaviour of the OPE coefficient
$C^{(\ell)}_{\phi [\phi\phi]_{0,J_1}[\phi\phi]_{0,J_2}}$ in the cross-channel. For the case of a leading twist exchange in the direct-channel, including the possibility of the stress  tensor 
exchange, we determined  the asymptotic behaviour of $C^{(\ell)}_{\phi [\phi\phi]_{0,J_1}[\phi{\cal O}_*]_{0,J_2}}$.
For the six-point function, in the case of a direct-channel identity exchange, we determined the 
large $J_i$ and $\ell_i$ behaviour of 
$C^{(\ell_i)}_{ [\phi\phi]_{0,J_1}[\phi\phi]_{0,J_2}[\phi\phi]_{0,J_3}}$. Subleading corrections to this OPE coefficient due to the direct-channel leading twist exchange were also
bootstrapped. An interesting interpretation of these results emerges in connection to the origin limit $U_i \to 0$. In this limit we observed that the correlation function diverges at most as $\log U_i ^3$ in contrast with the planar gauge theory case where the divergences can be an arbitrary power of $\log U_i$ \cite{Vieira:2020xfx,Bercini:2021jti}. The difference between these results follows from the existence or not of a twist gap in a CFT correlator.

Our knowledge of higher-point conformal blocks is still in its infancy. In particular, our work was limited to the leading order expansion of the blocks in the lightcone 
limit. In our notation this corresponds to the leading term in the limit $u_{\rm odd}\to0$ that defines the lightcone blocks. 
It would be very interesting to study subleading corrections to the blocks in this limit, which would allow us to bootstrap OPE coefficients with 
subleading double-twist operators of the form $ [\phi\phi]_{n,J}$ and $[\phi{\cal O}_*]_{n,J}$. Additionally, to simplify our analysis, we often took the origin limit $U_i \to 0$. It would also be interesting to compute subleading terms in this expansion, which can be done using only the available lightcone blocks.

In this paper we only considered the lightcone blocks in the snowflake channel. For the six-point function the comb channel block would lead  to a different expansion
involving the exchange of mixed symmetry operators, which we expect to be of triple-twist type. Such operators are expected to be degenerate at large spin, but this degeneracy should be lifted at finite spin. It is a very interesting question whether the bootstrap would be able to address this question in the large spin expansion. This could be a sign of 
analyticity in spin for each triple-twist family.

Analyticity is also an open question regarding the new OPE coefficients whose large spin behaviour we determined in this work. In this case, since there is a unique operator 
at each spin and analyticity has been proven in the  simpler case of the OPE coefficient $C_{\phi \phi  [\phi\phi]_{0,J}}$, we could also expect analyticity to hold.
However, the situation here is more subtle because in this case we also have the label $\ell_i$ that parametrizes  tensor structures and is basis dependent.  
This is an interesting question since in the case of $C^{(\ell_i)}_{ [\phi\phi]_{0,J_1}[\phi\phi]_{0,J_2}[\phi\phi]_{0,J_3}}$ it would connect to the 
OPE coefficients of the low spin contributions of this family of operators. In particular, for an appropriate choice of the external scalar operators, this will connect to the OPE coefficient between three energy-momentum tensors $C^{(\ell_i)}_{TTT}$.
In this case one would hope to derive reliable predictions by including the contributions from the 
first terms in the large $J$ expansion.  

Analyticity in spin is also important for Regge theory of higher-point functions. This is clear since Conformal Regge Theory relies on the analytic continuation 
in spin \cite{Costa:2012cb}. In the four-point case the Lorentzian inversion formula established such analyticity \cite{Caron-Huot:2017vep}. Thus, deriving a  Lorentzian inversion formula for 
higher-point functions would shed light in this problem and, most likely, sistematize the calculations reported in this  work.

A more ambitious problem is to set up the Euclidean numerical bootstrap for higher-point functions, with obvious gains in the available CFT data. As it is well known 
positivity is a key ingredient in the numerical bootstrap of four-point functions. In the case of the six-point function it is possible to choose reflection positive kinematics, 
however such positivity is not guaranteed term by term in the block expansion. The situation looks even worse in the case of the 
five-point function, since this correlator can not be seen as a positive norm of a state. One possibility would be to consider a positive semi-definite matrix whose matrix elements
would involve the four-, five- and six-point function. We hope to return to these questions in the future.

\section*{Acknowledgements}
We would like to thank Apratim Kaviraj, David Meltzer, Aaditya Salgarkar and Pedro Vieira for useful discussions and comments. 
This research received funding from the Simons Foundation grants \#488637 (Simons collaboration on the non-perturbative bootstrap).
Centro de F\'{i}sica do Porto is partially funded by Funda\c{c}\~{a}o para a Ci\^{e}ncia e Tecnologia (FCT) under the grant UID04650-FCUP. AA is funded by FCT under the IDPASC doctoral program with the fellowship PD/BD/135436/2017. JVB is funded by FCT with fellowship DFA/BD/5354/2020, co-funded by Norte Portugal Regional Operational Programme (NORTE 2020), under the PORTUGAL 2020 Partnership Agreement, through the European Social Fund (ESF).

\appendix

\section{Higher-point Conformal Blocks}
\label{app:Blocks}
 
\subsection{Mellin amplitudes}
\label{app:Mellin}
The Mellin amplitude of a connected $n$-point function of scalar conformal correlators can be defined as\cite{Mack:2009mi,Penedones:2010ue}
\begin{align}
\left\langle  \mathcal{O} _1\left(x_1\right)...\mathcal{O} _n\left(x_n\right) \right\rangle=\int 
\left[d\gamma\right]M\left(\gamma_{ij}\right)\prod_{1\leq i<j\leq n} \Gamma\left(\gamma_{ij}\right)\left(x_{ij}^2\right)^{-\gamma_{ij}}\ ,
\label{Mellinamplitude}
\end{align}
where $\left[d\gamma\right]$ denotes an integration with the constraints
\begin{align}
\label{eq:Mellinconditions}
\sum_{i=1}^n \gamma_{ij}=0\ ,  \ \ \ \ 
\gamma_{ij}=\gamma_{ji}\ ,  \ \ \ \ 
\gamma_{ii}=-\Delta_i\ . 
\end{align}

It is a well known fact by now that the OPE implies that the Mellin amplitude is a meromorphic function of the Mellin variables $\gamma_{ij}$. For each exchange of a primary operator with dimension $\Delta$ and spin $J$ there is an infinite set of poles in the Melllin amplitude, 
\begin{align}
M \approx 
\frac{\mathcal{Q}_m  }{\gamma_{LR}-\left(\Delta-J+2m\right)}\ ,\ \ \ \ \ \ \ 
m=0,1,2,\dots \ ,
\end{align} 
where
\begin{align}
\gamma_{LR}=-\left(\sum_{i=1}^{k}p_i\right)^2=\sum_{a=1}^{k}\sum_{i=k+1}^n\gamma_{ai}\,,
\end{align}
with the $p_i$ defined such that $p_i\cdot p_j= \gamma_{ij}$.
The residue $\mathcal{Q}_m$ is related to lower point functions and conformal blocks\cite{Goncalves:2014rfa}. The label $m$ is associated to the contribution of higher twist descendant operators. 

In particular, the equivalence between (\ref{Mellinamplitude}) and conformal block decompositions~(\ref{eq:5ptexpansion}) and~(\ref{eq:6ptexpansion}) imposes that the Mellin amplitude for the five and six-point correlator needs to have the following poles
\begin{align}
\label{eq:Mellinampbehav}
M_{5} &\approx \frac{\sum_{l}C_{12J_1}C_{34J_2}C_{5J_1J_2}^{(l)} F_{l}(\gamma)  \, }{\left(\gamma_{12}-\frac{J_1-\Delta_{J_1}+2 \Delta_\phi}{2}\right)\left(\gamma_{34}-\frac{J_2-\Delta_{J_2}+2 \Delta_\phi}{2}\right)}\,,\\
M_{6} &\approx \frac{\sum_{l_i}C_{12J_1}C_{34J_2}C_{56J_3}C_{J_1J_2J_3}^{(l_i)}  F_{l_1l_2l_3}(\gamma)  }{\left(\gamma_{12}-\frac{J_1-\Delta_{J_1}+2 \Delta_\phi}{2}\right)\left(\gamma_{34}-\frac{J_2-\Delta_{J_2}+2 \Delta_\phi}{2}\right)\left(\gamma_{56}-\frac{J_3-\Delta_{J_3}+2 \Delta_\phi}{2}\right)}\,,
\end{align}
where the functions $F_{l}$ and $F_{l_1l_2l_3}$ are computed by Mellin transforming the lightcone blocks used in this paper and $C_{XYZ}$ are OPE coefficients. In the following we will determine the form of $F_{l}$ and $F_{l_1l_2l_3}$ for some specific cases\footnote{It would be interesting to repeat the analysis of appendix A.1 of \cite{Costa:2012cb} for higher-point functions.}. 

Let us start with the five-point lightcone conformal block (\ref{eq:5ptlightconeblockdef}) with identical scalar operators $\mathcal{O}_i=\phi$, and write the numerator using the binomial formula
\begin{align}
&\sum_{i_1,i_2,j_1,j_2}{{J_1-l}\choose{i_1}}{{i_1}\choose{j_1}}{{J_2-l}\choose{i_2}}{{i_2}\choose{j_2}}\int  \frac{[dt_1][dt_2]t_1^{i_2-j_2}(1-t_1)^{j_2}t_2^{i_1-j_1}(1-t_2)^{j_1}}{ \big(1-(1-t_2)u_4\big)^{\frac{\Delta_2-\Delta_1+J_1+J_2-2l+\Delta_\phi}{2}}}\\
&\times \frac{u_1^{\frac{\Delta_1-J_1}{2}}u_3^{\frac{\Delta_2-J_2}{2}}(1-u_2)^{l}u_2^{j_1+j_2}u_5^{i_1}u_4^{i_2}}{\big(1-(1-t_1)(1-t_2)(1-u_2)\big)^{\frac{\Delta_1+\Delta_2+J_1+J_2-\Delta_\phi}{2}}\big(1-(1-t_1)u_5\big)^{\frac{\Delta_1-\Delta_2+J_1+J_2-2l+\Delta_\phi}{2}}}\nonumber\,.
\end{align}
Next we introduce three Mellin variables $s_1,s_2,s_3$ with respect to the cross-ratios $u_2,u_4$ and $u_5$,
\begin{align}
\label{eq:Mellin5ptss}
&\sum_{i_1,i_2,j_1,j_2}{{J_1-l}\choose{i_1}}{{i_1}\choose{j_1}}{{J_2-l}\choose{i_2}}{{i_2}\choose{j_2}}u_1^{\frac{\Delta_1-J_1}{2}}u_3^{\frac{\Delta_2-J_2}{2}}(1-u_2)^{l}\int ds_1ds_2ds_3 \Gamma(s_1)\Gamma(s_2)\Gamma(s_3)\nonumber\\
&u_2^{-s_1+j_1+j_2}u_4^{-s_2+i_2}u_5^{-s_3+i_1+\frac{\Delta_\phi}{2}}\left(\frac{\Delta_1+J_1+\Delta_2+J_2-\Delta_{\phi}}{2}\right)_{-s_1}\nonumber\\
&\left(\frac{\Delta_2-\Delta_1-2l+J_1+J_2+\Delta_{\phi}}{2}\right)_{-s_2}\left(\frac{\Delta_1-\Delta_2-2l+J_1+J_2+\Delta_{\phi}}{2}\right)_{-s_3}\mathcal{B}_{s_1,s_2,s_3}\,,
\end{align}
with the function $\mathcal{B}_{s_1,s_2,s_3}$ given by
\begin{align}
\mathcal{B}_{s_1,s_2,s_3} &= \int [dt_1] [dt_2] (1-t_1)^{i_2-j_2-s_3}t_1^{\frac{\Delta_2-\Delta_1-J_1-J_2+2(s_3-s_1)+2l-\Delta_{\phi}+2j_2}{2}}t_2^{i_1-j_1-s_2}\nonumber\\
&(1-t_2)^{\frac{\Delta_1-\Delta_2-J_1-J_2+2l+2(s_2-s_1)+2j_1-\Delta_{\phi}}{2}}\big(1-t_1(1-t_2)\big)^{\frac{2s_1-J_1-J_2-\Delta_1-\Delta_2+\Delta_{\phi}}{2}}.
\end{align}
For $J_1=J_2=0$ the function $\mathcal{B}_{s_1,s_2,s_3}$ can be integrated to 
\begin{align}
\mathcal{B}_{s_1,s_2,s_3} = \frac{\Gamma(\Delta_1)\Gamma(\Delta_2)\Gamma\big(\frac{\Delta_1-\Delta_{\phi}+2(s_2-s_1)}{2}\big)\Gamma\big(\frac{\Delta_2-\Delta_{\phi}+2(s_3-s_1)}{2}\big)\Gamma\big(\frac{2(s_1-s_2-s_3)+\Delta_{\phi}}{2}\big)}{\Gamma^2\big(\frac{\Delta_1}{2}\big)\Gamma^2\big(\frac{\Delta_2}{2}\big)\Gamma\big(\frac{\Delta_1+\Delta_2-2s_1-\Delta_{\phi}}{2}\big)}\,.
\end{align}
One of the advantages of this Mellin representation for the conformal block is that it makes it easier to study certain limits. For example, to get the leading term in the $u_2,u_4,u_5\rightarrow 0$ limit we just have to close each contour $s_1,s_2,s_3$ to the left picking all the poles along the way. Notice that $\mathcal{B}_{s_1,s_2,s_3}$ for generic spin can be written as a $\,_3F_2$ hypergeometric series
\begin{align}
&\mathcal{B}_{s_1,s_2,s_3} =  \frac{ \Gamma \left(\frac{J_1+\Delta_1+1}{2}\right) \Gamma \left(\frac{J_2+\Delta_2+1}{2}\right) \Gamma \left(i_2-j_2+\frac{J_1}{2}-s_3+\frac{\Delta
		_1}{2}\right) \Gamma \left(i_1-j_1+\frac{J_2}{2}-s_2+\frac{\Delta _2}{2}\right)  }{ 2^{2-\Delta _1-\Delta _2-J_1-J_2}\pi\,  \Gamma \left(\frac{J_1}{2}+\frac{\Delta
		_1}{2}\right) \Gamma \left(\frac{J_2}{2}+\frac{\Delta _2}{2}\right) } \nonumber\\
& \frac{\Gamma
	\left(\ell +j_1-\frac{J_1}{2}-s_1+s_2+\frac{\Delta _1}{2}-\frac{\Delta _{\phi
		}}{2}\right)\Gamma \left(\ell +j_2-\frac{J_2}{2}-s_1+s_3+\frac{\Delta
		_2}{2}-\frac{\Delta _{\phi }}{2}\right)}{ \Gamma \left(\ell +i_1-\frac{J_1}{2}+\frac{J_2}{2}-s_1+\frac{\Delta
		_1 +\Delta_2-\Delta_\phi}{2}\right) \Gamma \left(\ell
	+i_2+\frac{J_1}{2}-\frac{J_2}{2}-s_1+\frac{\Delta _1+\Delta_2-\Delta_\phi}{2}\right)}\\
& \setlength\arraycolsep{1pt}
{}_3 F_2\left(\begin{matrix}-\frac{\Delta _{\phi }}{2}+\frac{\tau _1}{2}+j_1-s_1+s_2+\ell\,,\, -\frac{\Delta _{\phi }}{2}+\frac{\tau _2}{2}+j_2-s_1+s_3+\ell\,,\,-\frac{\Delta _{\phi }}{2}+\frac{h_1}{2}+\frac{h
	_2}{2}-s_1\\-\frac{\Delta _{\phi }}{2}+\frac{\Delta
	_1}{2}+\frac{\Delta _2}{2}+i_2+\frac{J_1}{2}-\frac{J_2}{2}-s_1+\ell\,,\,-\frac{\Delta _{\phi
	}}{2}+\frac{\Delta _1}{2}+\frac{\Delta _2}{2}+i_1-\frac{J_1}{2}+\frac{J_2}{2}-s_1+\ell&\end{matrix};1\right)\,.\nonumber
\end{align}

To find $F_{l}$ one needs to relate the Mellin transform we have computed to the Mellin amplitude definition in (\ref{Mellinamplitude}). We use the conditions~(\ref{eq:Mellinconditions}) to write the Mellin amplitude in terms of five independent Mellin variables, namely:$\gamma_{12}, \gamma_{34}, \gamma_{13}, \gamma_{15}, \gamma_{35}$. After computing the integral in $\gamma_{12}$ and $\gamma_{34}$, we can relate the two sets of Mellin variables, $s_i$'s and $\gamma_{ij}$, by demanding the exponents of the cross-ratios to be the same on both expressions. To do so, we first expand $(1-u_2)^{l}=\sum_{k}\binom{l}{k} (-u_2)^{k}$. We find then the relation
\begin{align}
&s_1=\frac{2 j_1+ 2 j_2+2 k-J_2+\Delta_{J_2}-2 \gamma_{13}-2 \gamma_{35}}{2}\,, \qquad
s_3= \gamma_{15}+i_1\,,\nonumber\\
&s_2= \frac{2 i_2+ J_1-J_2-\Delta_{J_1}+\Delta_{J_2}+\Delta_\phi-2 \gamma_{35}}{2}
\,.
\end{align}
This relation depends on indices that are summed over. Thus, performing the change of variables in~(\ref{eq:Mellin5ptss}) leads us to finite sums of contour integrals. We would like to swap the order of sums and integrals to be able to write $F_{l}$ from those finite sums. This can be done if we are allowed to move, without crossing any poles, all the contours to the same region. Assuming this can be done~\footnote{To be rigorous one needs to study in detail the very complicated pole structure of the integrand. This is particularly challenging due to the possible presence of fake poles. As discussed in~\cite{EYYuan:2018}, gamma functions that depend on more than a single Mellin variable can naively suggest the presence of families of poles that differ depending on the order of integration of the Mellin variables. These poles are fake.}, to find $F_{l}$ is just simple algebra. For specific values of spin and scaling dimension of the exchanged operators, it is easy to see that $F_{l}$ defined in this way is, as expected, a polynomial in the Mellin variables $\gamma_{13}, \gamma_{15}, \gamma_{35}$ whose degree depends on $J_1, J_2, l$.

It is possible to repeat the same analysis for the six-point conformal block in the lightcone. Since the method is essentially the same we will just quote here the Mellin transform of the block for the exchange of scalar operators
\begin{align}
&\prod_{i=1}^3\frac{u_{2i-1}^{\frac{\Delta_{i}}{2}}\Gamma(\Delta_i)}{\Gamma(\frac{\sum_{j}\Delta_j-2\Delta_i}{2})\Gamma^2(\frac{\Delta_i}{2})}\int \prod_{i=1}^6 ds_i \Gamma(s_i)\prod_{i=1}^3 \frac{U_{2-i}}{u_{2i}U_{-i}}^{s_i}U_{i}^{-s_{3+i}}\Gamma\left(\frac{\Delta_i-2(s_i+\bar{s}_i)}{2}\right)\\
&\Gamma\left(\frac{\Delta_{21}-2(s_3+s_6-s_2)}{2}\right)\Gamma\left(\frac{\Delta_{13}-2(s_2+s_4-s_1)}{2}\right)\Gamma\left(\frac{\Delta_{32}-2(s_1+s_5-s_3)}{2}\right),
\nonumber
\end{align}
where $\bar{s}_1=s_5+s_6, \bar{s}_2=s_4+s_5, \bar{s}_3=s_4+s_6$  and $\Delta_{ij}=\Delta_i-\Delta_j$.
To relate this to $F_{000}$ we repeat the analysis above.  We write the usual Mellin amplitude definition~(\ref{Mellinamplitude}) in terms of 9 independent Mellin variables $\gamma_{ij}$. After integrating in $\gamma_{12},\gamma_{34}$ and 
$\gamma_{56}$, it is easy to relate the remaining $\gamma_{ij}$ to $s_i$'s by imposing the same power behaviour of the cross-ratios on both Mellin representations. We find:
\begin{equation}
s_1=\gamma_{23}\,, \quad s_2= \gamma_{45}\,, \quad s_3=\gamma_{16}\,, \quad s_4=\gamma_{46}\,, \quad s_5= \gamma_{24}\,, \quad s_6= \gamma_{26}\,.
\end{equation}

A simple computation shows that  $F_{000}$ is independent of $\gamma_{ij}$ as one would expect for scalar exchanges.

\subsection{Explicit computation of six-point blocks}
\label{app:HyperInt}
In the following we compute the leading lightcone limit contribution for the exchange of three minimal-twist operators in the snowflake channel of the six-point function. For simplicity, let us first consider that the corresponding operators are scalars. 
It will be useful to recall the definition of the block $g_{k* k* k*}\left(u_{2i},U_{i}\right)$ given in~(\ref{eq:6ptlightconeblockdef}). This is a complicated three-dimensional integral even in the simpler scalar case. 
One can show, however, that no divergences appear from the limit $u_{2 i}\to 0$~\footnote{This can be checked for example with the  HyperInt package \cite{panzer:2014hyp}. We find only logarithmic divergences in $U_{i}$ whenever $U_i\to 0$.}, since the $U_{i}$'s act as regulators of those possible divergences. This substantially simplifies our analysis.
 The situation  for the spinning operators is technically more involved but it is still free of divergences in the limit of $u_{2 i}\to 0$. 

As an example, consider the exchange of three leading-twist scalar operators with dimension 2 in terms of the cross-ratios $y_{u},y_v,y_w$\footnote{The appearance of these cross-ratios is not surprising given the duality between null polygon Wilson loops and correlation functions, see \cite{Bercini:2021jti} for recent development in this topic. In fact these cross-ratios have appeared before in the study of WL/scattering amplitudes in $\mathcal{N}=4$ SYM \cite{Dixon:2011pw}.} defined as
	\begin{equation}
	\label{eq:crossratiosUtoy}
	U_1=\frac{y_u \left(1-y_v\right)\left(1-y_w\right)}{\left(1-y_u y_v\right)\left(1-y_u y_w\right)}\,,\quad U_2=\frac{y_v \left(1-y_u\right)\left(1-y_w\right)}{\left(1-y_v y_u\right)\left(1-y_v y_w\right)}\,,\quad U_3=\frac{y_w \left(1-y_u\right)\left(1-y_v\right)}{\left(1-y_w y_u\right)\left(1-y_w y_v\right)}\,.
	\end{equation}
In these cross-ratios, the block becomes
\begin{equation}
\label{scscscdelta2}
g_{2 2 2}\left(0,U_{i}\right)=\prod_{i=0}^{3} \int_0^{\infty}  \frac{dt_i\,(y_i y_{i+1}-1)^2}{y_i (y_{i + 1}-1) (y_{i - 1}-1) + 
  t_i (1 + t_{i + 1}) (y_i y_{i + 1}-1) (y_i y_{i - 1}-1)}\,,
\end{equation}
where we have changed variables  $t_i\to t_i/(t_i+1)$ and identified $y_1=y_v,\, y_2=y_u$ and $y_3= y_w$. The subscripts should be understood $\mod 3$. These cross-ratios appear to be a more natural choice to compute these integrals, as the integrand factorizes into simpler pieces. The integration can be done exactly and written in terms of hyperlogarithmic functions as
\begin{align}
&g_{2 2 2}\left(0,U_{i}\right)=\nonumber\frac{\left(1-y_u y_w \right) \left(1-y_v y_w \right) \left(1-y_u y_v \right)}{\left(1-y_w \right) \left(1-y_u \right) \left(1-y_v \right) \left(y_u y_v y_w -1\right)}\bigg( \mathrm{H}_{0}(y_u)\left( \mathrm{H}_{0,1}(y_w)+ \mathrm{H}_{0,1}(y_v)-\mathrm{H}_{0,y_w^{-1}}(y_v)\right)\\
&\nonumber\left.-\mathrm{H}_{0}(y_v)\left(\mathrm{H}_{0,y_w^{-1}}(y_u)+\mathrm{H}_{0,(y_vy_w)^{-1}}(y_u)-\mathrm{H}_{0,1}(y_w)-\mathrm{H}_{0,y_v^{-1}}(y_u)-\mathrm{H}_{0,1}(y_u)  \right)+2 \mathrm{H}_{0,(y_vy_w)^{-1},y_v^{-1}}(y_u)\right.\\
&\nonumber\left.+\mathrm{H}_{0}(y_w)\left(\mathrm{H}_{0,y_w^{-1}}(y_v) +\mathrm{H}_{0,1}(y_v)+\mathrm{H}_{0,y_w^{-1}}(y_u)-\mathrm{H}_{0,y_v^{-1}}(y_u)+\mathrm{H}_{0,1}(y_u)\right.-\mathrm{H}_{0,(y_vy_w)^{-1}}(y_u) \right)\\
&\nonumber+2\mathrm{H}_{1}(y_v)\left(\mathrm{H}_{0,y_w^{-1}}(y_u)  -\mathrm{H}_{0,(y_vy_w)^{-1}}(y_u)\right) -2 \mathrm{H}_{0,y_w^{-1},y_w^{-1}}(y_v)+\mathrm{H}_{0,y_v^{-1},0}(y_u)-\mathrm{H}_{0,(y_vy_w)^{-1},0}(y_u)\\
&\nonumber+\mathrm{H}_{0,y_w^{-1},0}(y_v)+ 2\left(\mathrm{H}_{0,1,1}(y_u)+\mathrm{H}_{0,1,1}(y_v)+\mathrm{H}_{0,1,1}(y_w)\right)-2 \mathrm{H}_{0,(y_vy_w)^{-1},1}(y_u)-2 \mathrm{H}_{0,y_v^{-1},y_v^{-1}}(y_u)\\
&\nonumber\left.-\left(\mathrm{H}_{0,1,0}(y_u)+\mathrm{H}_{0,1,0}(y_v)+\mathrm{H}_{0,1,0}(y_w)\right)+2\mathrm{H}_{y_w^{-1}}(y_v)\left(\mathrm{H}_{0,(y_vy_w)^{-1}}(y_u) - \mathrm{H}_{0,1}(y_u)\right)\right.\\
&\nonumber\left.+2\mathrm{H}_{1}(y_w)\left(\mathrm{H}_{0,y_v^{-1}}(y_u)-\mathrm{H}_{0,(y_vy_w)^{-1}}(y_u)\right)+2 \mathrm{H}_{0,(y_vy_w)^{-1},y_w^{-1}}(y_u)-2 \mathrm{H}_{0,y_w^{-1},y_w^{-1}}(y_u)\right.\\
&+\mathrm{H}_{0}(y_u) \mathrm{H}_{0}(y_v)\mathrm{H}_{0}(y_w)+\mathrm{H}_{0,y_w^{-1},0}(y_u)+\zeta_{2}( \mathrm{H}_{0}(y_w)+ \mathrm{H}_{0}(y_u)+\mathrm{H}_{0}(y_v))\bigg) \,.
\end{align}
The hyperlogarithm functions $\mathrm{H}$ are defined recursively via the integral~\cite{panzer:2014hyp}
\begin{equation}
	\mathrm{H}_{\omega_1,\omega_2,\dots,\omega_n}\!(z)=\int_{0}^{z} \frac{dt}{t-\omega_1}\mathrm{H}_{\omega_2,\dots,\omega_n}\!(z), \quad \mathrm{H}_{0, 0,\dots, 0}(z)=\frac{\ln^{n} z}{n!}, \quad \mathrm{H}\!\left(z\right)=1.
\end{equation}
One can then check that in the limit where all $y_i\to 0$  (which corresponds to $U_{i}\to 0$), the integral~(\ref{scscscdelta2}) is given by
\begin{equation}
\lim_{y_i \to 0} \mathrm{(\ref{scscscdelta2})}\approx -\ln \! \left(y_u \right) \ln \! \left(y_v \right) \ln \! \left(y_w \right)-\zeta_{2} \ln \! \left(y_w \right)-\zeta_{2} \ln \! \left(y_u \right)-\zeta_{2} \ln \! \left(y_v \right)\,,
\end{equation}
which is consistent with the behaviour in~(\ref{eq:blockoriginlimitscalar}). In fact, one can repeat this computation for several even integer values of the dimension of the exchanged scalar operators. In this class of examples, the integral can be performed with the HyperInt package. We use several parameterizations of the block and guess its general form in the kinematic limit we consider in this paper, namely $u_{2i-1}\to 0$, followed by $u_{2i}\to 0$ and in last place $U_i \to 0$. This is~(\ref{eq:blockoriginlimitscalar}). We will later confirm these results by using a Mellin representation which we will define below.

For a stress tensor exchange, the form of the integrand is more complicated. Even for specific values of the $\ell_i$'s and of the space-time dimension $d$, we find that these computations extend in time and therefore this procedure becomes less useful. It is however worth stating that if we restrict ourselves to the case where $y_u=y_v=y_w$ these computations can be performed very quickly in HyperInt. We use these results as a sanity check for the Mellin method we now present.

In the kinematics relevant for the bootstrap calculation of section~\ref{sec:Bootstrap} we need to take $u_{2i}\to 0$, in which case we can derive a simplified Mellin representation. For that we consider the lightcone block~(\ref{eq:6ptlightconeblockdef}), set $u_{2i}\to 0$ in the integrand\footnote{This does not lead to any divergences as discussed above.} and then we Mellin transform with respect to the cross-ratios $U_i$. After some massaging we obtain
\begin{align}
&\nonumber g_{k*k*k*}^{\ell_1\ell_2\ell_3}= \prod_{i}^{3}\int \left[ds_i\right]\Gamma (s_i) \frac{ \Gamma (2 J+\tau )}{2^J\Gamma \left(\frac{2 J+\tau }{2}\right)^2 }\,  \sum_{n_i,m_i}(-1)^{m_i}U_i^{m_i+n_i-s_i+\ell_{2-i}}\\
   &\frac{\binom{J-\ell_{2-i}-\ell_{3-i}}{n_i}\binom{J-n_{i+1}-\ell_{1-i}-\ell_{2-i}}{m_i}\Gamma \left(s_i-n_i- \ell_{2-i}+ \ell_{1-i}\right)}{\left(2J- s_i- \ell_{1-i}-\ell_{3-i}+\frac{\tau}{2}\right)_{s_i}\left(J+m_{i + 1} + n_{i} - s_{i} - s_{i + 1}+\frac{\tau }{2}\right)_{s_i-n_i-\ell_{2-i}+\ell_{1-i}}}\,,
\end{align}
in the case where all the operators have the same twist and spin. The sums over $n_i$ and $m_i$ were introduced to reduce the binomials that appeared in the numerator into monomials of $U_i$.

We would like to make an expansion in the limit $U_i\to 0$. In Mellin language this is simply done by closing the $s_i$ contours to the left and picking the corresponding poles. At leading order only some poles contribute. We will call these the leading poles. The leading poles will only come from the gamma functions explicitly written above and which only depend on one of the Mellin variables. 

We observe that the position of the leading poles does not depend on the value of $m_i$. Therefore in the limit $U_i \to 0$, the leading contributions have to come from the terms with $m_i=0$. For fixed values of spin, twist and $\ell_i$, we perform the sum over $n_i$ and pick the residues of leading poles. These leading contributions are located at values of $s_i$ such that the exponent of the corresponding $U_i$ becomes 0, which leads to the expected logarithmic behaviour when there is a double pole\footnote{Other poles of the family will always contribute at subleading orders. In fact, if we have $s_i$ smaller than the required value, there will be a non-vanishing power $U_i$ which leads to a subleading contribution. On the other hand, if $s_i$ is instead larger, there is no corresponding pole and the residue is 0. In other words, leading poles are the rightmost poles of the family prescribed by the explicit gamma functions we wrote above.}. If we use this mechanism in the case of scalar minimal-twist exchange, we immediately reproduce the result of (\ref{eq:blockoriginlimitscalar})! Moreover, we can also check that this procedure for the leading poles nicely matches the results of direct integration using HyperInt in the limit $y_u=y_v=y_w$.

For a stress tensor exchange, we have three possible values of $\ell_i$'s, namely 0,1 and 2. If two or three $\ell_i$'s take value 1, those contributions will be subleading by powers of $U_i$. We thus list the results for the remaining cases
\begin{align}
&\nonumber g_{TTT}^{000}=- \frac{\Gamma (\tau+4)^3}{64 \Gamma \left(\frac{\tau+4}{2}\right)^6} \left[\prod_{i}^{3}\frac{\ln U_{i}}{3} 
+ \bigg(4\left(S_{\frac{\tau}{2}+1}\right)^2-S_{\frac{\tau}{2}+1}^{(2)}+\frac{8 \big(\tau  (\tau +6)+2\big)}{\tau  (\tau +2) (\tau +4) (\tau +6)}+\zeta_2\bigg)\ln U_1
\right.\\
&+2S_{\frac{\tau}{2}+1}\ln U_{1}\ln U_{2}-\frac{S_{\frac{\tau}{2}+1}}{3}\bigg(8\left(S_{\frac{\tau}{2}+1}\right)^2-6S_{\frac{\tau}{2}+1}^{(2)}\bigg)
-\frac{S_{\frac{\tau}{2}+1}\bigg(8(\tau(\tau+6)+2)+\zeta_2\bigg)}{2\tau  (\tau +2) (\tau +4) (\tau +6)}+\text{perm}\bigg]\,,\nonumber\\
&g_{TTT}^{100}=-\frac{\Gamma (\tau+4)^3 \big(\tau  (\tau +6)+4\big)}{16\Gamma \left(\frac{\tau+4}{2}\right)^6\tau  (\tau +2) (\tau +4) (\tau +6)}\Big[2S_{\frac{\tau}{2}+1}+\ln U_2\Big]\,\label{eq:blockstresstensorWL}\\
 &g_{TTT}^{200}= -\frac{ \Gamma (\tau +4)^3}{4 \Gamma \left(\frac{\tau +4}{2}\right)^6\tau  (\tau +2) (\tau +4) (\tau +6)}\Big[2 S_{\frac{\tau}{2}+1}+\ln U_2\Big]\,,\nonumber
\end{align}
where $\tau=d-2$ is the twist of the stress-tensor.
Notice  the result diverges for $\tau=0$. This is not a problem since we are considering the case where there is a twist gap which happens for $d>2$.  For other non-vanishing $\ell_i$, the result is obtained by permuting the cross-ratios.

\subsection{Euclidean expansion of six-point conformal blocks}
The results of the main part of the paper were derived using the leading term of the conformal blocks expanded around the lightcone. We will shift gears in this section and analyze the conformal blocks expanded around the Euclidean OPE limit in a similar approach to the one done for four- and five-point function conformal blocks \cite{Dolan:2003hv,Hogervorst:2013sma,Goncalves:2019znr}. 

The two key ingredients in the derivation of the blocks are that they satisfy the Casimir differential equation
\begin{equation}
\bigg[\frac{1}{2}\left(L_{AB}^{(i_1)}+L_{AB}^{(i_2)}\right)^2 -C_{\Delta,J}\bigg]f_{\Delta,J}(x_i)=0\,,
\end{equation}
with
\begin{equation}
	C_{\Delta,J} = \Delta(\Delta-d)+J(J+d-2)\,,
\end{equation}
where $L_{AB}$ are the generators of the conformal group and their boundary condition coming from the OPE 
\begin{align}
\mathcal{O}(x_{i_1})\mathcal{O}(x_{i_2}) = \sum_{k}C_{i_1i_2k}\frac{x_{i_1i_2}^{\mu_{1}}\dots x_{i_1i_2}^{\mu_{J}}}{(x_{i_1i_2}^{2})^{\frac{\Delta_{i_1}+\Delta_{i_2}-\Delta_k+J_k}{2}}}\mathcal{O}_{k,\mu_1\dots\mu_J}(x_{i_2})\label{eq:EuclideanOPEboundary}.
\end{align}

In the Euclidean OPE limit there are three cross-ratios that approach zero
\begin{align}
s_1^2= u_1\,,\qquad 
s_2^2= u_3\,, \qquad
s_3^2= u_5\,,
\end{align}
and six others that remain fixed
\begin{align}
&\xi_1 = \frac{U_1-u_2U_2}{s_1U_1}\,, \qquad  \xi_2 = \frac{U_3-u_4U_1}{s_2U_3}\,,\qquad  \xi_3= \frac{U_2-u_6U_3}{s_3U_2}\,,\nonumber\\
&\xi_4 = \frac{(u_2-U_1)U_2}{s_1s_2U_1}\,,\qquad  \xi_5 = \frac{(u_6-U_2)U_3}{s_1s_3U_2}\,,\qquad  \xi_6 = \frac{(u_4-U_3)U_1}{s_2s_3U_3}\,,
\end{align}
in a six-point correlation function and are analogous to the four-point cross-ratios written in equation (\ref{eq:sxi4pt}). The cross-ratios that remain fixed can be interpreted as measuring the angles that the points $2,4,6$ approach  $1,3,5$. It follows from the OPE (\ref{eq:EuclideanOPEboundary}) that the conformal block should behave as 
\begin{align}
G_{\Delta_i,J_i}(s_i,\xi_i)= \prod_{j=1}^3s_{j}^{\Delta_{j}} \,\, g_{J_i}(\xi_i)\,, \ \ \ \ \  \ s_i\rightarrow 0\,,
\end{align}
where  $g_{J_i}(\xi_i)$\footnote{This is the analogue of the Gengebauer polynomial that appears in the leading term of the OPE of a four-point function conformal block. Let us also remark that this function appears in the definition of the conformal block using the shadow formalism. }  is a polynomial function of the cross-ratios $\xi_i$ that satisfies three differential equations coming from the Casimir of the channel $(12)$ in the limit $s_i\rightarrow 0$,
\begin{align}
&\bigg[(4-\xi_1^2)\partial_{\xi_1}^2+(4-\xi_4^2)\partial_{\xi_4}^2+(4-\xi_5^2)\partial_{\xi_5}^2-2(2\xi_2+\xi_1\xi_4)\partial_{\xi_1}\partial_{\xi_4}
\nonumber\\
&
-2(2\xi_2+\xi_1\xi_4)\partial_{\xi_1}\partial_{\xi_4}-2(2\xi_3+\xi_1\xi_5)\partial_{\xi_1}\partial_{\xi_5}
+(1-d)(\xi_1\partial_{\xi_1}+\xi_4\partial_{\xi_4}+\xi_5\partial_{\xi_5})
\\
&+2(2\xi_2\xi_3-\xi_4\xi_5-2\xi_6)\partial_{\xi_4}\partial_{\xi_5}+J_1(J_1+d-2)\bigg]g_{J_i}(\xi_i) = 0\,,\nonumber
\end{align}
with similar equations for the channels $(34)$ and $(56)$. These three differential equations, together with the boundary condition for 
$\lambda\to 0$,
\begin{align}
g_{J_i}(\xi_i)  \rightarrow \xi_1^{J_1-\ell_2-\ell_3}\xi_2^{J_2-\ell_1-\ell_3}\xi_3^{J_3-\ell_1-\ell_2}\xi_4^{\ell_3}\xi_5^{\ell_2}\xi_6^{\ell_1}\,, \qquad
\xi_{1,2,3} \rightarrow \frac{\xi_{1,2,3}}{\lambda} \,,
\qquad
\xi_{4,5,6} \rightarrow \frac{\xi_{4,5,6}}{\lambda^2}\,, 
\label{eq:boundaryconditionEuclideanAngles}
\end{align}
fix completely the form of the function. It is possible (and easy) to get subleading corrections of $g_{J_i}(\xi_i)$ for any value of $J_i$ and $\ell_i$ from the differential equations. By analyzing these corrections we were able to check that the function $g_{J_i}(\xi_i)$ satisfies relations of the type
\begin{align}
\xi_k\, g_{J_i,\ell_i}(\xi_i) = \sum_{i_l=-1}^{1}c_{i_1\dots i_6}^{(k)}  g_{J_1+i_1,J_2+i_2,\dots,\ell_3+i_4,\dots \ell_1+i_6}(\xi_i)\,,
\end{align}
that can be used to define it recursively. One example of these relations is\footnote{The other relations as well as the definition of $g_{J_i,\ell_i}(\xi_i)$ in terms of a recurrence relation is provided in a auxiliary file.  }
\begin{align}
&
c_{-100000}^{(1)}= \frac{4(J_1-\ell_2-\ell_3)(J_1+\ell_2+\ell_3)}{(2J_1+d-4)(2J_1+d-2)}\,, 
\qquad 
c_{100000}^{(1)}=1\,,\\
&c_{-100-100}^{(1)}=-\frac{2\ell_3(d+2(\ell_2+\ell_3-2))}{(2J_1+d-4)(2J_1+d-2)}\,,
\qquad
c_{-1000-10}^{(1)}=- \frac{2\ell_2(d+2(\ell_2+\ell_3-2))}{(2J_1+d-4)(2J_1+d-2)} \,,\nonumber\\
&c_{-100-1-10}^{(1)}=-\frac{4\ell_2\ell_3}{(2J_1+d-4)(2J_1+d-2)}\,,
\qquad c_{-100-1-11}^{(1)}= \frac{4\ell_2\ell_3}{(2J_1+d-4)(2J_1+d-2)}\,.\nonumber
\end{align}
Let us remark that there are similar relations for the Gegenbauer polynomial and for the five-point analogue\cite{Goncalves:2019znr}. 

It is an interesting open problem to obtain a representation of the conformal block as a series expansion in $s_i$, as was done for four and five points\cite{Hogervorst:2013sma,Goncalves:2019znr}\footnote{It would also be interesting to see how the recent and new approaches to the conformal blocks\cite{Buric:2020dyz,Buric:2021ywo,PolandValentina:2021} can help in this problem.}.

\section{D-functions}
\label{app:Dfuncs}
In this appendix we analyze five- and six-point D-functions using standard technology from perturbation theory in $AdS$ \cite{Jepsen:2019svc,Meltzer:2019}.
\subsection{Five Points}
We start from a five-point contact Witten diagram with a non-derivative interaction
\begin{equation}
W^{\text{ctc}}_{\Delta_1,\dots,\Delta_5}(x_1,\dots,x_5)= \int_{AdS_{d+1}} d^{d+1}y K_{\Delta_1}(x_1,y) \dots K_{\Delta_5}(x_5,y) = D_{\Delta_1,\dots,\Delta_5}\,,
\end{equation}
where the bulk-boundary propagator is defined as
\begin{equation}
K_{\Delta}(x_i,y)= \left( \frac{z}{(\vec{x}_i-\vec{y})^2+z^2}\right) ^\Delta\,.
\end{equation}
We can expand this in five-point conformal blocks without knowing their explicit form, using Harmonic analysis and the conformal partial waves. We will do this in the $(12)(34)$ channel, but other channels can be obtained with the same method. Start by introducing auxiliary $1= \int_{AdS} dy' \delta(y'-y)$ and attach the bulk to boundary propagators to the auxiliary points in the desired (12)(34) structure, i.e.
\begin{equation}
W^{\text{ctc}}= \int dy dy' dy'' K_{\Delta_1}(x_1,y')K_{\Delta_2}(x_2,y')K_{\Delta_3}(x_3,y'')K_{\Delta_4}(x_4,y'')K_{\Delta_5}(x_5,y) \delta(y'-y)\delta(y''-y)\,.
\end{equation}
Next, we use the spectral representation of the AdS delta function and the split representation of the harmonic function to obtain
\begin{equation}
\delta(y_1-y_2)= \int dx' \int_{-i \infty}^{+i\infty} \frac{dc}{2\pi i} \rho_\delta(c) K_{h+c}(x',y_1) K_{h-c}(x',y_2)\,,
\end{equation}
where $c$ is the imaginary spectral parameter, $h=d/2$ and the spectral function for the Dirac delta is
\begin{equation}
\rho_\delta(c)= \frac{\Gamma\left(\frac{d}{2}+c\right) \Gamma\left(\frac{d}{2}-c\right)}{2 \pi^{d} \Gamma(-c) \Gamma(c)}\,.
\end{equation}
Now, all three bulk integrals can be performed, since they are of the AdS three-point function type
\begin{equation}
\int dy K_{\Delta_1}(x_1,y)K_{\Delta_2}(x_2,y)K_{\Delta_3}(x_3,y)= a_{\Delta_1,\Delta_2,\Delta_3} \langle \mathcal{O}_1 (x_1)\mathcal{O}_2 (x_2)\mathcal{O}_3 (x_3)\rangle\,,
\end{equation}
where 
\begin{equation}
\langle \mathcal{O}_1 (x_1)\mathcal{O}_2 (x_2)\mathcal{O}_3 (x_3)\rangle\ =  \frac{1}{x_{12}^{\Delta_{12,3}}x_{23}^{\Delta_{23,1}}x_{13}^{\Delta_{13,2}}}\,
\end{equation}
is the kinematical three-point function without OPE coefficient, and
\begin{equation}
a_{\Delta_1,\Delta_2,\Delta_3}= \frac{\pi^{\frac{d}{2}} \Gamma\big(\frac{\Delta_{1}+\Delta_{2}-\Delta_{3}}{2}\big) \Gamma\big(\frac{\Delta_{1}+\Delta_{3}-\Delta_{2}}{2}\big) \Gamma\big(\frac{\Delta_{2}+\Delta_{3}-\Delta_{1}}{2}\big)}{2 \Gamma\left(\Delta_{1}\right) \Gamma\left(\Delta_{2}\right) \Gamma\left(\Delta_{3}\right)}\, \Gamma\bigg(\frac{\Delta_{1}+\Delta_{2}+\Delta_{3}-d}{2}\bigg)\,.
\end{equation}
We are then left with two spectral integrals and two boundary integrals
\begin{align}
W^{\text{ctc}}&= \int [dc'][dc'']dx'dx''\rho_\delta(c')\rho_\delta(c'') a_{\Delta_1,\Delta_2,h+c'} a_{h-c',\Delta_5,h-c''}a_{h+c'',\Delta_3,\Delta_4}\\&\qquad
\langle \mathcal{O}_1(x_1)\mathcal{O}_2(x_2)\mathcal{O}_{h+c'}(x') \rangle \langle \mathcal{O}_{h-c'}(x')\mathcal{O}_5(x_5)\mathcal{O}_{h-c''}(x'') \rangle\langle \mathcal{O}_{h+c''}(x'') \mathcal{O}_3(x_3)\mathcal{O}_4(x_4) \rangle\,,\nonumber
\end{align}
where $[dc]=dc/2\pi i$. The position space integrals precisely coincide with the  definition of the five-point conformal partial wave for the exchange of two scalar operators of dimension $h+c'$ and $h+c''$
\begin{equation}
\Psi^{\Delta_1 \dots \Delta_5}_{h+c',h+c''}(x_i)=\int dx dx' \langle \mathcal{O}_1\mathcal{O}_2\mathcal{O}_{h+c'}(x') \rangle \langle \mathcal{O}_{h-c'}(x')\mathcal{O}_5\mathcal{O}_{h-c''}(x'') \rangle\langle \mathcal{O}_{h+c''}(x'') \mathcal{O}_3\mathcal{O}_4 \rangle\,.
\end{equation}
Thus, we find the partial have expansion for the five-point contact Witten diagram
\begin{equation}
W^{ctc}= \int [dc'][dc''] \tilde{\rho}_5(c',c'') \Psi^{\Delta_1 \dots \Delta_5}_{h+c',h+c''}(x_i)\,,
\end{equation}
with 
\begin{equation}
\tilde{\rho}_5(c',c'') =\rho_\delta(c')\rho_\delta(c'') a_{\Delta_1,\Delta_2,h+c'} a_{h-c',\Delta_5,h-c''}a_{h+c'',\Delta_3,\Delta_4}\,.
\end{equation}
To obtain the conformal block expansion we deform the contours towards the real axis and pick up the physical poles. To do this we need the relation between the conformal partial waves and the conformal blocks. Since they solve the same Casimir equations, the conformal partial waves must be a linear combination of the blocks for the exchanged operators and their shadows. We provide a detailed analysis of this relation in Appendix \ref{app:Harmonic}. The coefficients can be obtained in the OPE limits and are given in terms of shadow factors $K$ ($h-c$ appears since it is the shadow of $h+c$)
\begin{equation}
\Psi^{\Delta_1 \dots \Delta_5}_{h+c',h+c''}(x_i) = K^{\Delta_5 , h-c''}_{h-c'} K^{\Delta_5,h+c'}_{h-c''} G_{h+c',h+c''}^{\Delta_1,\dots,\Delta_5}(x_i) + \text{3 shadow terms}\,
\end{equation}
With
\begin{equation}
K_{\Delta,J}^{\Delta_{1}, \Delta_{2}}=\left(-\frac{1}{2}\right)^{J} \frac{\pi^{\frac{d}{2}} \Gamma\big(\Delta-\frac{d}{2}\big)\, \Gamma(\Delta+J-1)\, \Gamma\big(\frac{\widetilde{\Delta}+\Delta_{1}-\Delta_{2}+J}{2}\big)\, \Gamma\big(\frac{\widetilde{\Delta}+\Delta_{2}-\Delta_{1}+J}{2}\big)}{\Gamma(\Delta-1)\, \Gamma(d-\Delta+J)\, \Gamma\big(\frac{\Delta+\Delta_{1}-\Delta_{2}+J}{2}\big)\, \Gamma\big(\frac{\Delta+\Delta_{2}-\Delta_{1}+J}{2}\big)}\,,
\end{equation}
which are related to the shadow factors $S$ we will compute below by $K_{\Delta,J}^{\Delta_{1}, \Delta_{2}}=(-\frac{1}{2})^J S_{\Delta,J}^{\Delta_{1}, \Delta_{2}}$. We will carefully describe these factors in Appendix \ref{app:Harmonic}.
Note that since we only exchange scalar operators we always have $J=0$ so we suppress that label. We now have the block expansion in contour integral form
\begin{equation}
W^{ctc}= \int [dc'][dc''] \rho_5(c',c'') G^{\Delta_1 \dots \Delta_5}_{h+c',h+c''}(x_i)\,,
\end{equation}
where
\begin{equation}
\rho_5(c',c'')=4 K^{\Delta_5 , h-c''}_{h-c'} K^{\Delta_5,h+c'}_{h-c''} \tilde{ \rho}_5(c',c'')\,
\end{equation}
and the factor of 4 comes from the shadow combinations. The function $\rho_5$ contains three families of poles corresponding to the exchanged operators. Introducing the notation $\Delta'=h+c'$, we have
\begin{align}
\text{Family 1:}&\quad \Delta'= \Delta_1 +\Delta_2 +2n_1\,, \quad \Delta''= \Delta_3 +\Delta_4 +2m_1\,,\\
\text{Family 2:}&\quad \Delta' = \Delta_1 +\Delta_2 +2n_2\,, \quad \Delta'' = \Delta_1+\Delta_2+\Delta_5+2n_2+2m_2\,,\\
\text{Family 3:}&\quad \Delta'=\Delta_3+\Delta_4+\Delta_5+ 2n_3+2m_3 \,, \quad \Delta''= \Delta_3 +\Delta_4 +2m_3\,.
\end{align}
Thus we can write the block expansion as
\begin{align}
W^{\text{ctc}}&= \sum_{n_1,m_1=0}^\infty P_{[12]_{n_1}[34]_{m_1}} G^{\Delta_1 \dots \Delta_5}_{[12]_{n_1},[34]_{m_1}} +\sum_{n_2,m_2=0}^\infty P_{[12]_{n_2}[125]_{n_2+m_2}} G^{\Delta_1 \dots \Delta_5}_{[12]_{n_2},[125]_{n_2+m_2}}\nonumber\\&+\sum_{n_3,m_3=0}^\infty P_{[345]_{n_3+m_3}[34]_{m_3}} G^{\Delta_1 \dots \Delta_5}_{[345]_{n_3+m_3},[34]_{m_3}}\,,
\end{align}
where $[ij]_n$ denotes the scalar double-twist $[\mathcal{O}_i \mathcal{O}_j]_n$ with $n$ laplacians, and similarly for the triple-twists $[ijk]_{n+m}$. The $P_{ab}$ are related to the OPE coefficients through (\ref{eq:P5pt}) with $\ell=0$. Finally, we specify how to obtain the $P_{ab}$ from the residues of $\rho_5$
\begin{align}
P_{[12]_{n_1}[34]_{m_1}}&= \text{Res}_{\Delta''=\Delta_3+\Delta_4+2m_1}\text{Res}_{\Delta'=\Delta_1+\Delta_2+2n_1}\rho_5(\Delta',\Delta'')\,,\nonumber\\
P_{[12]_{n_2}[125]_{n_2+m_2}}&=\text{Res}_{\Delta''=\Delta_1+\Delta_2+\Delta_5+2n_2+2m_2}\text{Res}_{\Delta'=\Delta_1+\Delta_2+2n_2}\rho_5(\Delta',\Delta'')\,,\nonumber\\
P_{[345]_{n_3+m_3}[34]_{m_3}}&=\text{Res}_{\Delta''=\Delta_3+\Delta_4+2m_3}\text{Res}_{\Delta'=\Delta''+\Delta_5+2n_3}\rho_5(\Delta',\Delta'')\,.
\end{align}
Some comments on this block expansion are in order:
\begin{itemize}
	\item{We have exchange of both double-twist and triple-twist operators. Unlike the double-twist operators, of which there is only one of a given dimension, triple-twist operators are degenerate at leading order in $1/N$. Since we have operators of dimension $\Delta_1+\Delta_2+\Delta_5 + 2(n+m)$, and we sum over both $n$ and $m$ this means that there are $p+1$ triple-twist operators of dimension $\Delta_1+\Delta_2+\Delta_5 + 2p$.}
	\item{Large $N$ counting determines that a connected five-point function has a leading behaviour $\sim 1/N^3$. (One can have factorized three-point $\times$ two-point functions at order $1/N$ but let's ignore those). We can check this large $N$ behaviour in the OPE coefficients. For family 1 we have \begin{equation}
		P_{[12]_{n_1}[34]_{m_1}}= C_{12[12]_{n_1}}C_{[12]_{n_1}5[34]_{m_1}}C_{[34]_{m_1}34}\,
		\end{equation}
		where the first and last OPE coefficient are the MFT ones, so we are accessing the $1/N^3$ information in $C_{[12]_{n_1}5[34]_{m_1}}$. For the second family we have
		\begin{equation}
		P_{[12]_{n_2}[125]_{n_2+m_2}}=C_{12[12]_{n_2}}C_{[12]_{n_2}5[125]_{n_2+m_2}}C_{[125]_{n_2+m_2}34}\,,
		\end{equation} where now the first two OPE coefficients are MFT (although the second one is single-twist/double-twist/triple-twist), and the $1/N^3$ data we are probing is $C_{[125]_{n_2+m_2}34}$.The third family is similar to the second one.}
	\item{For generic dimensions we have an expansion in terms of blocks, however when the exchanged operators in different families have dimensions that differ by an even integer, we find that the OPE coefficients naively diverge. This happens when
		\begin{equation}
		\Delta_1+\Delta_2+\Delta_5-\Delta_3-\Delta_4= 2p\, \quad \text{or} \quad \Delta_1+\Delta_2-\Delta_5-\Delta_3-\Delta_4= 2q\,
		\end{equation} for some $p,q \in \mathbb{Z}$. By carefully regulating the external dimensions and taking the limit, one finds that the divergences in OPE coefficients cancel, and we get instead derivatives of the blocks with respect to the exchanged dimension. This is the tell-tale sign of anomalous dimensions for the exchanged operators. We will see this explicitly in the $D_{11112}$ example that we will analyze below. Equivalently, we can take the integer separated dimensions at the level of the spectral function, which will then have double poles. Picking their residues also leads to the derivatives of the blocks.
		In particular, recall that the D functions which admit a closed form expression are the ones where the total dimension is an even integer. This means that either $\Delta_1+\Delta_2+\Delta_5$ and $\Delta_3+\Delta_4$ are both odd or both even. In any case, their difference is an even number, and will therefore satisfy the above condition. Therefore, we learn that explicitly computable D-functions must always contain derivatives of blocks.}
\end{itemize}

\subsubsection{The case of $D_{11112}$}
The simplest computable (in terms of ladder integrals) five-point D-function is $D_{11112}$. As argued above, this D-function contains blocks and derivatives of blocks corresponding to anomalous dimensions in its expansion. Following the limiting procedure described in the previous section, the coefficients in the expansion can be read off. We can organize the sum into two integers corresponding to the two exchanged operators. It is actually more convenient to pick the two integers to parametrize the dimension of one of the operators and the difference between the two. We separate the cases with same dimension and positive difference, since they are qualitatively different. Therefore we write
\begin{align}
&W^{\text{ctc}}= \sum_{n_1=0}^{\infty} \frac{ \Gamma_{2 n_1+1} \Gamma_{n_1+1}^2\,\Gamma_{-\frac{d}{2}+n_1+2}^2\,\Gamma
	_{-\frac{d}{2}+2 n_1+3}\left(1-\frac{3 \delta
		_{0,n_1}}{4}\right)}{2\pi ^{-d/2} \,\Gamma_{2 n_1+2}^2\, \Gamma_{
	-\frac{d}{2}+2 n_1+2}^2}\, G_{2+2n_1,2+2n_1} \\
&+  \sum_{n_1=0,\delta =1}^\infty \left( \frac{\pi ^{d/2} \delta\,  \Gamma_{n_1+1} 
	\Gamma_{\delta +n_1+1} \Gamma_{\delta +2 n_1+1} \Gamma_{-\frac{d}{2}+n_1+2}  \Gamma_{
		-\frac{d}{2}+\delta +n_1+2} }{\Gamma_{-\frac{d}{2}+\delta +2
		n_1+3}^{-1}\Gamma_{2 n_1+2} \Gamma_{ -\frac{d}{2}+2 n_1+2} \Gamma_{
	2 \left(\delta +n_1+1\right)} \Gamma_{ 2 \left(\delta
	+n_1+1\right)-\frac{d}{2}}} \,\partial_{\Delta_1} G_{2+2(n_1+\delta ),2+2n_1} \right. \nonumber\\  
&+\Bigg[ \frac{ \delta  \left(S_{-\frac{d}{2}+\delta
		+n_1+1}+S_{-\frac{d}{2}+\delta +2 n_1+2}-2 \left(S_{-\frac{d}{2}+2 \delta +2 n_1+1}+S_{2
		\delta +2 n_1+1}\right)+S_{\delta +n_1}+S_{\delta +2 n_1}\right)+1}{2\pi ^{-d/2} \Gamma _{2 n_1+2}
	\Gamma _{-\frac{d}{2}+2 n_1+2} \Gamma _{2 \left(\delta +n_1+1\right)} \Gamma _{2
		\left(\delta +n_1+1\right)-\frac{d}{2}}}\nonumber\\& \Gamma _{n_1+1} \Gamma _{-\frac{d}{2}+n_1+2} \Gamma _{\delta +n_1+1} \Gamma
_{\delta +2 n_1+1} \Gamma
_{-\frac{d}{2}+\delta +n_1+2} \Gamma _{-\frac{d}{2}+\delta +2 n_1+3}G_{2+2(n_1+\delta ),2+2n_1}\Bigg]+(\Delta_1 \leftrightarrow \Delta_2)\Bigg)\,,\nonumber
\end{align}
where we introduced the shorthand notation $\Gamma_a\equiv \Gamma(a)$.
 Specializing for concreteness to the case $d=4$ and explicitly writing the block expansion for the first few operators, we have
\begin{align}
8W^{\text{ctc}}&=4 \pi ^2 G_{2,2}-\frac{10}{9} \pi ^2 G_{2,4}-\frac{134}{675} \pi ^2 G_{2,6}-\frac{10}{9}
\pi ^2 G_{4,2}+\frac{4}{9} \pi ^2 G_{4,4}-\frac{16}{225} \pi ^2 G_{4,6}-\frac{134}{675}
\pi ^2 G_{6,2}\nonumber\\&-\frac{16}{225} \pi ^2 G_{6,4}+\frac{4}{225} \pi ^2 G_{6,6}+\frac{4}{3}
\pi ^2 G_{2,4}{}^{(0,1)}+\frac{8}{45} \pi ^2 G_{2,6}{}^{(0,1)}+\frac{2}{15} \pi ^2
G_{4,6}{}^{(0,1)}\nonumber\\
&+\frac{4}{3} \pi ^2 G_{4,2}{}^{(1,0)}+\frac{8}{45} \pi ^2
G_{6,2}{}^{(1,0)}+\frac{2}{15} \pi ^2 G_{6,4}{}^{(1,0)} + \text{higher dimension operators}\,,
\end{align}
which has the expected left-right symmetry.
On the other hand, $D_{11112}$ admits an explicit position space expression in terms of a linear combination of products of rational functions of the five cross-ratios and one-loop ladder functions $\Phi(z,\bar{z})$ with the arguments being all possible five-point cross-ratios. In practice, we have to invert to the variables $u,v$ and use
\begin{align}
&\Phi(u,v)=\frac{2 \text{Li}_2(1-v)+\log (u) \log (v)}{1-v}+\\&\frac{u (2 (v+1) \text{Li}_2(1-v)+\log (u)
	(-2 v+v \log (v)+\log (v)+2)+2 (v+v \log (v)-1))}{(1-v)^3} + O(u^2)\,.\nonumber
\end{align}
Using the radial expansion for the five-point blocks described in \cite{Goncalves:2019znr}
\begin{equation}
G_{\Delta',\Delta''}= \sum_{n',n''} a_{n',n''} s_1^{\Delta'+n'}s_2^{\Delta''+n''} \mathcal{H}_{n',n''}(\chi_1,\chi_2,\chi_3)\,,
\end{equation}
Where $a_{n',n''}$ are kinematically fixed coefficients, $s_1,s_2$ are radial variables which are small in the double $(12)(34)$ OPE limit and $\mathcal{H}$ is a polynomial in the $\chi_1,\chi_2,\chi_3$ angular variables,\footnote{We have $2 \chi_1= \xi_1$,$2 \chi_2=\xi_3$ and $-2\chi_3=\xi_1$ in terms of the $\xi_i$ variables introduced in \cite{Goncalves:2019znr}.} which are fixed in this limit. As an example we have:
\begin{equation}
G_{2,2}=s_2^2 s_1^2+s_2^2 s_1^3 \chi _1-s_2^3 s_1^2 \chi _2+
 \frac{1}{3} s_2^2 s_1^4 \left(4 \chi
_1^2-1\right)+\frac{1}{2}s_2^3 s_1^3(\chi _3-2\chi_1\chi_2)+\frac{1}{3} s_2^4 s_1^2 \left(4
\chi _2^2-1\right)+ O(s^7)\,.
\end{equation}
Using the explicit blocks and the expression in terms of ladder functions, we can form an expansion in the small $s_1,s_2$ limit, and we precisely reproduce the block expansion derived through harmonic analysis in the previous section.
\subsection{Six Points}
It is not hard to generalize the previous analysis to the six-point D-function. We will consider the expansion in terms of the snowflake partial wave
\begin{equation}
\Psi_{A,B,C}^{\text{sf}}= \int dx_{7,8,9} \langle \mathcal{O}_1\mathcal{O}_2 \mathcal{O}_A(x_7) \rangle\langle \mathcal{O}_3\mathcal{O}_4 \mathcal{O}_B(x_8) \rangle\langle \mathcal{O}_5\mathcal{O}_6 \mathcal{O}_C(x_9) \rangle \langle\tilde{\mathcal{O}}_A^\dagger(x_7)\tilde{\mathcal{O}}_B^\dagger(x_8)\tilde{\mathcal{O}}_C^\dagger(x_9)\rangle\,,
\end{equation}
A similar analysis to the five-point case leads to the spectral function 
\begin{equation}
	\tilde{\rho}_6(c_1,c_2,c_3) =\rho_\delta(c_1)\rho_\delta(c_2) \rho_\delta(c_3) a_{\Delta_1,\Delta_2,h+c_1}a_{\Delta_3,\Delta_4,h+c_2}a_{\Delta_5,\Delta_6,h+c_3} a_{h-c_1,h-c_2,h-c_3}\,.
\end{equation}
Using the OPE limits discussed in Appendix \ref{app:Harmonic}, we can then determine the proportionality factor between the partial wave and the block
\begin{equation}
	\Psi_{h+c_1,h+c_2,h+c_3}(x_i) = K^{h-c_2 , h-c_3}_{h-c_1} K^{h+c_1,h-c_3}_{h-c_2}
	K^{h+c_1,h+c_2}_{h-c_3}G_{h+c_1,h+c_2,h+c_3}(x_i) + \text{7 shadow terms}\,
\end{equation}
Such that we can represent the six-point function by
\begin{equation}
	W^{ctc}= \int [dc_{1,2,3}] \rho_6(c_{1,2,3}) G_{h+c_1,h+c_2,h+c_3}(x_i)\,,
\end{equation}
with 
\begin{equation}
	\rho_6(c_{1,2,3})= 8  K^{h-c_2 , h-c_3}_{h-c_1} K^{h+c_1,h-c_3}_{h-c_2}
	K^{h+c_1,h+c_2}_{h-c_3}\tilde{\rho}_6 (c_{1,2,3})\,.
\end{equation}
This spectral function leads to the following families of exchanged operators
\begin{align}
\text{1:}&\quad \Delta_A= \Delta_1 +\Delta_2 +2n_1\,, \, \Delta_B= \Delta_3 +\Delta_4 +2n_2\,, \, \Delta_C= \Delta_5 +\Delta_6 +2n_3\,,\\
\text{2:}&\quad \Delta_A= \Delta_3 +\Delta_4+\Delta_5+\Delta_6 +2m_t\,, \, \Delta_B= \Delta_3 +\Delta_4 +2m_2\,, \, \Delta_C= \Delta_5 +\Delta_6 +2m_3\,,\nonumber\\
\text{3:}&\quad \Delta_A= \Delta_1 +\Delta_2 +2p_1\,, \, \Delta_B=\Delta_1 +\Delta_2+\Delta_5+\Delta_6 +2p_t \,,\, \Delta_C= \Delta_5 +\Delta_6 +2p_3\,,\nonumber\\
\text{4:}&\quad \Delta_A= \Delta_1 +\Delta_2 +2q_1\,, \, \Delta_B= \Delta_3 +\Delta_4 +2q_2\,, \, \Delta_C= \Delta_1 +\Delta_2+\Delta_3+\Delta_4 +2q_t\,,\nonumber\\\nonumber
\end{align}
where $m_t = m_1+m_2+m_3$ and similarly for the other indices.
Note that we identify double- and quadruple-twist operator families in the spectrum.
\subsubsection{The case of $D_{111111}$}
Once again we consider integer valued D-functions, the simplest of which has all dimensions equal to 1. They are particularly useful in the study of $\phi^3$ theory in $6-\epsilon$ dimensions.
 On the lightcone $(12)(34)(56)$, the $D$-function $D_{111111}$ has been computed in \cite{DelDuca:2011wh}.
 The fact that all dimensions are identical and furthermore integer, leads to the usual degeneracies, and pole collisions, which are responsible for generating derivatives of blocks, and therefore tree level anomalous dimensions.

Note that for poles to collide, we must have that some double-twist operators in family 1 have the same dimension as a quadruple trace operator in families 2,3 or 4. Therefore, the sum of operators naturally organizes in terms of a triangle function. If the three dimensions satisfy the triangle inequality, then there are no pole collisions, and the contributions can only come from family 1. If the triangle inequality is violated by some exchanged operator (and of course this can only happen to one operator at a time), then we must consider the poles in family 1 along with the family who has that operator as a quadruple trace (e.g. if $\Delta_A \geq \Delta_B+\Delta_C$ then we take family 2).
We write
\begin{align}
	&W^{\text{ctc}}= \sum_{n_1,n_2,n_3=0}^\infty \frac{\pi^{d/2}}{2} \Gamma_{3-\frac{d}{2}+n_1+n_2+n_3} \prod_{i=1}^{3} \frac{\Gamma_{n_i+1} \Gamma_{2-\frac{d}{2}+n_1} \Gamma_{1-n_i+n_j+n_k}}{\Gamma_{2+2n_i} \Gamma_{2-\frac{d}{2} +2 n_i}} \,G_{2+2n_1,2+2n_2,2+2n_3}+\nonumber\\
	& +\left(\sum_{n_1,n_2,\delta}^\infty \frac{ \Gamma_{n_1+1}
		\Gamma_{n_2+1}
		\Gamma_{-\frac{d}{2}+n_1+2}
		\Gamma_{-\frac{d}{2}+n_2+2}
		\Gamma_{\delta +2 n_1+1}
		\Gamma_{\delta +2 n_2+1}
		}{
		\Gamma_{2 n_1+2}
		\Gamma_{2 n_2+2}
		\Gamma_{-\frac{d}{2}+2 n_1+2}
		\Gamma_{-\frac{d}{2}+2 n_2+2}
	}\right.\nonumber\\
	& \times \frac{ \pi ^{d/2}\Gamma_{-\frac{d}{2}+n_t+2}\Gamma_{n_t+1} \Gamma_{-\frac{d}{2}+\delta +2
			n_1+2 n_2+3}}{\Gamma_{\delta } 	\Gamma_{2 \left(n_t+1\right)}
		\Gamma_{-\frac{d}{2}+2n_t +2}}\,\partial_{\Delta_3}G_{2+2n_1,2+2n_2,2+2n_t}+\nonumber\\
	&\sum_{n_1,n_2,\delta}^\infty \frac{
		-\psi _{-\frac{d}{2}-\delta +2
			n_t+3}-\psi _{-\frac{d}{2}+n_t+2}+2
		\psi _{-\frac{d}{2}+2 n_t+2}+\psi
		_{\delta }-\psi _{\delta +2 n_1+1}+2
		\psi _{2 n_t}-\psi _{\delta +2
			n_2+1}-\psi _{n_t+1}}{ \Gamma
		_{n_2+1}^{-1} \Gamma _{-\frac{d}{2}+n_1+2}^{-1}
		\Gamma _{-\frac{d}{2}+n_2+2}^{-1} \Gamma
		_{\delta +2 n_1+1}^{-1} \Gamma
		_{\delta } \Gamma _{2 n_1+2} \Gamma
		_{2 n_2+2} \Gamma _{-\frac{d}{2}+2
			n_1+2} \Gamma _{-\frac{d}{2}+2 n_2+2}
		\Gamma _{2 \left(n_t+1\right)} }\nonumber\\
	&\left. \times \frac{-\pi ^{d/2} \Gamma _{n_1+1} \Gamma _{n_t+1}  \Gamma
		_{-\frac{d}{2}-\delta +2 n_t+3}}{2  \Gamma
		_{-\frac{d}{2}+2 n_t+2}\Gamma _{\delta +2
			n_2+1}^{-1}  \Gamma
		_{-\frac{d}{2}+n_t+2}^{-1}}\,G_{2+2n_1,2+2n_2,2+2n_t} +(\Delta_3 \leftrightarrow \Delta_1)+(\Delta_3 \leftrightarrow \Delta_2)\right)\,,
\end{align}
where $n_t=n_1+n_2+\delta$ and $\psi_a=S_a-a^{-1}-\gamma_E$.

\section{Higher-point correlators and Harmonic Analysis}
\label{app:Harmonic}

Harmonic analysis of the conformal group leads to the Euclidean inversion formula, which extracts the CFT data from the full correlator. This tool is available even for higher-point functions, but is generically not a useful apparatus for computations. A notable exception is the case of MFT correlators where the inversion can be performed rather explicitly in the case of four-pt functions \cite{karateev:2018}. In this appendix we derive some of the results needed to generalize this procedure to higher-point functions.
\subsection{MFT six-point function from Harmonic Analysis}

 We will study the six-point function of identical real scalar operators $\phi$ of dimension $\Delta_\phi$ presented previously in (\ref{eq:6ptMFT}).
Before moving on, it is important to point out that depending on the OPE channel (snowflake vs comb), we can have different amounts of identity operator exchanges which must be accounted separately in the conformal partial wave expansion, since they are non-normalizable with respect to the Euclidean inversion formula.
To analyze this we recall the definition of the six-point partial waves. The snowflake partial wave is
\begin{equation}
\Psi_{A,B,C}^{\text{sf},1\dots6,abcd}= \int_{7,8,9} \langle \mathcal{O}_1\mathcal{O}_2 \mathcal{O}_A(x_7) \rangle^a\langle \mathcal{O}_3\mathcal{O}_4 \mathcal{O}_B(x_8) \rangle^b\langle \mathcal{O}_5\mathcal{O}_6 \mathcal{O}_C(x_9) \rangle^c \langle\tilde{\mathcal{O}}_A^\dagger(x_7)\tilde{\mathcal{O}}_B^\dagger(x_8)\tilde{\mathcal{O}}_C^\dagger(x_9)\rangle^d\,,
\label{eq:6ptsfpartialwave}
\end{equation}
where we introduced the notation $\int_{i,j,\dots}=\int dx_i dx_j \dots$ to make the equations more compact, $a,b,c,d$ are tensor structure labels and the daggers denote the dual representation, meaning the indices of the $A,B,C$ exchanged operators are contracted. We can now identify the problematic identity exchanges. The $12-34-56$ contraction corresponds to the exchange of three identity operators, which is non-normalizable but can trivially be written as the conventional prefactor times 1. We can also have the exchange of one identity operator and two non-trivial double-twists. This will be the case, for example in the Wick contraction $12-35-46$. Pulling out the prefactor, we will be able to expand this in a factorized form, as a two-point function times a four-point function, and of course the block expansion of the four-pt function will be the non-trivial, but well-known MFT one. In total, we have one wick contraction with three identities and six with one identity. Below, we will therefore focus on the eight remaining non-trivial ones.
On the other hand, we have the comb channel partial wave:
\begin{equation}
\Psi_{A,B,C}^{\text{c},1\dots6,abcd}= \int_{7,8,9} \langle \mathcal{O}_1\mathcal{O}_2 \mathcal{O}_A(x_7) \rangle^a\langle \tilde{\mathcal{O}}_A^\dagger(x_7)\mathcal{O}_3 \mathcal{O}_B(x_8) \rangle^b
\langle\tilde{\mathcal{O}}_B^\dagger(x_8)\mathcal{O}_4\mathcal{O}_C(x_9) \rangle^c
\langle \tilde{\mathcal{O}}_C^\dagger(x_9)\mathcal{O}_5\mathcal{O}_6 \rangle^d \,.
\end{equation}
We can now have two identity exchanges (which is again a factor of 1 with the conventional prefactor choice), or one identity exchange (four choices). We must account for $15-34-26$ and $16-34-25$ Wick contractions which exchanged an identity in the snowflake channel, but do not do so in the comb channel. The remaining eight non-trivial contractions are the same as before.

To obtain the OPE coefficients, we will be using the euclidean inversion formula, which amounts to integrating the euclidean correlator multiplied by an appropriate conformal partial wave. This works because of the orthogonality property of partial waves. The appropriate inner product is given by
\begin{equation}
\left(\left\langle O_{1} \cdots O_{n}\right\rangle,\langle\widetilde{O}_{1}^{\dagger} \cdots \widetilde{O}_{n}^{\dagger}\rangle\right)=\int \frac{d^{d} x_{1} \cdots d^{d} x_{n}}{\operatorname{vol} \mathrm{SO}(d+1,1)}\left\langle O_{1} \cdots O_{n}\right\rangle\langle\widetilde{O}_{1}^{\dagger} \cdots \widetilde{O}_{n}^{\dagger}\rangle\,.
\end{equation}
\subsubsection{Snowflake channel}
For the snowflake partial waves we find the orthogonality property
\begin{align}
&\left(\Psi^{\text{sf},1\dots6,abcd}_{ABC} ,\Psi^{\text{sf},\tilde{1}^\dagger\dots\tilde{6}^\dagger,efgh}_{\tilde{A}'^\dagger \tilde{B}'^\dagger \tilde{C}'^\dagger}\right)= \frac{\delta_{A,A'}\delta_{B,B'}\delta_{C,C'}}{\mu(\Delta_A,J_A)\mu(\Delta_B,J_B)\mu(\Delta_C,J_C)}\times \\
&\left(\langle 12A \rangle^a ,\langle \tilde{1}^\dagger \tilde{2}^\dagger \tilde{A}^\dagger  \rangle^e\right) \left(\langle 34B \rangle^b ,\langle \tilde{3}^\dagger \tilde{4}^\dagger \tilde{B}^\dagger  \rangle^f\right)\left(\langle 56C \rangle^c ,\langle \tilde{5}^\dagger \tilde{6}^\dagger \tilde{C}^\dagger  \rangle^g\right)\left(\langle \tilde{A}^\dagger \tilde{B}^\dagger \tilde{C}^\dagger  \rangle^d ,\langle ABC\rangle^h\right)\,,\nonumber
\end{align}
where $\delta_{X,X'}= 2\pi \delta(\nu_X-\nu_{X'}) \delta_{J_X,J_{X'}}$ and we adopted the shorthand notation $X\equiv\mathcal{O}_X$.
The snowflake partial wave expansion is given by
\begin{equation}
\langle\mathcal{O}_1 \dots \mathcal{O}_6 \rangle = \sum_{J_A,J_B,J_C}\int d\nu_A d\nu_B d\nu_C I^{\text{sf}}_{abcd}(\nu_A,J_A,\nu_B,J_B,\nu_C,J_C) \Psi^{\text{sf},1\dots6}_{A,B,C}(x_i)\,,
\label{eq:6ptsnowflakepwexp}
\end{equation}
and we invert this with the orthogonality relation
\begin{align}
I^{efgh}\equiv\left(\langle\mathcal{O}_1 \dots \mathcal{O}_6 \rangle ,\Psi^{\text{sf},\tilde{1}^\dagger\dots\tilde{6}^\dagger,efgh}_{\tilde{A}'^\dagger \tilde{B}'^\dagger \tilde{C}'^\dagger}\right) = \frac{I^{\text{sf}}_{abcd}(\nu_A,J_A,\nu_B,J_B,\nu_C,J_C)}{\mu(\Delta_A,J_A)\mu(\Delta_B,J_B)\mu(\Delta_C,J_C)} \times \nonumber\\
\left(\langle 12A \rangle^a ,\langle \tilde{1}^\dagger \tilde{2}^\dagger \tilde{A}^\dagger  \rangle^e\right) \left(\langle 34B \rangle^b ,\langle \tilde{3}^\dagger \tilde{4}^\dagger \tilde{B}^\dagger  \rangle^f\right)\left(\langle 56C \rangle^c ,\langle \tilde{5}^\dagger \tilde{6}^\dagger \tilde{C}^\dagger  \rangle^g\right)\left(\langle \tilde{A}^\dagger \tilde{B}^\dagger \tilde{C}^\dagger  \rangle^d ,\langle ABC\rangle^h\right)
\end{align}
Taking identical real scalars $\mathcal{O}_i=\mathcal{O}=\mathcal{O}^\dagger$, this reduces the calculation of the spectral function to the calculation of the integral on the left hand side of the above equation, which is given by
\begin{align}
I^{a}=\int \frac{dx_{1,\dots,9}}{\text{Vol}} \langle\tilde{\mathcal{O}}(x_1)\tilde{\mathcal{O}}(x_2)\tilde{\mathcal{O}}^\dagger_A(x_7) \rangle\langle\tilde{\mathcal{O}}(x_3)\tilde{\mathcal{O}}(x_4)\tilde{\mathcal{O}}^\dagger_B(x_8) \rangle\langle\tilde{\mathcal{O}}(x_5)\tilde{\mathcal{O}}(x_6)\tilde{\mathcal{O}}^\dagger_C(x_9) \rangle\times\nonumber\\ \langle\mathcal{O}_A(x_7)\mathcal{O}_B(x_8)\mathcal{O}_C(x_9)\rangle^a \langle \mathcal{O}(x_1) \dots \mathcal{O}(x_6) \rangle_{\text{MFT}}\,.
\end{align}
As discussed above, the MFT correlator consists of fifteen triplets of Wick contractions. Clearly, when either of the pairs are $12$, $34$ or $56$, we can integrate one of the variables, and this will shadow transform one of the three-point functions. However, we will then have a three-point function with two coincident points, integrated over this point, which is badly divergent. This is the reason why such contributions are non-normalizable and need to be accounted for separately. Therefore, we henceforth focus on a representative contribution, and the remaining ones can be obtained in an identical manner (in fact some of them give a manifestly equal result).
Let us take for concreteness $\langle\mathcal{O}(x_1)\mathcal{O}(x_3)\rangle\langle\mathcal{O}(x_2)\mathcal{O}(x_5)\rangle\langle\mathcal{O}(x_4)\mathcal{O}(x_6)\rangle \subset \langle \mathcal{O}(x_1)\dots\mathcal{O}(x_6)\rangle_{\text{MFT}}$ Performing the integration over $x_{3,5,6}$ applies shadow transforms on the 3-pt functions:
\begin{align}
I^a&=\int \frac{dx_{1,2,4,7,8,9}}{\text{Vol}} \langle\tilde{\mathcal{O}}(x_1)\tilde{\mathcal{O}}(x_2)\tilde{\mathcal{O}}^\dagger_A(x_7) \rangle \langle S[\tilde{\mathcal{O}}](x_1)\tilde{\mathcal{O}}(x_4)\tilde{\mathcal{O}}^\dagger_B(x_8) \rangle \langle S[\tilde{\mathcal{O}}](x_2)S[\tilde{\mathcal{O}}](x_4)\tilde{\mathcal{O}}^\dagger_C(x_9) \rangle \times\nonumber\\
&\times\langle\mathcal{O}_A(x_7)\mathcal{O}_B(x_8)\mathcal{O}_C(x_9)\rangle^a\,,
\end{align}
with the shadow transform for the scalar defined as
\begin{equation}
\langle S[\mathcal{O}](x)\dots \rangle=\int dy \langle\tilde{\mathcal{O}}(x)\tilde{\mathcal{O}}(y) \rangle \langle \mathcal{O}(y)\dots\rangle\,.
\end{equation}
We also define the shadow factor for the three-point functions, which is the fundamental building block for the following calculations
\begin{equation}
\langle S[\mathcal{O}] \mathcal{O}_I \mathcal{O}_J \rangle^a = S([\mathcal{O}]\mathcal{O}_I \mathcal{O}_J)^a_{\,b} \langle \tilde{\mathcal{O}} \mathcal{O}_I \mathcal{O}_J \rangle^b\,.
\end{equation}
We can now write the spectral function as
\begin{align}
I^a=\int \frac{dx_{1,2,4,7,8,9}}{\text{Vol}}\langle\tilde{\mathcal{O}}(x_1)\tilde{\mathcal{O}}(x_2)\tilde{\mathcal{O}}^\dagger_A(x_7) \rangle \langle\mathcal{O}(x_1)\tilde{\mathcal{O}}(x_4)\tilde{\mathcal{O}}^\dagger_B(x_8) \rangle \langle \mathcal{O}(x_2) \mathcal{O}(x_4)\tilde{\mathcal{O}}^\dagger_C(x_9) \rangle \times\nonumber\\
S([\tilde{\mathcal{O}}]\tilde{\mathcal{O}}\tilde{\mathcal{O}}^\dagger_C)S(\mathcal{O}[\tilde{\mathcal{O}}]\tilde{\mathcal{O}}^\dagger_C)S([\tilde{\mathcal{O}}]\tilde{\mathcal{O}}\tilde{\mathcal{O}}^\dagger_B)\langle\mathcal{O}_A(x_7)\mathcal{O}_B(x_8)\mathcal{O}_C(x_9)\rangle^a\,.
\end{align}
Let us make a few comments. First note that there is some freedom in choosing what operators we actually shadow transform, and in the case where we transform two in the same three-point function, we can also choose the order. This leads to apparently different expressions, which presumably give the same result in the end. We should also point out that independently of these choices, the shadow factors only include one spinning operator and are therefore known in closed form for any $J$ and $d$. Additionally, it is clear that each three-point function has exactly one point in common with the other ones, and therefore the position space integrals remain non-trivial. 

To address this, we note that an integral of two three-point functions integrated by a common point is just a four-point partial wave, which admits well-known crossing relations, whose kernel are the $6j$ symbols of the conformal group. There is now some freedom in choosing over what integration point to perform crossing. Crossing over the scalar corresponds to a $6j$ symbol with three spinning operators. Crossing over a spinning one will lead to a similar result. Let us first define the $6j$ symbol\footnote{Our convention for the 6j symbol differs from others in the literature by a normalization factor.} through the crossing relation
\begin{equation}
\Psi^{3214,ab}_{\Delta',J'}(x_3,x_2,x_1,x_4)= \sum_J \int [d\Delta] \left\{\begin{array}{ccc}
{[\Delta_{1},J_1]} & {[\Delta_{2},J_2]} & {\left[\Delta^{\prime}, J^{\prime}\right]} \\
{[\Delta_{3},J_3]} & {[\Delta_{4},J_4]} & {[\Delta, J]}
\end{array}\right\}^{abcd} \Psi^{1234,cd}_{\Delta,J}(x_1,x_2,x_3,x_4)\,.
\end{equation}
Let us cross through the scalar at $x_4$ using
\begin{align}
&\int dx_4
\langle  \tilde{\mathcal{O}}^\dagger_C(x_9)\mathcal{O}(x_2) \mathcal{O}(x_4)\rangle	 \langle\tilde{\mathcal{O}}(x_4)\mathcal{O}(x_1)\tilde{\mathcal{O}}^\dagger_B(x_8) \rangle =\sum_{J'}\int [d\Delta']\\
& \left\{\begin{array}{ccc}
\Delta & \Delta & \Delta \\
{[\tilde{\Delta}_{C},J_C]} & {[\tilde{\Delta}_{B},J_B]} & {[\Delta', J']}
\end{array}\right\}^{b} \int dx_4 \langle\mathcal{O}(x_1)\mathcal{O}(x_2)\mathcal{O}'(x_4) \rangle\langle\tilde{\mathcal{O}}'^\dagger(x_4) \tilde{\mathcal{O}}^\dagger_C(x_9) \tilde{\mathcal{O}}^\dagger_B(x_8)\rangle^b\,.\nonumber
\end{align}
With this, we can easily perform the $x_1,x_2$ integrals using the bubble integral formula
\begin{equation}
\int dx_{1,2} \langle \tilde{\mathcal{O}}(x_1)\tilde{\mathcal{O}}(x_2)\tilde{\mathcal{O}}^\dagger_A(x_7)\rangle \langle\mathcal{O}(x_1)\mathcal{O}(x_2)\mathcal{O}'(x_4) \rangle = \frac{\delta_{A,\mathcal{O}'}}{\mu(\Delta_A,J_A)}\delta(x_{74})\left(\langle\tilde{\mathcal{O}}\tilde{\mathcal{O}}\tilde{\mathcal{O}}_A^\dagger \rangle ,\langle\mathcal{O}\mathcal{O}\mathcal{O}_A \rangle\right)\,.
\end{equation}
The delta function between operators $\mathcal{O}_A$ and $\mathcal{O}'$ removes the auxiliary spectral integral, and the position space delta function gives a final pairing between $A,B,C$ three-point functions. Collecting everything, we obtain
\begin{align}
I^a= S([\tilde{\mathcal{O}}]\tilde{\mathcal{O}}\tilde{\mathcal{O}}^\dagger_C)S(\mathcal{O}[\tilde{\mathcal{O}}]\tilde{\mathcal{O}}^\dagger_C)S([\tilde{\mathcal{O}}]\tilde{\mathcal{O}}\tilde{\mathcal{O}}^\dagger_B)   \left\{\begin{array}{ccc}
\Delta & \Delta & \Delta \\
{[\tilde{\Delta}_{C},J_C]} & {[\tilde{\Delta}_{B},J_B]} & {[\Delta_A, J_A]}
\end{array}\right\}^{b} \times\nonumber\\
\frac{\left(\langle\tilde{\mathcal{O}}\tilde{\mathcal{O}}\tilde{\mathcal{O}}_A^\dagger \rangle ,\langle\mathcal{O}\mathcal{O}\mathcal{O}_A \rangle\right) }{\mu(\Delta_A,J_A)} \left(\langle\tilde{\mathcal{O}}_A^\dagger\tilde{\mathcal{O}}_B^\dagger\tilde{\mathcal{O}}_C^\dagger \rangle^b ,\langle\mathcal{O}_A\mathcal{O}_B\mathcal{O}_C \rangle^a\right)\,.
\end{align}
Note that we have a $6j$ symbol with three spinning operators. When one or two of these operators are scalars, this should be related to well-known $6j$ symbols through the tetrahedral $S_4$ symmetry. Otherwise, this is a non-trivial object to be obtained either through weight-shifting operators, or more directly from the Euclidean inversion formula applied to the cross-channel partial wave with the appropriate tensor structures. 
\subsubsection{Comb channel}
In the comb channel we have slight modifications to the orthogonality properties. The orthogonality relation now reads
\begin{align}
&\left(\Psi^{\text{c},1\dots6,abcd}_{ABC} ,\Psi^{\text{c},\tilde{1}^\dagger\dots\tilde{6}^\dagger,efgh}_{\tilde{A}'^\dagger \tilde{B}'^\dagger \tilde{C}'^\dagger}\right)= \frac{\delta_{A,A'}\delta_{B,B'}\delta_{C,C'}}{\mu(\Delta_A,J_A)\mu(\Delta_B,J_B)\mu(\Delta_C,J_C)}\times \\&\left(\langle 12A \rangle^a ,\langle \tilde{1}^\dagger \tilde{2}^\dagger \tilde{A}^\dagger  \rangle^e\right) \left(\langle \tilde{A}^\dagger3B \rangle^b ,\langle A\tilde{3}^\dagger \tilde{B}^\dagger  \rangle^f\right)\left(\langle  \tilde{B}^\dagger 4 C  \rangle^c ,\langle B \tilde{4}^\dagger\tilde{C}^\dagger\rangle^g\right)\left(\langle \tilde{C}^\dagger 56 \rangle^d ,\langle C \tilde{5}^\dagger \tilde{6}^\dagger   \rangle^h\right)\,,\nonumber
\end{align}
from which the spectral function now follows from the Euclidean inversion integral
\begin{align}
&I^{efgh}\equiv\left(\langle\mathcal{O}_1 \dots \mathcal{O}_6 \rangle ,\Psi^{\text{c},\tilde{1}^\dagger\dots\tilde{6}^\dagger,efgh}_{\tilde{A}'^\dagger \tilde{B}'^\dagger \tilde{C}'^\dagger}\right) = \frac{I^{\text{c}}_{abcd}(\nu_A,J_A,\nu_B,J_B,\nu_C,J_C)}{\mu(\Delta_A,J_A)\mu(\Delta_B,J_B)\mu(\Delta_C,J_C)} \times \\
&\left(\langle 12A \rangle^a ,\langle \tilde{1}^\dagger \tilde{2}^\dagger \tilde{A}^\dagger  \rangle^e\right) \left(\langle \tilde{A}^\dagger3B \rangle^b ,\langle A\tilde{3}^\dagger \tilde{B}^\dagger  \rangle^f\right)\left(\langle  \tilde{B}^\dagger 4 C  \rangle^c ,\langle B \tilde{4}^\dagger\tilde{C}^\dagger\rangle^g\right)\left(\langle \tilde{C}^\dagger 56 \rangle^d ,\langle C \tilde{5}^\dagger \tilde{6}^\dagger   \rangle^h\right)\,,\nonumber
\end{align}
Once again, we specialize to the case of identical external scalars $\mathcal{O}$, such that the spectral function can be obtained from the integral
\begin{align}
I^{ab}=&\int \frac{dx_{1,\dots,9}}{\text{Vol}} \langle\tilde{\mathcal{O}}(x_1)\tilde{\mathcal{O}}(x_2)\tilde{\mathcal{O}}^\dagger_A(x_7) \rangle\langle\mathcal{O}_A(x_7)\tilde{\mathcal{O}}(x_3)\tilde{\mathcal{O}}^\dagger_B(x_8) \rangle^a\langle\mathcal{O}_B(x_8)\tilde{\mathcal{O}}(x_4)\tilde{\mathcal{O}}^\dagger_C(x_9) \rangle^b\times\nonumber\\ 
&\langle\mathcal{O}_C(x_9)\tilde{\mathcal{O}}(x_5)\tilde{\mathcal{O}}(x_6)\rangle \langle \mathcal{O}(x_1) \dots \mathcal{O}(x_6) \rangle_{\text{MFT}}\,.
\end{align}
\subsubsection*{34 Identity}
As discussed above, in the Comb channel there are two qualitatively different types of terms without an identity exchange. The non-trivial contractions in the snowflake channel are also non-trivial in the comb channel. However, the $\langle \mathcal{O}(x_3) \mathcal{O}(x_4) \rangle$ Wick contraction, which is an identity exchange in the snowflake OPE, now becomes a non-trivial contribution. Let us take the $15-34-26$ contraction. This gives a contribution
\begin{align}
I^{ab}\supset\int& \frac{dx_{1,2,3,7,8,9}}{\text{Vol}} \langle\tilde{\mathcal{O}}(x_1)\tilde{\mathcal{O}}(x_2)\tilde{\mathcal{O}}^\dagger_A(x_7) \rangle\langle\mathcal{O}_A(x_7)\tilde{\mathcal{O}}(x_3)\tilde{\mathcal{O}}^\dagger_B(x_8) \rangle^a\langle\mathcal{O}_B(x_8)S[\tilde{\mathcal{O}}](x_3)\tilde{\mathcal{O}}^\dagger_C(x_9) \rangle^b\times\nonumber\\
& \langle\mathcal{O}_C(x_9)S[\tilde{\mathcal{O}}](x_1)S[\tilde{\mathcal{O}}](x_2)\rangle \,.
\end{align}
Note that there is again a lot of freedom in what operator to take the shadow transform, and in the subsequent steps. However, it is unavoidable to obtain a shadow transform on a three-point function with two spinning operators, which gives a complicated (matrix) shadow factor
\begin{align}
I^{ab}\supset\int \frac{dx_{1,2,3,7,8,9}}{\text{Vol}} &\langle\tilde{\mathcal{O}}(x_1)\tilde{\mathcal{O}}(x_2)\tilde{\mathcal{O}}^\dagger_A(x_7) \rangle\langle\mathcal{O}_A(x_7)\tilde{\mathcal{O}}(x_3)\tilde{\mathcal{O}}^\dagger_B(x_8) \rangle^a\langle\mathcal{O}_B(x_8)\mathcal{O}(x_3)\tilde{\mathcal{O}}^\dagger_C(x_9) \rangle^c\times\nonumber\\
&S(\mathcal{O}_C[\tilde{\mathcal{O}}]\tilde{\mathcal{O}})S(\mathcal{O}_C\mathcal{O}[\tilde{\mathcal{O}}])S(\mathcal{O}_B [\tilde{\mathcal{O}}]\tilde{\mathcal{O}}_C)^b_{\,\,c} \langle\mathcal{O}_C(x_9)\mathcal{O}(x_1)\mathcal{O}(x_2)\rangle \,.
\end{align}
We can now apply the bubble integral formula for the $x_{1,2}$ integrals. This imposes a delta function between operators $A$ and $C$, and also on their positions, $x_7-x_9$. In the end, we obtain
\begin{align}
I^{ab}\supset \frac{\delta_{A,C}}{\mu(\Delta_A,J_A)}S(\mathcal{O}_C[\tilde{\mathcal{O}}]\tilde{\mathcal{O}})S(\mathcal{O}_C\mathcal{O}[\tilde{\mathcal{O}}])S(\mathcal{O}_B [\tilde{\mathcal{O}}]\tilde{\mathcal{O}}_C)^b_{\,\,c} \left( \langle\tilde{\mathcal{O}}\tilde{\mathcal{O}}\tilde{\mathcal{O}}_A  \rangle,\langle\mathcal{O}_A \mathcal{O}\mathcal{O}\rangle\right)\times\nonumber\\\left( \langle\mathcal{O}_A\tilde{\mathcal{O}}\tilde{\mathcal{O}}_B  \rangle^a,\langle\mathcal{O}_B \mathcal{O}\tilde{\mathcal{O}}_A\rangle^c\right)\,.
\end{align}
We again emphasize that this depends on a non-trivial shadow factor.
\subsubsection*{Non-trivial contractions: one point in common}
Now, we have to consider again the eight non-trivial Wick contractions, which contain no identity operators in any channel. 
There are two further classes of Wick contractions, ones which will induce two common points between two pairs of three-point functions, and ones where all three-point functions will have one point in common with each other.
A representative example of the second type is the Wick contraction $14-25-36$. Its contribution to the spectral function is given by
\begin{align}
I^{ab}\supset\int \frac{dx_{1,\dots,9}}{\text{Vol}} &\langle\tilde{\mathcal{O}}(x_1)\tilde{\mathcal{O}}(x_2)\tilde{\mathcal{O}}^\dagger_A(x_7) \rangle\langle\mathcal{O}_A(x_7)\tilde{\mathcal{O}}(x_3)\tilde{\mathcal{O}}^\dagger_B(x_8) \rangle^a\langle\mathcal{O}_B(x_8)\tilde{\mathcal{O}}(x_4)\tilde{\mathcal{O}}^\dagger_C(x_9) \rangle^b\times\nonumber\\ 
&\langle\mathcal{O}_C(x_9)\tilde{\mathcal{O}}(x_5)\tilde{\mathcal{O}}(x_6)\rangle \langle \mathcal{O}(x_1)\mathcal{O}(x_4) \rangle\langle \mathcal{O}(x_2)\mathcal{O}(x_5) \rangle\langle \mathcal{O}(x_3)\mathcal{O}(x_6) \rangle\,.
\end{align}
As usual we have some freedom in what operators to shadow transform. In this case, this is particularly relevant, since out of the three shadow factors, we can have either zero, one or two "difficult" shadow factors, depending on what operators we transform. Sticking to the easiest possibility, we
inevitably get only one common point per three-point function, which means that once again we need to use crossing relations or $6j$ symbols to proceed with the position space integrals. It is convenient to cross through $\mathcal{O}_A(x_7)$ and then  do the $x_{2,3}$ integrals using the bubble formula. In the end we get
\begin{align}
I^{ab}\supset \left\{\begin{array}{ccc}
\Delta & {[\tilde{\Delta}_{B},J_B]} & {[\Delta_{A},J_A]} \\
\tilde{\Delta} & \tilde{\Delta} & {[\Delta_{C},J_C]}
\end{array}\right\}^{ac} &S([\tilde{\mathcal{O}}]\tilde{\mathcal{O}} \tilde{\mathcal{O}}_A)S(\mathcal{O}_C[\tilde{\mathcal{O}}] \tilde{\mathcal{O}})S(\mathcal{O}_C\mathcal{O}  [\tilde{\mathcal{O}}]) \times\\
&\frac{\left(\langle\tilde{\mathcal{O}}\tilde{\mathcal{O}}\tilde{\mathcal{O}}_C^\dagger \rangle ,\langle\mathcal{O}\mathcal{O}\mathcal{O}_C \rangle\right)}{\mu(\Delta_C,J_C)}\left(\langle\mathcal{O}\tilde{\mathcal{O}}_B^\dagger \mathcal{O}_C \rangle^c ,\langle \tilde{\mathcal{O}}\mathcal{O}_B\tilde{\mathcal{O}}_C^\dagger \rangle^b\right)\nonumber
\end{align}
There is just one more class of Wick contractions to analyze.
\subsubsection*{Non-trivial contractions: two points in common}
We can also have two-point functions connecting the adjacent three-point functions of the partial wave. A representative example for this case is the Wick contraction $16-23-45$. The contribution to the spectral function is given by
\begin{align}
I^{ab}\supset\int \frac{dx_{1,\dots,9}}{\text{Vol}}& \langle\tilde{\mathcal{O}}(x_1)\tilde{\mathcal{O}}(x_2)\tilde{\mathcal{O}}^\dagger_A(x_7) \rangle\langle\mathcal{O}_A(x_7)\tilde{\mathcal{O}}(x_3)\tilde{\mathcal{O}}^\dagger_B(x_8) \rangle^a\langle\mathcal{O}_B(x_8)\tilde{\mathcal{O}}(x_4)\tilde{\mathcal{O}}^\dagger_C(x_9) \rangle^b\times\nonumber\\ 
&\langle\mathcal{O}_C(x_9)\tilde{\mathcal{O}}(x_5)\tilde{\mathcal{O}}(x_6)\rangle \langle \mathcal{O}(x_1)\mathcal{O}(x_6) \rangle\langle \mathcal{O}(x_2)\mathcal{O}(x_3) \rangle\langle \mathcal{O}(x_4)\mathcal{O}(x_5) \rangle\,.
\end{align}
Once again, we have the freedom to perform the shadow transforms, and we can get either zero, one or two hard factors. Let us get all simple factors by making the choice
\begin{align}
I^{ab}\supset\int \frac{dx_{1,\dots,9}}{\text{Vol}} &\langle\tilde{\mathcal{O}}(x_1)\mathcal{O}(x_3)\tilde{\mathcal{O}}^\dagger_A(x_7) \rangle\langle\mathcal{O}_A(x_7)\tilde{\mathcal{O}}(x_3)\tilde{\mathcal{O}}^\dagger_B(x_8) \rangle^a\langle\mathcal{O}_B(x_8)\tilde{\mathcal{O}}(x_4)\tilde{\mathcal{O}}^\dagger_C(x_9) \rangle^b\times\nonumber\\
&S(\mathcal{O}_C[\tilde{\mathcal{O}}]\tilde{\mathcal{O}})S(\mathcal{O}_C\mathcal{O}[\tilde{\mathcal{O}}])S(\tilde{\mathcal{O}}[\tilde{\mathcal{O}}]\tilde{\mathcal{O}}_A) \langle\mathcal{O}_C(x_9)\mathcal{O}(x_4)\mathcal{O}(x_1)\rangle \,.
\end{align}
There are now two possible approaches. We can try to do, for example the $x_3,x_7$ integrals, which would involve a bubble integral with a spinning operator integrated over
\begin{equation}
\int_{3,7} \langle\tilde{\mathcal{O}}(x_1)\mathcal{O}(x_3)\tilde{\mathcal{O}}^\dagger_A(x_7) \rangle\langle\mathcal{O}_A(x_7)\tilde{\mathcal{O}}(x_3)\tilde{\mathcal{O}}^\dagger_B(x_8) \rangle^a = \frac{\delta_{\tilde{\mathcal{O}}_B,\mathcal{O}} \delta(x_1-x_8)}{\mu(\Delta,0)}  \left(\langle\tilde{\mathcal{O}}\mathcal{O}\tilde{\mathcal{O}}_A \rangle ,\langle\mathcal{O}\tilde{\mathcal{O}}\mathcal{O}_A \rangle\right)\,.
\end{equation}
This would mean that the operator exchanged at $\mathcal{O}_B(x_8)$ would need to be the same as the external operator. It is not hard to argue that this is possible in MFT. We are then able to do the final three-point pairing and obtain
\begin{equation}
I\supset S(\mathcal{O}_C[\tilde{\mathcal{O}}]\tilde{\mathcal{O}})S(\mathcal{O}_C\mathcal{O}[\tilde{\mathcal{O}}])S(\tilde{\mathcal{O}}[\tilde{\mathcal{O}}]\tilde{\mathcal{O}}_A) \frac{\left(\langle\tilde{\mathcal{O}}\mathcal{O}\tilde{\mathcal{O}}_A \rangle ,\langle\mathcal{O}\tilde{\mathcal{O}}\mathcal{O}_A \rangle\right)}{\mu(\Delta,0)}  \left(\langle\tilde{\mathcal{O}}\tilde{\mathcal{O}}\tilde{\mathcal{O}}_C \rangle ,\langle\mathcal{O}\mathcal{O}\mathcal{O}_C \rangle\right)\,.
\end{equation}
Note that the tensor structure indices went away, since $\mathcal{O}_B$ became a scalar operator, and therefore all tensor structures became unique.

\subsection{Partial wave decompositon and conformal blocks}
In the previous section we formally derived the partial wave decomposition of MFT six-point functions. However, to obtain the actual CFT data, we need to write down the conformal block decomposition and read-off the OPE coefficients.
In this subsection, we establish a relation between the partial wave decomposition and the conformal block expansion.
We quickly review the case of the four-point function which can be expanded in partial waves as
\begin{equation}
\label{pwdecomp}
\langle \mathcal{O}_1\mathcal{O}_2\mathcal{O}_3\mathcal{O}_4\rangle = \sum_{\rho} \int_{\frac{d}{2}}^{\frac{d}{2}+i \infty}\frac{d\Delta}{2\pi i} I_{ab}(\Delta, \rho) \Psi_{\mathcal{O}}^{\mathcal{O}_i(ab)} (x_i)+ \text{discrete}\,.
\end{equation}
Here discrete is associated with possible additional isolated contributions, notably including the identity. 
The partial wave is defined in terms of a conformally-invariant integral involving two three-point structures 
\begin{equation}
\label{pw4pt}
\Psi_{\mathcal{O}}^{\mathcal{O}_i(ab)} (x_i)=\int d^{d}x \langle\mathcal{O}_1\mathcal{O}_2\mathcal{O}(x) \rangle^{(a)} \langle\mathcal{O}_3\mathcal{O}_4\widetilde{\mathcal{O}}^{\dagger}(x) \rangle^{(b)}\,.
\end{equation}

In order to relate the partial wave decomposition to conformal blocks we follow the strategy of~\cite{karateev:2018}. The partial wave in~(\ref{pw4pt}) is a solution of the Casimir equation and therefore one can establish its relation to conformal blocks by uniquely estimating its form in the OPE limit $x_1\rightarrow x_2$. Obviously the Euclidean OPE limit cannot be taken simply inside the integral as the integrand probes regions where the OPE in the pair (12) is no longer valid. However, understanding the leading behaviour outside this region is enough to match those contributions to a given conformal block.
For concreteness, consider the replacement
\begin{equation}
\langle \mathcal{O}_1 \mathcal{O}_2 \mathcal{O}(x)\rangle^{(a)}\rightarrow C_{12\mathcal{O}}^{(a)}\langle \mathcal{O}^{\dagger}(x_2)\mathcal{O}(x)\rangle\,,
\end{equation}
where $ C_{12\mathcal{O}}^{(a)}$ encodes leading terms in the OPE $\mathcal{O}_1 \times \mathcal{O}_2$. With this replacement the integral in~(\ref{pw4pt}) becomes a shadow transform of $\widetilde{\mathcal{O}}^{\dagger}$, 
\begin{equation}
\Psi_{\mathcal{O}}^{\mathcal{O}_i(ab)}\sim C_{12\mathcal{O}}^{(a)}\langle \mathcal{O}_3\mathcal{O}_4\mathbf{S}[\widetilde{\mathcal{O}}^{\dagger}]\rangle^{(b)}=S(\mathcal{O}_3\mathcal{O}_4[\widetilde{\mathcal{O}}^{\dagger}])^{b}_{c}C_{12\mathcal{O}}^{(a)}\langle\mathcal{O}_3\mathcal{O}_4\mathcal{O}^{\dagger}\rangle^{(c)}\,.
\end{equation}
On the other hand, the conformal block $G_{\mathcal{O}}^{(ab)}$ is a solution of the Casimir equation, which in the OPE limit of $\mathcal{O}_1 \times \mathcal{O}_2$ behaves as
\begin{equation}
G_{\mathcal{O}}^{(ab)}\sim  C_{12\mathcal{O}}^{(a)}\langle\mathcal{O}_3\mathcal{O}_4\mathcal{O}^{\dagger}\rangle^{(b)}\,,\qquad (x_1\rightarrow x_2)\,.
\end{equation}
It is thus clear that the partial wave must contain a term 
\begin{equation}
\Psi_{\mathcal{O}}^{\mathcal{O}_i(ab)}\supset S(\mathcal{O}_3\mathcal{O}_4[\widetilde{\mathcal{O}}^{\dagger}])^{b}_{c}G_{\mathcal{O}}^{(ac)}\,.
\end{equation}
Similarly, if one performs an OPE on $\mathcal{O}_3\times\mathcal{O}_4$ instead, it is straightforward to show that the partial wave contains a term
\begin{equation}
\Psi_{\mathcal{O}}^{\mathcal{O}_i(ab)}\supset S(\mathcal{O}_1\mathcal{O}_2[\mathcal{O}])^{a}_{c}G_{\widetilde{\mathcal{O}}}^{(cb)}\,.
\end{equation}
Putting everything together we conclude that
\begin{equation}
\Psi_{\mathcal{O}}^{\mathcal{O}_i(ab)}=S(\mathcal{O}_3\mathcal{O}_4[\widetilde{\mathcal{O}}^{\dagger}])^{b}_{c}G_{\mathcal{O}}^{(ac)}+S(\mathcal{O}_1\mathcal{O}_2[\mathcal{O}])^{a}_{c}G_{\widetilde{\mathcal{O}}}^{(cb)}\,,
\end{equation}
which reflects the fact that the Casimir equation is invariant under $\Delta\rightarrow d-\Delta$. Inserting this relation on~(\ref{pwdecomp}), extending the integration region along the entire imaginary axis and using shadow symmetry, allows us to write
\begin{equation}
\langle \mathcal{O}_1\dots\mathcal{O}_4\rangle=\sum_{\rho} \int_{\frac{d}{2}-i\infty}^{\frac{d}{2}+i \infty}\frac{d\Delta}{2\pi i} C_{ac}(\Delta, \rho) G_{\mathcal{O}}^{(ac)}\,,
\end{equation}
where $C_{ac}(\Delta, \rho)\equiv I_{ab}(\Delta, \rho)S(\mathcal{O}_3\mathcal{O}_4[\widetilde{\mathcal{O}}^{\dagger})^{b}_{c}$. As usual we can then deform the contour integration away from the principal series and pick up poles of $C_{ac}(\Delta, \rho)$ on the real line, which have residues that encode CFT data. For a particular exchanged operator $\mathcal{O}_*$, we have
\begin{equation}
	C_{12*}C_{34*}= - \textrm{Res}_{\Delta=\Delta_*} C(\Delta,\rho_*)\,.
\end{equation}

This formalism can straightforwardly be adapted to the case of higher-point functions. For five-point functions, the discussion has already been presented in~\cite{Meltzer:2019}, but we also review it here.
We consider the partial wave
\begin{equation}
\Psi_{A, B}^{\mathcal{O}_i(abc)}(x_i)=\int d^{d}x_A d^{d}x_B \langle \mathcal{O}_1\mathcal{O}_2 \mathcal{O}_A \rangle ^{(a)}\langle\widetilde{\mathcal{O}}^{\dagger}_A\mathcal{O}_5 \widetilde{\mathcal{O}}^{\dagger}_B \rangle ^{(b)}\langle \mathcal{O}_B\mathcal{O}_3 \mathcal{O}_4 \rangle ^{(c)}
\end{equation}
where $\mathcal{O}_{A, B}$ are exchanged operators.
A five-point function can be decomposed in terms of this partial wave 
\begin{equation}
\langle \mathcal{O}_1\dots\mathcal{O}_5 \rangle = \sum_{\rho_A, \rho_B} \int_{\frac{d}{2}}^{\frac{d}{2}+ i \infty}\frac{d\Delta_A}{2\pi i} \int_{\frac{d}{2}}^{\frac{d}{2}+ i \infty}\frac{d\Delta_B}{2\pi i} I_{abc}(\Delta_A, \rho_A; \Delta_B, \rho_B) \Psi_{A, B}^{\mathcal{O}_i(abc)}(x_i)\,.
\end{equation}
To expand this partial wave in terms of conformal blocks we again consider OPE limits. In particular, we take $x_1\rightarrow x_2$ and $x_3 \rightarrow x_4$ at the level of the integrand and we observe that the partial wave must contain the term
\begin{equation}
\Psi_{A, B}^{\mathcal{O}_i(abc)}(x_i)\supset C_{12A}^{(a)}C_{34B}^{(c)}\langle \mathbf{S}[\widetilde{\mathcal{O}}^{\dagger}_A]\mathcal{O}_5\mathbf{S}[\widetilde{\mathcal{O}}^{\dagger}_B]\rangle^{(b)}=(S_{\widetilde{A}}^{5\widetilde{B}})^{b}_{d}(S_{\widetilde{B}}^{A 5})^{d}_{e}\underbrace{C_{12A}^{(a)}C_{34B}^{(c)}\langle \mathcal{O}^{\dagger}_A\mathcal{O}_5\mathcal{O}^{\dagger}_B \rangle^{(e)}}_{\propto G_{AB}^{(aec)}}\,,
\end{equation}
where we have used the shorthand notation $S_{A}^{BC}=S([\mathcal{O}_A]\mathcal{O}_B\mathcal{O}_c)$ and recognized the leading behaviour of the conformal block $G_{AB}^{(aec)}$ in the OPE limits $x_1\rightarrow x_2$ and $x_3 \rightarrow x_4$. As above, we notice that the partial wave $\Psi_{A, B}^{\mathcal{O}_i(abc)}(x_i)$ is a solution of the Casimir equations, one for each OPE exchange, and therefore it enjoys the invariance $\Delta \leftrightarrow d- \Delta$. We can then propose the decomposition
\begin{equation}
\Psi_{A, B}^{\mathcal{O}_i}(x_i)={R_1} G_{AB}(x_i)+{R_2} G_{\widetilde{A} B}(x_i)+{R_3} G_{A \widetilde{B}}(x_i)+{R_4} G_{\widetilde{A}\widetilde{B}}(x_i)\,,
\end{equation}
where, as we have seen, ${R_1}^{a}_{b}=(S_{\widetilde{A}}^{5\widetilde{B}})^{a}_{c}(S_{\widetilde{B}}^{A 5})^{c}_{b}$. In order to find the remaining $R_i$'s we explore the symmetry of the partial wave:
\begin{equation}
\label{pw5ptsym}
\begin{aligned}
\Psi_{A, B}^{\mathcal{O}_i(abc)}(x_i)&=\int d^{d}x_A d^{d}x_B \langle \mathcal{O}_1\mathcal{O}_2 \mathcal{O}_A \rangle ^{(a)}\langle\widetilde{\mathcal{O}}^{\dagger}_A\mathcal{O}_5 \widetilde{\mathcal{O}}^{\dagger}_B \rangle ^{(b)}\langle \mathcal{O}_B\mathcal{O}_3 \mathcal{O}_4 \rangle ^{(c)}\\
&=\int d^{d}x_A d^{d}x_A^{\prime} d^{d}x_B (({S_{A}^{5\widetilde{B}}})^{-1})^{b}_{d}\langle \mathcal{O}_1\mathcal{O}_2 \mathcal{O}_A \rangle ^{(a)}\langle \widetilde{\mathcal{O}}^{\dagger}_A\widetilde{\mathcal{O}}_{A^{\prime}}\rangle\langle\mathcal{O}^{\dagger}_{A^{\prime}}\mathcal{O}_5 \widetilde{\mathcal{O}}^{\dagger}_B \rangle ^{(d)}\langle \mathcal{O}_B\mathcal{O}_3 \mathcal{O}_4 \rangle ^{(c)}\\
&= \int d^{d}x_A d^{d}x_B (S_{A}^{12})^{a}_{d}(({S_{A}^{5\widetilde{B}}})^{-1})^{b}_{e}\langle \mathcal{O}_1\mathcal{O}_2 \widetilde{\mathcal{O}}_A \rangle ^{(d)}\langle\mathcal{O}^{\dagger}_A\mathcal{O}_5 \widetilde{\mathcal{O}}^{\dagger}_B \rangle ^{(e)}\langle \mathcal{O}_B\mathcal{O}_3 \mathcal{O}_4 \rangle ^{(c)}\\
&= (S_{A}^{12})^{a}_{d}(({S_{A}^{5\widetilde{B}}})^{-1})^{b}_{e}\Psi_{\widetilde{A}, B}^{\mathcal{O}_i(dec)}(x_i)\,.
\end{aligned}
\end{equation}
Performing an OPE expansion on the $\Psi_{\widetilde{A}, B}^{\mathcal{O}_i(abc)}(x_i)$, we observe
\begin{equation}
\Psi_{\widetilde{A}, B}^{\mathcal{O}_i (abc)}(x_i)\supset (S_{A}^{5 \widetilde{B}})^{b}_{d} (S_{\widetilde{B}}^{\widetilde{A}5})^{d}_{e} G_{\widetilde{A} B}^{(aec)}(x_i)\,,
\end{equation}
from which follows that
\begin{equation}
{R_2}^{ab}_{de}=(S_{A}^{12})^{a}_{d}(S_{\widetilde{B}}^{\widetilde{A}5})^{b}_{e}\,.
\end{equation}
Similarly, one can show that
\begin{equation}
R_3=S_{\widetilde{A}}^{5\widetilde{B}}S_{B}^{34}\,,\quad
R_4=S_{A}^{12}S_{B}^{34}\,.
\end{equation}
Just as we have shown in the 4-point case, one can use the shadow symmetry of $I_{abc}$ to extend the region of integration such that
\begin{equation}
\langle \mathcal{O}_1\dots\mathcal{O}_5 \rangle = \sum_{\rho_A, \rho_B} \int_{\frac{d}{2}-i\infty}^{\frac{d}{2}+ i \infty}\frac{d\Delta_A}{2\pi i} \int_{\frac{d}{2}-i\infty}^{\frac{d}{2}+ i \infty}\frac{d\Delta_B}{2\pi i} I_{abc}(\Delta_A, \rho_A; \Delta_B, \rho_B)(S_{\widetilde{A}}^{5\widetilde{B}})^{b}_{d}(S_{\widetilde{B}}^{A 5})^{d}_{e}G_{AB}^{aec}\,.
\end{equation}

The exact same techniques can be applied to six-point functions. Here, we focus on the snowflake decomposition which admits the partial wave expansion (\ref{eq:6ptsnowflakepwexp}), where the snowflake partial wave is defined in (\ref{eq:6ptsfpartialwave}).
In a completely analogous procedure as discussed above, we can relate this partial wave to conformal blocks. In particular, from the shadow invariance of the Casimir equations it is natural to expand the partial wave as
\begin{align}
\label{pwtoblocks}
\Psi_{A,B,C}^{\mathcal{O}_i}(x_i)&= R_1 G_{ABC}+R_2 G_{\widetilde{A}BC}+ R_3 G_{A\widetilde{B}C}+ R_4 G_{AB\widetilde{C}}\nonumber\\&+ R_5 G_{\widetilde{A}\widetilde{B}C}+ R_6 G_{A\widetilde{B}\widetilde{C}}+ R_7 G_{\widetilde{A} B \widetilde{C}}+ R_8 G_{\widetilde{A} \widetilde{B} \widetilde{C}}\,,
\end{align}
where
\begin{align}
R_1 &= S_{\widetilde{A}}^{\widetilde{B}\widetilde{C}}S_{\widetilde{B}}^{A\widetilde{C}}S_{\widetilde{C}}^{AB}\,, & R_2&= S_{A}^{12} S_{\widetilde{B}}^{\widetilde{A}\widetilde{C}} S_{\widetilde{C}}^{\widetilde{A}B}\,, & R_3 &= S_{B}^{34} S_{\widetilde{A}}^{\widetilde{B}\widetilde{C}} S_{\widetilde{C}}^{A\widetilde{B}}\,, & R_4 &= S_{C}^{56} S_{\widetilde{A}}^{\widetilde{B}\widetilde{C}} S_{\widetilde{B}}^{A\widetilde{C}}\,,\nonumber\\
R_5 &= S_{A}^{12} S_{B}^{34} S_{\widetilde{C}}^{\widetilde{A}\widetilde{B}}
\,, & R_6&= S_{B}^{34} S_{C}^{56} S_{\widetilde{A}}^{\widetilde{B}\widetilde{C}}\,, & R_7 &= S_{A}^{12} S_{C}^{56} S_{\widetilde{B}}^{\widetilde{A}\widetilde{C}}
\,, & R_8 &= S_{A}^{12} S_{B}^{34} S_{C}^{56} \,.
\end{align}
The computation of these coefficients exactly mimics the computations in (\ref{pw5ptsym}) and below. One can now insert~(\ref{pwtoblocks}) on the partial wave expansion and extend the region of integration to the whole imaginary axis, keeping only one term which reads
\begin{align}
\langle \mathcal{O}_1 \dots \mathcal{O}_6 \rangle = \sum_{\rho_A, \rho_B, \rho_C} \int_{\frac{d}{2}-i\infty}^{\frac{d}{2}+ i \infty}\frac{d\Delta_A}{2\pi i}\frac{d\Delta_B}{2\pi i}\frac{d\Delta_C}{2\pi i}&I_{abcd}(\Delta_A, \rho_A; \Delta_B, \rho_B; \Delta_C, \rho_C)\times\nonumber\\&{S_{\widetilde{A}}^{\widetilde{B}\widetilde{C}}}^{d}_{e}{S_{\widetilde{B}}^{A\widetilde{C}}}^{e}_{f}{S_{\widetilde{C}}^{AB}}^{f}_{g} G_{ABC}^{(abcg)}\,.
\end{align}

\subsection{Direct computation of spinning shadow coefficients}

In the previous subsections, we have repeatedly come across shadow coefficients involving multiple spinning operators but the computation of these shadow coefficients is an important question on its own. In this subsection, we will derive some of them using the shadow formalism.
In \cite{karateev:2018} some of these coefficients were computed using weight-shifting operators from which recursion relations were derived~\cite{karateev:2017}. Here, we extend these results and compute directly the explicit integration  involved in the definition of these coefficients. We can write the shadow transform of an operator in a three-point structure as
\begin{equation}
    \langle  \mathcal{O}_1 \mathcal{O}_2 \mathcal{S}[\mathcal{O}_3] \rangle^{(a)}=\int d^{d}x_0 \langle \widetilde{\mathcal{O}}_3\widetilde{\mathcal{O}}^{\dagger}_0\rangle \langle\mathcal{O}_1 \mathcal{O}_2 \mathcal{O}_0\rangle^{(a)}\,,
\end{equation}
where we have an implicit contraction of indices. Here we only consider symmetric and traceless representations of the conformal group and so the two- and three-point structures can be written in terms of the two fundamental building blocks~\cite{costa:2011} that appeared in (\ref{eq:estruturas3pf}).
In particular we choose the normalization of the two-point structure to take the form
\begin{equation}
    \begin{split}
        &\langle \mathcal{O}(x_1, z_1)\mathcal{O}(x_2, z_2) \rangle =\dfrac{H_{12}^{J}}{x_{12}^{\Delta+J}}\,.
    \end{split}
\end{equation}
On the other hand, the three-point structure is given by (\ref{eq:SpinningThreepointFunction}) once we omit the OPE coefficients. As in the main text, we use here the index-free notation of~\cite{costa:2011,costa:2011scb}. In particular, in what follows we will use the formula
\begin{equation}
\label{contraction}
    (a\cdot \mathcal{D}_{z})^J(b\cdot z)^J=\frac{(J!)^2}{2^J}(a^2 b^2)^{\frac{J}{2}} C^{h-1}_{J}\left(\frac{a\cdot b}{(a^2 b^2)^{\frac{1}{2}}}\right)\,,
\end{equation}
where $C^{h-1}_{J}$ is a Gegenbauer polynomial and $h=d/2$.

Before moving on to more complicated examples, let us, as a warm-up, compute the shadow integral for three scalar operators. In this case, we can use the well-known star-triangle formula~\cite{Fradkin:1978}
\begin{equation}
\label{startriangle}
    \int d^d x_{0} \frac{1}{(x_{10}^2)^{a}(x_{20}^2)^{b}(x_{30}^2)^{c}}=\underbrace{\frac{\pi^{h}\Gamma(h-a)\Gamma(h-b)\Gamma(h-c)}{\Gamma(a)\Gamma(b)\Gamma(c)}}_{\equiv\, {
    \textstyle G(a,b,c)}}\frac{1}{(x_{12}^2)^{h-c}(x_{13}^2)^{h-b}(x_{23}^2)^{h-a}}\,,
\end{equation}
with $ a+b+c=2h$ to get
\begin{equation}
\begin{split}
    \langle \phi_{\Delta_1} \phi_{\Delta_2} \mathcal{S}[\phi_{\Delta_3}]\rangle&=\int d^d x_{0} \frac{1}{x_{30}^{2(d-\Delta_{3})}}\frac{1}{(x_{12}^2)^{\frac{\Delta_1+\Delta_2-\Delta_3}{2}}(x_{10}^2)^{\frac{\Delta_1-\Delta_2+\Delta_3}{2}}(x_{20}^2)^{\frac{-\Delta_1+\Delta_2+\Delta_3}{2}}}\\
    &=\dfrac{\pi^{h}\Gamma(\Delta_3-h)\Gamma(\frac{\tilde{\Delta}_3+\Delta_1-\Delta_2}{2})\Gamma(\frac{\tilde{\Delta}_3+\Delta_2-\Delta_1}{2})}{\Gamma(2h-\Delta_3)\Gamma(\frac{\Delta_3+\Delta_1-\Delta_2}{2})\Gamma(\frac{\Delta_3+\Delta_2-\Delta_1}{2})}\langle\phi_{\Delta_1} \phi_{\Delta_2} \phi_{\tilde{\Delta}_3} \rangle\,,
\end{split}
\end{equation}
from which we can easily read the shadow coefficient $S(\phi_{\Delta_1}\phi_{\Delta_2}[\phi_{\Delta_3}])$.

In~\cite{karateev:2018} the authors computed the shadow coefficients for the case where two of the operators were scalars and one of them had spin $J$. Here we compute the coefficients corresponding to two spinning operators and a scalar and we shall recover their results as a restriction. Let us take the operators at $x_1$ and $x_3$ to be spinning operators whereas the operator at $x_2$ is a scalar. In this case the three-point structure simplifies and we are left just with the label $\ell_{2}=\ell$. We first do a shadow transform of the operator at $x_3$
\begin{equation}
\begin{aligned}
    &\langle \mathcal{O}_{\Delta_1,J_1}\phi_{\Delta_2}\mathcal{S}[\mathcal{O}_{\Delta_3,J_3}] \rangle^{(\ell)}=\\
    &=\int d^d x_{0} \langle \widetilde{\mathcal{O}}_{\Delta_3, J_3}(x_3,z_3)\widetilde{\mathcal{O}}_{\Delta_3\,\mu_1\dots \mu_{J_3}}^{\dagger}(x_0)\rangle\langle \mathcal{O}_{\Delta_1, J_1}(x_1,z_1)\phi_{\Delta_2}(x_2)\mathcal{O}_{\Delta_3}^{\mu_1\dots\mu_{J_3}}(x_0)\rangle^{(\ell)}\,,
\end{aligned}
\end{equation}
where the indices to be contracted are explicitly shown. In light of the results of~\cite{costa:2011}, this contraction can be simply done in terms of encoding polynomials that depend on the buildings blocks $H_{ij}$ and $V_{i,jk}$. By doing so, one immediately recognizes that the term associated with the two-point function is already of the desired form $(a\cdot \mathcal{D}_{z_0})^{J_3}$ with $a^{\mu}=(x_{03}\cdot z_{3}) x_{03}^{\mu}-\frac{1}{2} x_{03}^2 z_{3}^{\mu}$.\footnote{Notice that $a^2=0$. We may then just keep the term $k=0$ in the series definition of the Gegenbauer polynomial, $C_{J}^{\lambda}(z)=\sum_{k=0}^{\left\lfloor{\frac{J}{2}}\right\rfloor}\dfrac{(-1)^k (\lambda)_{J-k}(2z)^{J-2k}}{k!(J-k)!}$.} The terms in the three-point structure require some additional care. It is easy to see however that the $z_0$-dependent terms can be completed to a binomial of degree $J_3$ of form $\left(b\cdot z_0\right)^{J_3}$, as appears in~(\ref{contraction}). After using this equation, one then needs to expand back the binomial and collect only the term we have started with. The computation is straightforward and leads to the following expression for our integral
\begin{align}
     \int d^d x_0\,&\frac{ (x_{12}^2)^{-\frac{1}{2}(\Delta_1+J_1+\Delta_2-\Delta_3+J_3-2\ell)}}{2^{J3} (x_{01}^{2})^{\frac{1}{2}(\Delta_1+J_1-\Delta_2+\Delta_3-J_3)}(x_{02}^{2})^{\frac{1}{2}(-\Delta_1-J_1+\Delta_2+\Delta_3-J_3+2\ell)}(x_{03}^{2})^{\widetilde{\Delta}_3+J_3}}\times\nonumber\\[4mm]&
    \times V_{1,20}^{J_1-\ell} \left(V_{3,01}+V_{3,20}\right)^{J_3-\ell} \big(V_{3,01} (x_{01}\cdot z_{1})-\mathcal{H}_{0,3,1}\big)^{\ell}\,,
\end{align}
where for compactness we defined $\mathcal{H}_{i,j,k}=(x_{ij}\cdot z_{j}) (x_{k j}\cdot z_k)-\frac{1}{2} (z_j\cdot z_k) x_{ij}^2$.

After performing the expansion of the integrand, one observes that all the terms to be integrated take the simple form 
\begin{equation}
\label{intterms}
    \frac{(x_{01}\cdot z_1)^{\alpha}(x_{03}\cdot z_3)^{\beta}}{ (x_{01}^2)^{a}(x_{02}^2)^{b}(x_{03}^2)^{c}}\,.
\end{equation}
The terms in the numerator can be found from taking derivatives of the denominator as
\begin{equation}
    (z_j \cdot \partial_{x_j})^{\alpha} (x_{ij}^2)^{-a}= 2^{\alpha}\frac{\Gamma(a+\alpha)}{\Gamma(a)} \frac{(x_{ij}\cdot z_j)^{\alpha}}{(x_{ij}^2)^{a+\alpha}}\,.
\end{equation}
It is then easy to integrate the terms in (\ref{intterms}) by swapping the order of integration and differentiation
\begin{equation}
\begin{aligned}
\int d^d x_0\, \frac{(x_{01}\cdot z_1)^{\alpha}(x_{03}\cdot z_3)^{\beta}}{ (x_{01}^2)^{a}(x_{02}^2)^{b}(x_{03}^2)^{c}}=&\frac{\Gamma(a-\alpha)}{2^{\alpha}\Gamma(a)}\frac{\Gamma(c-\beta)}{2^{\beta}\Gamma(c)} G(a-\alpha, b, c-\beta)\times \\[4mm]
&\times \left(z_1\cdot \partial_{x_1}\right)^{\alpha}\left(z_3\cdot \partial_{x_3}\right)^{\beta} (x_{12}^2)^{c-h-\beta}(x_{13}^2)^{b-h}(x_{23}^2)^{a-h-\alpha}\,,
\end{aligned}
\end{equation}
where $a+b+c=2h+\alpha+\beta$ and $G(a,b,c)$ was defined in (\ref{startriangle}).

We can use a conformal transformation to  fix the position of the scalar operator $x_2$ at infinity. For a scalar, this can be safely done without loss of information. Indeed, there is only one nonzero $\ell_{i}$ which controls both $z_1$ and $z_3$ and there is no $z_2$-dependence. If one does so, the integrand simplifies and the $x_{i2}^2$ drop out. The action of the derivatives can then be given in terms of known functions,
\begin{align}
    \left(z_1\cdot \partial_{x_1}\right)^{\alpha}\left(z_3\cdot \partial_{x_3}\right)^{\beta}(x_{13}^2)^{b-h}=&\,2^{\alpha} 2^{\beta}\frac{\Gamma(h+\alpha+\beta-b)}{\Gamma(h-b)}(x_{31}\cdot z_1)^{\alpha}(x_{13}\cdot z_3)^{\beta} (x_{13}^2)^{b-h-\alpha-\beta}\times\nonumber\\ &\times\,_2F_1\bigg(-\alpha,-\beta,b+1-h-\alpha-\beta; \frac{z_1\cdot z_3\, x_{13}^2}{2 x_{13}\cdot z_3 x_{13}\cdot z_1}\bigg)\,.
\end{align}
Putting everything together, we find 
\begin{align}
    &\langle \mathcal{O}_{\Delta_1,J_1} \phi_{\Delta_2} \mathcal{S}[\mathcal{O}_{\Delta_3, J_3}]\rangle ^{(\ell)}=\nonumber\\[4mm]
    &=\sum_{p=0}^{J_3-\ell}\sum_{q=0}^{\ell}\sum_{s=0}^{\ell-q}\sum_{t=0}^{p}\sum_{r=0}^{q}\sum_{w=0}^{\infty}\sum_{m=0}^{s+w}\binom{J_3-\ell}{p}\binom{\ell}{q}\binom{\ell-q}{s}\binom{p}{t}\binom{q}{r}\binom{s+w}{m}\times\nonumber\\[4mm]
    &(-1)^{J_3+r+s+t+2w-m}\, 2^{-J_3}\frac{\pi^{h}\Gamma\left(\frac{J_1+J_3+2r-2s+2t+\Delta_1-\Delta_2+\widetilde{\Delta}_3}{2}\right)\Gamma\left(\frac{J_1-J_3+2p-2t-\Delta_1+\Delta_2+\widetilde{\Delta}_3}{2}\right)}{\Gamma\left(\frac{J_1+J_3-2p-2q+2r-2s+2t+\Delta_1-\Delta_2+\Delta_3}{2}\right)\Gamma\left(\frac{J_1-J_3+2p-2t-\Delta_1+\Delta_2+\Delta_3}{2}\right)}\times\nonumber\\[4mm]
    &\frac{\Gamma\left(\Delta_3-h\right)}{\Gamma\left(1+w\right)\Gamma\left(p+q+\widetilde{\Delta}_3\right)}\frac{\left(-p-q\right)_{w}\left(-J_1+q-r+s\right)_w}{\left(\frac{2-J_1-J_3-2r+2s-2t-\Delta_1+\Delta_2-\widetilde{\Delta}_3}{2}\right)_{w}}\underbrace{\frac{H_{13}^{m}V_{1,23}^{J_1-m}V_{3,12}^{J_3-m}}{(x_{13}^{2})^{\frac{\Delta_1+J_1-\Delta_2+\widetilde{\Delta}_3+J_3}{2}}}}_{\textstyle \langle\mathcal{O}_{\Delta_1,J_1}\phi_{\Delta_2}\mathcal{O}_{\widetilde{\Delta}_3,J_3}\rangle^{(m)}}\,,
\end{align}
from which we can easily read the shadow coefficients associated with each possible three-point structure. One can check that this expression reproduces the results of~\cite{karateev:2018} as a special case.\footnote{Strictly speaking there is a $2^{-J_3}$ difference which follows from a different normalization of the two-point function.} It is worth stating that all the sums here have indeed a finite number of terms. This can be seen from the expression above by noticing that for sufficiently large $w$ the Pochhammer symbols in the numerator will vanish. 

One could have wanted to do instead the shadow transform of the scalar operator. That case is simpler as there is no need to deal with the contractions of indices as we did in the beginning of this subsection. Keeping $x_2$ at infinity, we have the following integral to do 
\begin{align}
&    \int d^d x_0  \frac{(x_{13}^2)^{\frac{-\Delta_1-J_1+\Delta_2-\Delta_3-J_3}{2}}}{(x_{01}^2)^{\frac{\Delta_1+J_1+\Delta_2-\Delta_3-J_3}{2}}(x_{03}^2)^{\frac{-\Delta_1-J_1+\Delta_2+\Delta_3+J_3}{2}}}V_{1,03}^{J_1-\ell}V_{3,10}^{J_3-\ell}H_{13}^{\ell}\,,
\end{align}
which can be integrated in the exact same way as before. This is a straightforward computation and we find
\begin{align}
    &\langle \mathcal{O}_{\Delta_1,J_1} \mathcal{S}[\phi_{\Delta_2}] \mathcal{O}_{\Delta_3, J_3}\rangle ^{(\ell)}=\nonumber\\[4mm]
    &\sum_{p=0}^{J_1-\ell}\sum_{q=0}^{J_3-\ell}\sum_{r=0}^{\ell}\sum_{s=0}^{\infty}\sum_{m=0}^{\ell+s-r}\binom{J_1-\ell}{p}\binom{J_3-\ell}{q}\binom{\ell}{r}\binom{\ell+s-r}{m}(-1)^{J_1+J_3-p+q-r+2s+\ell-m}\times\nonumber\\[4mm]
    &\frac{\pi^{h}\Gamma\left(J_1+J_3-p-q-2\ell+\Delta_2-h \right)\Gamma\left(\frac{-J_1+J_3+\Delta_1+\widetilde{\Delta}_2-\Delta_3}{2}\right)\Gamma\left(\frac{J_1-J_3-\Delta_1+\widetilde{\Delta}_2+\Delta_3}{2}\right)}{\Gamma\left(1+s\right)\Gamma\left(\widetilde{\Delta}_2\right)\Gamma\left(\frac{J_1+J_3-2p-2\ell+\Delta_1+\Delta_2-\Delta_3}{2}\right)\Gamma\left(\frac{J_1+J_3-2q-2\ell-\Delta_1+\Delta_2+\Delta_3}{2}\right)}\times\nonumber\\[4mm]
    &\frac{(-J_1+p+\ell)_s(-J_3+q+\ell)_s}{(1+h+p+q+2\ell-J_1-J_3-\Delta_2)_s}\underbrace{\frac{H_{13}^{m}V_{1,23}^{J_1-m}V_{3,12}^{J_3-m}}{(x_{13}^{2})^{\frac{\Delta_1+J_1-\widetilde{\Delta}_2+\Delta_3+J_3}{2}}}}_{\textstyle \langle\mathcal{O}_{\Delta_1,J_1}\phi_{\widetilde{\Delta}_2}\mathcal{O}_{\Delta_3,J_3}\rangle^{(m)}}\,.
\end{align}
The shadow coefficients computed in this way also reproduce the known results of~\cite{karateev:2018} in the appropriate restriction.

Lastly, let us comment on the more generic situation where all operators have spin, which is, of course, more complicated. Note that we were only able to write the action of the derivatives in such a compact form because we fixed $x_2$ to infinity. In the more general case, we are no longer able to naively set $x_2$ to infinity since we would lose control of $\ell_{1}$ and $\ell_{3}$. On the other hand, we can still successfully integrate the shadow transform in a case-by-case basis, but this becomes cumbersome for large values of spin. For completeness, let us write down the integral that remains after having dealt with the contraction of indices 
\begin{align}
    \int d^d x_0\, &\frac{(-1)^{\ell_1+\ell_2}(x_{12}^2)^{\frac{-\Delta_1-J_1-\Delta_2-J_2+\Delta_3-J_3+2\ell_{2}}{2}}}{2^{J_3}(x_{01}^2)^{\frac{\Delta_1+J_1-\Delta_2-J_2+\Delta_3+J_3}{2}}(x_{02}^2)^{\frac{-\Delta_1-J_1+\Delta_2+J_2+\Delta_3-J_3+2\ell_{2}}{2}}(x_{03}^2)^{\widetilde{\Delta}_3+J_3-\ell_2} (x_{13}^2)^{\ell_2}(x_{23}^2)^{J_3-\ell_1}}\times\nonumber\\[4mm]
    & \times H_{12}^{\ell_3} V_{1,20}^{J_1-\ell_2-\ell_3} V_{2,01}^{J_2-\ell_1-\ell_3} \Big(V_{3,02}\left(V_{2,01} x_{01}^2-x_{12}\cdot z_2 x_{02}^2\right)+\mathcal{H}_{0,3,2}x_{12}^2\Big)^{\ell_1}\times\nonumber\\[4mm]
    &\times \Big(V_{1,03}\left( V_{3,02}x_{02}^2 x_{13}^2+ V_{3,21}x_{03}^2 x_{12}^2-x_{13}\cdot z_3 x_{03}^2 x_{23}^2 \right)+\mathcal{H}_{0,1,3}x_{13}^2 x_{23}^2\Big)^{\ell_2}\nonumber\times\\[4mm]
    &\times \Big(V_{3,21}x_{03}^2 x_{12}^2+ V_{3,02}\left(x_{02}^2 x_{13}^2-x_{01}^2 x_{23}^2\right)\Big)^{J_3-\ell_1-\ell_2}\,,
\end{align}
where we assume that the shadow transform is done in the operator at $x_3$.
One can easily see that all the terms can be integrated in the same way as before
\begin{align}
\int d^d x_0\,& \frac{(x_{01}\cdot z_1)^{\alpha} (x_{02}\cdot z_2)^{\beta} (x_{03}\cdot z_3)^{\gamma}}{ (x_{01}^2)^{a}(x_{02}^2)^{b}(x_{03}^2)^{c}}=\frac{\Gamma(a-\alpha)}{2^{\alpha}\Gamma(a)}\frac{\Gamma(b-\beta)}{2^{\beta}\Gamma(b)}\frac{\Gamma(c-\gamma)}{2^{\gamma}\Gamma(c)} \,G(a-\alpha, b-\beta, c-\gamma)\nonumber \\[4mm]
&\times \left(z_1\cdot \partial_{x_1}\right)^{\alpha}\left(z_2\cdot \partial_{x_2}\right)^{\beta}\left(z_3\cdot \partial_{x_3}\right)^{\gamma} (x_{12}^2)^{c-h-\gamma}(x_{13}^2)^{b-h-\beta}(x_{23}^2)^{a-h-\alpha}\,,
\end{align}
where $a+b+c=2h+\alpha+\beta+\gamma$. 

This is all we need to successfully compute any shadow coefficient of a three-point function of three operators in spin $J_i$ representation, but we did not manage to find a simple and compact formula for the action of derivatives in the above expression. 
While one can use this formalism to compute the shadow coefficients of three spinning operators, in practice the procedure becomes too computationally expensive at large spin. It would be interesting to investigate if the weight-shifting formalism of~\cite{karateev:2018} offers a more efficient alternative.

\end{document}